\documentclass[numberedappendix]{emulateapj}
\usepackage{natbib}
\usepackage{bm,amsmath}
\usepackage{graphicx}
\singlespace
\shorttitle{\map\ 7-year Cosmological Interpretation}
\shortauthors{Komatsu et al.}
\newcommand{\map}    {{\sl WMAP}}

\newcommand{\fnlKS}  {f_{NL}^{\rm local}}
\newcommand{\fnleq}  {f_{NL}^{\rm equil}} 
\newcommand{\fnlor}  {f_{NL}^{\rm orthog}}

\newcommand{\bsrc}   {b_{src}}
\def\Nchan{N_{\rm chan}}
\def\Npix{N_{\rm pix}}
\def\Nt{N_{\rm tmpl}}
\def\hp{{\hat p}}
\def\L{{\mathcal L}}
\tabletypesize{\scriptsize}

\slugcomment{Accepted for Publication in the Astrophysical Journal
Supplement Series}
\begin{document}
\title{Seven-Year Wilkinson Microwave Anisotropy Probe 
(\map\altaffilmark{1}) Observations:\\
Cosmological Interpretation}
\author{
{{E. Komatsu}}\altaffilmark{2}, 
{{K. M. Smith}}\altaffilmark{3},
{{J. Dunkley}}\altaffilmark{4}, 
{{C. L. Bennett}}\altaffilmark{5},
{{B. Gold}}\altaffilmark{5},
{{G. Hinshaw}}\altaffilmark{6},
{{N. Jarosik}}\altaffilmark{7},
{{D. Larson}}\altaffilmark{5},
{{M. R. Nolta}}\altaffilmark{8},
{{L. Page}}\altaffilmark{7},
{{D. N. Spergel}}\altaffilmark{3,9},
{{M. Halpern}}\altaffilmark{10},
{{R. S. Hill}}\altaffilmark{11},
{{A. Kogut}}\altaffilmark{6},
{{M. Limon}}\altaffilmark{12},
{{S. S. Meyer}}\altaffilmark{13},
{{N. Odegard}}\altaffilmark{11},
{{G. S. Tucker}}\altaffilmark{14},
{{J. L. Weiland}}\altaffilmark{11},
{{E. Wollack}}\altaffilmark{6},
and 
{{E. L. Wright}}\altaffilmark{15}
}
\altaffiltext{1}{\map\ is the result of a partnership between Princeton 
                 University and NASA's Goddard Space Flight Center. Scientific 
                 guidance is provided by the \map\ Science Team.}
\altaffiltext{2}{Texas Cosmology Center and Department of Astronomy,
University of Texas, Austin, 2511 Speedway, RLM 15.306, Austin, TX 78712;
{komatsu@astro.as.utexas.edu}}
\altaffiltext{3}{{Dept. of Astrophysical Sciences, %
                    Peyton Hall, Princeton University, Princeton, NJ 08544-1001}}
\altaffiltext{4}{{Astrophysics, University of Oxford, %
                    Keble Road, Oxford, OX1 3RH, UK}}
\altaffiltext{5}{{Dept. of Physics \& Astronomy, %
                    The Johns Hopkins University, 3400 N. Charles St., %
                    Baltimore, MD  21218-2686}}
\altaffiltext{6}{{Code 665, NASA/Goddard Space Flight Center, %
                    Greenbelt, MD 20771}}
\altaffiltext{7}{{Dept. of Physics, Jadwin Hall, %
                    Princeton University, Princeton, NJ 08544-0708}}
\altaffiltext{8}{{Canadian Institute for Theoretical Astrophysics, %
                    60 St. George St, University of Toronto, %
                    Toronto, ON  Canada M5S 3H8}}
\altaffiltext{9}{{Princeton Center for Theoretical Physics, %
                    Princeton University, Princeton, NJ 08544}}
\altaffiltext{10}{{Dept. of Physics and Astronomy, University of %
                    British Columbia, Vancouver, BC  Canada V6T 1Z1}}
\altaffiltext{11}{{ADNET Systems, Inc., %
                    7515 Mission Dr., Suite A100 Lanham, Maryland 20706}}
\altaffiltext{12}{{Columbia Astrophysics Laboratory, %
                    550 W. 120th St., Mail Code 5247, New York, NY  10027-6902}}
\altaffiltext{13}{{Depts. of Astrophysics and Physics, KICP and EFI, %
                    University of Chicago, Chicago, IL 60637}}
\altaffiltext{14}{{Dept. of Physics, Brown University, %
                    182 Hope St., Providence, RI 02912-1843}}
\altaffiltext{15}{{UCLA Physics \& Astronomy, PO Box 951547, %
                    Los Angeles, CA 90095--1547}}
\begin{abstract}
 The combination of 7-year data from \map\  and improved
 astrophysical data rigorously 
 tests the standard cosmological model and places new constraints on its
 basic parameters and extensions. By combining the \map\ data with
 the latest distance measurements from the Baryon Acoustic Oscillations
 (BAO) in the distribution of galaxies \citep{percival/etal:2009} and
 the Hubble constant ($H_0$) measurement \citep{riess/etal:2009}, we
 determine the  parameters of the simplest 6-parameter
 $\Lambda$CDM model. The power-law index of the primordial
 power spectrum  is
 $n_s=0.968\pm 0.012$~(68\%~CL) for this
 data combination, a measurement that excludes the
 Harrison-Zel'dovich-Peebles spectrum by 99.5\%~CL. 
The other
 parameters, including those beyond the minimal set, are also consistent
 with, and improved  from, the 5-year results. We find
 no convincing deviations from the minimal model.
 The 7-year temperature
 power spectrum gives a better determination of the third acoustic peak,
 which results in a better determination of the redshift of the
 matter-radiation equality epoch. Notable examples of
 improved parameters are the total  mass of neutrinos,
 \ensuremath{\sum m_\nu < 0.58\ \mbox{eV}\ \mbox{(95\% CL)}}, and the
 effective number of neutrino species,
 \ensuremath{N_{\rm eff} = 4.34^{+ 0.86}_{- 0.88}}~(68\%~CL),
 which benefit from better determinations of the third peak and $H_0$. 
 The limit on a constant dark energy equation of state
 parameter from \map+BAO+$H_0$, without high-redshift Type Ia
 supernovae, is
 \ensuremath{w = -1.10\pm 0.14}~(68\%~CL). 
 We detect
 the effect of primordial helium on the temperature power spectrum and
 provide a new test of big bang nucleosynthesis by measuring
 $Y_p=0.326\pm 0.075$~(68\%~CL).
 We detect, and show on the map for the first time, the tangential and
 radial polarization patterns around hot and cold spots of temperature
 fluctuations, an important test of physical processes at $z=1090$ and
 the dominance of adiabatic scalar fluctuations. 
 The 7-year polarization data have
 significantly improved: we now detect the temperature-$E$-mode
 polarization cross power spectrum at 21$\sigma$, compared to
 13$\sigma$ from the 5-year data. With the 7-year temperature-$B$-mode
 cross power spectrum, the limit on a rotation of the polarization plane due
 to potential parity-violating effects has improved by 38\% to
 $\Delta\alpha=-1.1^\circ\pm 1.4^\circ~(\rm statistical)\pm
 1.5^\circ~(\rm systematic)$ (68\%~CL). 
 We report significant detections of the
 Sunyaev-Zel'dovich (SZ) effect at the locations of known clusters of
 galaxies. The measured SZ signal agrees well with the expected signal
 from the X-ray data on a cluster-by-cluster basis. However, it is
 a factor of 
 0.5 to 0.7 times the predictions from ``universal profile'' of Arnaud et al., 
 analytical models, and hydrodynamical simulations.
 We find, for the first time in the SZ effect, a significant difference
 between the 
 cooling-flow and non-cooling-flow clusters (or relaxed and non-relaxed
 clusters),  which can explain some of the discrepancy.  
 This lower amplitude is consistent
 with the lower-than-theoretically-expected 
 SZ power spectrum recently measured by the South Pole Telescope
 collaboration. 
\end{abstract}
\keywords{cosmic microwave background, cosmology: observations, early
universe, dark matter, space vehicles, space vehicles: instruments, 
instrumentation: detectors, telescopes}

\section{Introduction}
\label{sec:intro}
A simple cosmological model, a flat universe with nearly
scale-invariant adiabatic Gaussian fluctuations, has proven to be a
remarkably good fit to ever improving cosmic microwave background (CMB)
data \citep{hinshaw/etal:2009,reichardt/etal:2009,brown/etal:2009},
large-scale structure data \citep{reid/etal:2010,percival/etal:2009},
supernova data \citep{hicken/etal:2009,kessler/etal:2009}, cluster measurements
\citep{vikhlinin/etal:2009,mantz/etal:2010}, distance measurements
\citep{riess/etal:2009}, measurements of strong
\citep{suyu/etal:2010,fadely/etal:prep} 
and weak \citep{massey/etal:2007,fu/etal:2008,schrabback/etal:prep} gravitational lensing effects.

Observations of CMB have been playing an essential
role in testing the model and constraining its basic parameters.  
The \map\ satellite
\citep{bennett/etal:2003,bennett/etal:2003b} has been measuring
temperature and polarization anisotropies of the CMB over the full sky since
2001. With
7 years of integration, 
the errors in the temperature spectrum at each multipole  are dominated by
cosmic variance  (rather than by noise) up to $l\approx 550$, and
the signal-to-noise at each multipole exceeds unity up to $l\approx 900$
\citep{larson/etal:prep}.  
The power spectrum of primary CMB on smaller angular scales has been
measured by other experiments up to $l\approx 3000$
\citep{reichardt/etal:2009,brown/etal:2009,lueker/etal:2010,fowler/etal:2010}.  

The polarization data show the most dramatic improvements over our
earlier \map\ results: the 
temperature-polarization cross power spectra measured by \map\ at
$l\gtrsim 10$ are still dominated by noise, and the errors in the 7-year
cross power spectra have improved by nearly 40\% compared to the 5-year
cross power spectra.
While the error in the power 
spectrum of the cosmological $E$-mode polarization
\citep{seljak/zaldarriaga:1997,kamionkowski/kosowsky/stebbins:1997}
averaged over $l=2-7$ is cosmic-variance limited, individual multipoles
are not yet cosmic-variance limited. Moreover, the
cosmological $B$-mode 
polarization has not been detected 
\citep{nolta/etal:2009,komatsu/etal:2009,brown/etal:2009,chiang/etal:2010}. 

The temperature-polarization (TE and TB) power spectra offer
unique tests of the standard model. The TE spectrum can be predicted
given the cosmological constraints from the temperature power spectrum,
and the TB spectrum is predicted to vanish in a parity-conserving
universe. They also provide a clear physical picture of how the CMB polarization
is created from quadrupole temperature anisotropy. We show the
success of the standard model in an even 
more striking way by measuring this correlation in map space, rather
than in harmonic space. 

The constraints on the basic 6 parameters of a flat $\Lambda$CDM model
(see Table~\ref{tab:summary}),
as well as those on the parameters beyond the minimal set (see
Table~\ref{tab:deviation}), continue to 
improve with the 7-year \map\ temperature and polarization data,
combined with improved external astrophysical data sets.
In this paper, we shall give an update on the cosmological
parameters, as determined from the latest cosmological data set.
Our best estimates
of the cosmological parameters are presented in the last 
columns of Table~\ref{tab:summary} and \ref{tab:deviation} under the name
``\map+BAO+$H_0$.'' While this is the minimal combination of
robust data sets such that adding other data sets does not significantly
improve most parameters, the other data combinations
provide better limits than \map+BAO+$H_0$ in some cases.
For example, adding the small-scale CMB data improves the limit on the
primordial helium abundance, $Y_p$ 
(see Table~\ref{tab:yhe} and Section~\ref{sec:helium}),
the supernova data are needed to improve limits on properties of dark energy
(see Table~\ref{tab:darkenergy} and Section~\ref{sec:darkenergy}), and
the power spectrum of Luminous Red Galaxies (LRGs; see
Section~\ref{sec:lrg}) improves limits on properties of neutrinos (see
footnotes g, h, and i in Table~\ref{tab:deviation} and
Sections~\ref{sec:massnu} and \ref{sec:neff}). 

The CMB can also be used to probe the abundance as well as the
physics of clusters of galaxies, via the SZ
effect \citep{zeldovich/sunyaev:1969,sunyaev/zeldovich:1972}. In this
paper, we shall present the \map\ measurement of the averaged profile of
SZ effect measured towards the directions of known clusters of galaxies,
and discuss implications of the \map\ measurement for the very
small-scale ($l\gtrsim 3000$) power spectrum recently measured by the
South Pole Telescope \citep[SPT;][]{lueker/etal:2010} and 
Atacama Cosmology Telescope \citep[ACT;][]{fowler/etal:2010}
collaborations. 

This paper is one of six papers on the analysis of the
\map\ 7-year data: \citet{jarosik/etal:prep} report on the data
processing, map-making, and systematic error limits; 
\citet{gold/etal:prep} on the modeling, understanding, and subtraction of the
temperature and polarized foreground emission; 
\citet{larson/etal:prep} on the measurements of
the temperature and polarization power spectra, 
extensive testing of the parameter estimation methodology by Monte Carlo
simulations, and the
cosmological parameters inferred from the \map\ data alone;
\citet{bennett/etal:prep} on the assessments of statistical significance
of various ``anomalies'' in the \map\ temperature map reported in the
literature; and \citet{weiland/etal:prep} on \map's measurements of the
brightnesses of planets and various celestial calibrators. 

This paper is organized as follows. 
In Section~\ref{sec:pol}, we present results from the new method of
analyzing the polarization patterns around temperature hot and cold spots.
In Section~\ref{sec:analysis}, we briefly
summarize new aspects of our analysis of the \map\ 7-year temperature
and polarization data, as well as improvements from the 
5-year data. 
In Section~\ref{sec:parameters}, we present updates on various
cosmological parameters, except for dark energy. We explore the nature
of dark energy in Section~\ref{sec:darkenergy}.
In Section~\ref{sec:NG}, we present limits on primordial non-Gaussianity
parameters $f_{\rm NL}$. 
In Section~\ref{sec:SZ}, we report  detection, characterization, and
interpretation of the SZ effect toward locations of
known clusters of galaxies. We conclude in Section~\ref{sec:conclusion}.

\begin{deluxetable*}{lccccc}
\tablecolumns{6}
\small
\tablewidth{0pt}
\tablecaption{%
Summary of the cosmological parameters of $\Lambda$CDM model\tablenotemark{a}
}
\tablehead{\colhead{Class} &
\colhead{Parameter}
&\colhead{\map\ 7-year ML\tablenotemark{b}}
&\colhead{\map+BAO+$H_0$ ML}
&\colhead{\map\ 7-year Mean\tablenotemark{c}}
&\colhead{\map+BAO+$H_0$ Mean}
}
\startdata
Primary &
$100\Omega_bh^2$
&2.227 
&2.253 
&$2.249^{+0.056}_{-0.057}$ 
&$2.255\pm 0.054$ 
\nl
&
$\Omega_ch^2$
&0.1116 
&0.1122 
&$0.1120\pm 0.0056$ 
&$0.1126\pm 0.0036$ 
\nl
&
$\Omega_\Lambda$
&0.729 
&0.728 
&$0.727^{+0.030}_{-0.029}$ 
&$0.725\pm 0.016$ 
\nl
&
$n_s$
&0.966 
&0.967 
&$0.967\pm 0.014$ 
&$0.968\pm 0.012$ 
\nl
&
$\tau$
&0.085 
&0.085 
&$0.088\pm 0.015$ 
&$0.088\pm 0.014$ 
\nl
&
$\Delta^2_{\cal R}(k_0)\tablenotemark{d}$
&$2.42\times 10^{-9}$ 
&$2.42\times 10^{-9}$ 
&$(2.43\pm 0.11)\times 10^{-9}$ 
&$(2.430\pm 0.091)\times 10^{-9}$ 
\nl
\hline
Derived &
$\sigma_8$
&0.809 
&0.810 
&$0.811^{+0.030}_{-0.031}$ 
&$0.816\pm 0.024$ 
\nl
&
$H_0$
&$70.3~{\rm km/s/Mpc}$ 
&$70.4~{\rm km/s/Mpc}$ 
&$70.4\pm 2.5$~{km/s/Mpc} 
&$70.2\pm 1.4$~{km/s/Mpc} 
\nl
&
$\Omega_b$
&0.0451 
&0.0455 
&$0.0455\pm 0.0028$ 
&$0.0458\pm 0.0016$ 
\nl
&
$\Omega_c$
&0.226 
&0.226 
&$0.228\pm 0.027$ 
&$0.229\pm 0.015$ 
\nl
&
$\Omega_mh^2$
&0.1338 
&0.1347 
&$0.1345^{+0.0056}_{-0.0055}$ 
&$0.1352\pm 0.0036$ 
\nl
&
$z_{\rm reion}$\tablenotemark{e}
&10.4 
&10.3 
&$10.6\pm 1.2$ 
&$10.6\pm 1.2$ 
\nl 
&
$t_0$\tablenotemark{f}
&13.79~Gyr 
&13.76~Gyr 
&$13.77\pm 0.13$~Gyr 
&$13.76\pm 0.11$~Gyr 
\enddata
\tablenotetext{a}{The parameters listed here are derived using
 the RECFAST 1.5 and version 4.1 of the \map\ likelihood code. All the
 other parameters in the other tables are derived using the RECFAST
 1.4.2 and version 4.0 of the \map\ likelihood code, unless stated otherwise.
 The difference is small. See Appendix~\ref{app:comparison} for comparison.} 
\tablenotetext{b}{\citet{larson/etal:prep}. ``ML'' refers to the Maximum Likelihood parameters.}
\tablenotetext{c}{\citet{larson/etal:prep}. ``Mean'' refers to the mean
 of the posterior distribution of each parameter. The quoted errors show
 the 68\%~confidence levels (CL).} 
\tablenotetext{d}{$\Delta^2_{\cal R}(k)=k^3P_{\cal R}(k)/(2\pi^2)$ and
 $k_0=0.002~{\rm Mpc}^{-1}$.}
\tablenotetext{e}{``Redshift of reionization,'' if the universe was
 reionized instantaneously from the neutral state to the fully ionized
 state at $z_{\rm reion}$. Note that these values are somewhat different
 from those in Table 1 of \citet{komatsu/etal:2009}, largely because of
 the changes in the treatment of reionization history in the Boltzmann code
 {\sf CAMB} \citep{lewis:2008}.} 
\tablenotetext{f}{The present-day age of the universe.}
\label{tab:summary}
\end{deluxetable*}

\begin{deluxetable*}{lccccc}
\tablecolumns{6}
\small
\tablewidth{0pt}
\tablecaption{%
Summary of the 95\% confidence limits on deviations from the simple 
(flat, Gaussian, adiabatic, power-law) $\Lambda$CDM model except for
 dark energy parameters
}
\tablehead{
\colhead{Section}&
\colhead{Name}
&\colhead{Case}
&\colhead{\map\ 7-year}
&\colhead{\map+BAO+SN\tablenotemark{a}}
&\colhead{\map+BAO+$H_0$}}
\startdata
Section~\ref{sec:GW}&
Grav. Wave\tablenotemark{b}
& No Running Ind.
& $r<0.36$\tablenotemark{c}
& $r<0.20$
& $r<0.24$
\nl
Section~\ref{sec:running}&
Running Index
& No Grav. Wave
& $-0.084<dn_s/d\ln k<0.020$\tablenotemark{c}
& $-0.065<dn_s/d\ln k<0.010$
& $-0.061<dn_s/d\ln k<0.017$
\nl
Section~\ref{sec:OK}&
Curvature 
&$w=-1$
&N/A
&$-0.0178<\Omega_k<0.0063$
&$-0.0133<\Omega_k< 0.0084$
\nl
Section~\ref{sec:AD} &
Adiabaticity
&Axion
&$\alpha_0<0.13$\tablenotemark{c}
&$\alpha_0<0.064$
&$\alpha_0<0.077$
\nl 
&
&Curvaton
&$\alpha_{-1}<0.011$\tablenotemark{c}
&$\alpha_{-1}<0.0037$
&$\alpha_{-1}<0.0047$
\nl
Section~\ref{sec:TB} &
Parity Violation
& Chern-Simons\tablenotemark{d}
& $-5.0^\circ<\Delta\alpha<2.8^\circ$\tablenotemark{e}
&N/A
&N/A
\nl
Section~\ref{sec:massnu}&
Neutrino Mass\tablenotemark{f}
& $w=-1$
& $\sum m_\nu<1.3~{\rm eV}$\tablenotemark{c}
& $\sum m_\nu<0.71~{\rm eV}$
& $\sum m_\nu<0.58~{\rm eV}$\tablenotemark{g} 
\nl 
&
& $w\neq-1$
& $\sum m_\nu<1.4~{\rm eV}$\tablenotemark{c}
& $\sum m_\nu<0.91~{\rm eV}$
& $\sum m_\nu<1.3~{\rm eV}$\tablenotemark{h}
\nl 
Section~\ref{sec:neff}&
Relativistic Species
& $w=-1$
& $N_{\rm eff}>2.7$\tablenotemark{c}
& N/A
& \ensuremath{4.34^{+ 0.86}_{- 0.88}}~\mbox{(68\%~CL)}\tablenotemark{i}
\nl
Section~\ref{sec:NG} &
Gaussianity\tablenotemark{j}
&Local
&$-10<\fnlKS<74$\tablenotemark{k}
&N/A
&N/A
\nl
&
&Equilateral
&$-214<\fnleq<266$
&N/A
&N/A
\nl
&
&Orthogonal
&$-410<\fnlor<6$
&N/A
&N/A
\enddata
\tablenotetext{a}{``SN'' denotes the ``Constitution'' sample of Type
 Ia supernovae compiled by \citet{hicken/etal:2009}, which is an
 extension of the ``Union'' sample \citep{kowalski/etal:2008} that we
 used for the 5-year ``\map+BAO+SN'' parameters presented in
 \citet{komatsu/etal:2009}. Systematic errors in the
 supernova data are not included. While the parameters in this column can be
 compared directly to the 5-year \map+BAO+SN parameters, they may not be
 as robust as the ``\map+BAO+$H_0$'' parameters, as the other compilations of
 the supernova data do not give the same answers
 \citep{hicken/etal:2009,kessler/etal:2009}. See Section~\ref{sec:sn}
 for more discussion. The SN data will be used to put limits on dark
 energy properties. See Section~\ref{sec:darkenergy} and
 Table~\ref{tab:darkenergy}.} 
\tablenotetext{b}{In the form of the tensor-to-scalar
 ratio, $r$, at $k=0.002~{\rm Mpc}^{-1}$.}
\tablenotetext{c}{\citet{larson/etal:prep}.}
\tablenotetext{d}{For an interaction of the form given by
$[\phi(t)/M]F_{\alpha\beta}\tilde{F}^{\alpha\beta}$, the polarization
 rotation angle is $\Delta\alpha=M^{-1}\int \frac{dt}{a} \dot{\phi}$.}
\tablenotetext{e}{The 68\%~CL limit is $\Delta\alpha= -1.1^\circ \pm 1.4^\circ~({\rm stat.}) \pm 1.5^\circ~({\rm syst.})$, where the first error is
 statistical and the second error is  systematic.}
\tablenotetext{f}{$\sum m_\nu=94(\Omega_\nu h^2)~{\rm eV}$.}
\tablenotetext{g}{For \map+LRG+$H_0$, $\sum m_\nu<0.44~{\rm eV}$.}
\tablenotetext{h}{For \map+LRG+$H_0$, $\sum m_\nu<0.71~{\rm eV}$.}
\tablenotetext{i}{The 95\% limit is $2.7<N_{\rm eff}<6.2$. For \map+LRG+$H_0$,
 $N_{\rm eff}= 4.25\pm 0.80$ (68\%) and $2.8<N_{\rm eff}<5.9$ (95\%).}
\tablenotetext{j}{V$+$W map masked by the {\it KQ75y7} mask. The Galactic
 foreground templates are marginalized over.}
\tablenotetext{k}{When combined with the limit on $\fnlKS$ from {\sl SDSS},
 $-29<\fnlKS<70$ \citep{slosar/etal:2008}, we find $-5<\fnlKS<59$.}
\label{tab:deviation}
\end{deluxetable*}

\begin{deluxetable}{lcc}
\tablecolumns{3}
\small
\tablewidth{0pt}
\tablecaption{%
Primordial helium Abundance\tablenotemark{a}
}
\tablehead{
&\colhead{\map\ only}
&\colhead{\map+ACBAR+QUaD}
}
\startdata
$Y_p$
& $<0.51$~(95\%~CL)
& $Y_p=0.326\pm 0.075$~(68\%~CL)\tablenotemark{b} 
\enddata
\tablenotetext{a}{See Section~\ref{sec:helium}.}
\tablenotetext{b}{
The 95\% CL limit is $0.16<Y_p<0.46$. 
For \map+ACBAR+QUaD+LRG+$H_0$,
 \ensuremath{Y_{\rm He} = 0.349\pm 0.064}~(68\%~CL)
 and $0.20<Y_p<0.46$~(95\%~CL).}
\label{tab:yhe}
\end{deluxetable}

\begin{deluxetable*}{lccccc}
\tablecolumns{6}
\small
\tablewidth{0pt}
\tablecaption{%
Summary of the 68\% limits on dark energy properties from \map\ combined with
 other data sets
}
\tablehead{
\colhead{Section}&
\colhead{Curvature}&
\colhead{Parameter}&
\colhead{+BAO+$H_0$}&
\colhead{+BAO+$H_0$+$D_{\Delta t}$\tablenotemark{a}}&
\colhead{+BAO+SN\tablenotemark{b}}
}
\startdata
Section~\ref{sec:w_flat}
&$\Omega_k=0$
&Constant $w$
&\ensuremath{-1.10\pm 0.14}
&\ensuremath{-1.08\pm 0.13}
&\ensuremath{-0.980\pm 0.053}
\nl
Section~\ref{sec:w_curve}
&$\Omega_k\neq 0$
&Constant $w$
&\ensuremath{-1.44\pm 0.27}
&\ensuremath{-1.39\pm 0.25}
&\ensuremath{-0.999^{+ 0.057}_{- 0.056}}
\nl
&
& $\Omega_k$
&\ensuremath{-0.0125^{+ 0.0064}_{- 0.0067}}
&\ensuremath{-0.0111^{+ 0.0060}_{- 0.0063}}
&\ensuremath{-0.0057^{+ 0.0067}_{- 0.0068}}
\nl
\hline
&
&
&+$H_0$+SN
&+BAO+$H_0$+SN
&+BAO+$H_0$+$D_{\Delta t}$+SN
\nl
\hline
Section~\ref{sec:w0wa}
&
$\Omega_k=0$
&$w_0$
& $-0.83\pm 0.16$
& $-0.93\pm 0.13$
& $-0.93\pm 0.12$
\nl
&
&$w_a$
& $-0.80^{+0.84}_{-0.83}$
& $-0.41^{+0.72}_{-0.71}$
& $-0.38^{+0.66}_{-0.65}$
\enddata
\tablenotetext{a}{``$D_{\Delta t}$'' denotes the time-delay distance to
 the lens system B1608+656 at $z=0.63$ measured by
 \citet{suyu/etal:2010}. See Section~\ref{sec:timedelay} for details.}
\tablenotetext{b}{``SN'' denotes the ``Constitution'' sample of Type
 Ia supernovae compiled by \citet{hicken/etal:2009}, which is an
 extension of the ``Union'' sample \citep{kowalski/etal:2008} that we
 used for the 5-year ``\map+BAO+SN'' parameters presented in
 \citet{komatsu/etal:2009}. Systematic errors in the
 supernova data are not included.}
\label{tab:darkenergy}
\end{deluxetable*}

\section{CMB Polarization on the Map}
\label{sec:pol}

\subsection{Motivation}
Electron-photon scattering converts quadrupole temperature anisotropy in
the CMB at the decoupling epoch, $z=1090$, into linear polarization
\citep{rees:1968,basko/polnarev:1980,kaiser:1983,bond/efstathiou:1984,polnarev:1985,bond/efstathiou:1987,harari/zaldarriaga:1993}. This
produces a correlation between the temperature pattern and the
polarization pattern
\citep{coulson/crittenden/turok:1994,crittenden/coulson/turok:1995}. 
Different mechanisms for generating fluctuations produce distinctive
correlated patterns in temperature and polarization:
\begin{itemize}
\item[1.] Adiabatic scalar fluctuations predict a radial polarization
      pattern around temperature cold spots and a tangential pattern
      around temperature hot spots on angular scales greater than the
	  horizon size at the decoupling epoch, $\gtrsim 2^\circ$. 
On angular scales smaller than the sound horizon size at the
decoupling epoch, {\it both} radial {\it and} tangential patterns are
	  formed around both hot and cold spots, as the acoustic
	  oscillation of the CMB modulates the polarization pattern
	  \citep{coulson/crittenden/turok:1994}. 
 As we have not seen any evidence for
non-adiabatic fluctuations \citep[][see Section~\ref{sec:AD} for the
7-year limits]{komatsu/etal:2009}, in this section we
shall assume that fluctuations are purely adiabatic.  
\item[2.] Tensor fluctuations predict the opposite pattern: the
	  temperature cold spots are surrounded by a tangential
	  polarization pattern, while the hot spots are surrounded by a
	  radial pattern \citep{crittenden/coulson/turok:1995}. Since
	  there is no acoustic oscillation for tensor modes, there is no
	  modulation of polarization patterns around temperature spots
	  on small angular 
	  scales. We do not expect this contribution to be visible in
	  the \map\ data, given the noise level.
\item[3.] Defect models predict that there should be minimal correlations
      between temperature and polarization on $2^\circ \lesssim \theta\lesssim
10^\circ$ \citep{seljak/pen/turok:1997}. The detection of large-scale
      temperature polarization fluctuations rules out any causal models
      as the primary mechanism for generating the CMB fluctuations
      \citep{spergel/zaldarriaga:1997}. This implies that the
      fluctuations were either generated during an accelerating phase in
      the early universe or were present at the time of the initial singularity.
\end{itemize}

This section presents the first direct measurement of the predicted
pattern of adiabatic scalar fluctuations in CMB polarization maps. We stack
maps of Stokes $Q$ and $U$ around temperature hot and cold spots to show the
expected polarization pattern at the statistical significance level of
8$\sigma$.  While we have detected the TE correlations in the first year
data \citep{kogut/etal:2003}, we present here the direct real space
pattern around hot and cold spots.  In Section~\ref{sec:polarizationdiscussion},
we discuss the relationship between the two measurements. 

\subsection{Measuring Peak-Polarization Correlation}

We first identify temperature hot (or cold) spots, and then stack the
polarization data (i.e., Stokes $Q$ and $U$) on the locations of the
spots. As we shall show below, the resulting polarization data is
equivalent to the temperature {\it peak}-polarization correlation function
which is similar to, but different in an important way from, the
temperature-polarization correlation function.  

\subsubsection{$Q_r$ and   $U_r$: Transformed Stokes Parameters}

\begin{figure}[t]
\centering \noindent
\includegraphics[width=7cm]{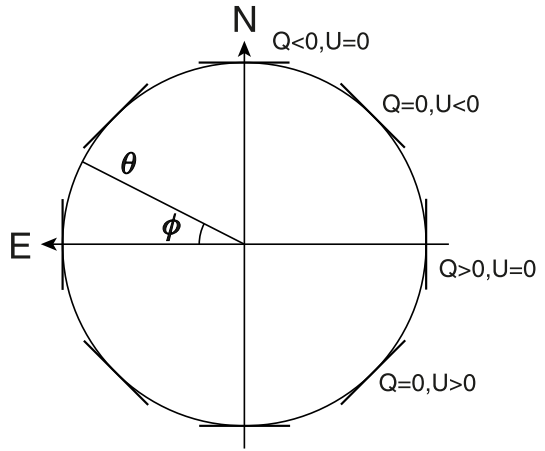}
\caption{%
Coordinate system for Stokes $Q$ and $U$. We use Galactic coordinates with north
 up and east left. In this example, $Q_r$ is always negative, and $U_r$
 is always zero. When $Q_r>0$ and $U_r=0$, the polarization pattern is radial.
} 
\label{fig:directions}
\end{figure}

Our definitions of Stokes $Q$ and $U$ follow that of
\citet{kogut/etal:2003}: the polarization that is parallel to the
Galactic meridian is $Q>0$ and $U=0$. Starting from this, the
polarization that is rotated  by $45^\circ$ from east to west 
(clockwise, as seen by an observer on Earth looking up at the sky)
has $Q=0$
and $U>0$,  that perpendicular to the Galactic meridian has $Q<0$ and
$U=0$, and that rotated further by $45^\circ$ from east to west has
$Q=0$ and $U<0$. With one more rotation we go back to $Q>0$ and
$U=0$. We show this in Figure~\ref{fig:directions}.  

As the predicted polarization pattern around temperature spots is either
radial or tangential, we find it most convenient to work with $Q_r$
and $U_r$ first introduced by
\citet{kamionkowski/kosowsky/stebbins:1997}: 
\begin{eqnarray}
\label{eq:Qrdef}
Q_r(\bm{\theta}) &=&
 -Q(\bm{\theta})\cos(2\phi)-U(\bm{\theta})\sin(2\phi),\\
\label{eq:Urdef}
U_r(\bm{\theta}) &=&
 Q(\bm{\theta})\sin(2\phi)-U(\bm{\theta})\cos(2\phi).
\end{eqnarray}
These transformed Stokes parameters are defined with respect to the new
coordinate system that is rotated by $\phi$, and thus they are defined
with respect to the line connecting the temperature spot at the center
of the coordinate and the polarization at an angular distance $\theta$
from the center (also see
Figure~\ref{fig:directions}). 
Note that we have used the small-angle (flat-sky) approximation for
simplicity of the algebra. This approximation is justified as we are
interested in relatively small angular scales, $\theta <5^\circ$. 

The above definition of $Q_r$ is equivalent to the so-called
``tangential shear'' statistic used by the weak gravitational lensing
community. By following what has been already done for the tangential
shear, we can find the necessary formulae for $Q_r$ and
$U_r$. Specifically, we shall follow the derivations given in
\citet{jeong/komatsu/jain:2009}. 

With the small-angle approximation, $Q$ and $U$ are related to the $E$-
and $B$-mode polarization in Fourier space 
\citep{seljak/zaldarriaga:1997,kamionkowski/kosowsky/stebbins:1997b} as
\begin{eqnarray}
-Q(\bm{\theta}) &=& \int\frac{d^2{\mathbf l}}{(2\pi)^2}
\left[E_{\mathbf l}\cos(2\varphi)-B_{\mathbf
 l}\sin(2\varphi)\right]e^{i{\mathbf l}\cdot{\bm{\theta}}},\\
-U(\bm{\theta}) &=& \int\frac{d^2{\mathbf l}}{(2\pi)^2}
\left[E_{\mathbf l}\sin(2\varphi)+B_{\mathbf
 l}\cos(2\varphi)\right]e^{i{\mathbf l}\cdot{\bm{\theta}}},
\end{eqnarray}
where $\varphi$ is the angle between ${\mathbf l}$ and the line of
Galactic latitude, ${\mathbf l}=(l\cos\varphi,l\sin\varphi)$. 
Note that we have included the negative signs on the left hand side because
our sign convention for the Stokes parameters is opposite of that
used in equation~(38) of \citet{zaldarriaga/seljak:1997}.
The transformed Stokes parameters are
given by
\begin{eqnarray}
\nonumber
- Q_r(\bm{\theta}) &=&
-\int\frac{d^2{\mathbf l}}{(2\pi)^2}
\left\{E_{\mathbf l}\cos[2(\phi-\varphi)]\right.\\
& &\qquad\qquad \left.+B_{\mathbf
 l}\sin[2(\phi-\varphi)]\right\}e^{i{\mathbf l}\cdot{\bm{\theta}}},\\
\nonumber
-U_r(\bm{\theta}) &=&
\int\frac{d^2{\mathbf l}}{(2\pi)^2}
\left\{E_{\mathbf l}\sin[2(\phi-\varphi)]\right.\\
& &\qquad\qquad \left.-B_{\mathbf
 l}\cos[2(\phi-\varphi)]\right\}e^{i{\mathbf l}\cdot{\bm{\theta}}}.
\end{eqnarray}

The stacking of $Q_r$ and $U_r$ at the locations of temperature {\it
peaks} can be written as
\begin{eqnarray}
 \langle Q_r\rangle(\bm{\theta}) = \frac1{N_{\rm pk}}
\int d^2\hat{\mathbf n}M(\hat{\mathbf n})\langle n_{\rm
pk}(\hat{\mathbf{n}})Q_r(\hat{\mathbf{n}}+\hat{\bm{\theta}})\rangle,\\
\langle U_r\rangle(\bm{\theta}) = \frac1{N_{\rm pk}}
\int d^2\hat{\mathbf n}M(\hat{\mathbf n})\langle n_{\rm
pk}(\hat{\mathbf{n}})U_r(\hat{\mathbf{n}}+\hat{\bm{\theta}})\rangle,
\end{eqnarray}
where the angle bracket, $\langle\dots \rangle$, denotes the average
over the locations of peaks, 
$n_{\rm pk}(\hat{\mathbf{n}})$ is the surface number density of
peaks (of the temperature fluctuation) at the location
$\hat{\mathbf{n}}$, $N_{\rm pk}$ is the 
total number of temperature peaks used in the stacking analysis,
and $M(\hat{\mathbf n})$ is equal to 0 at the masked pixels and 1
otherwise. Defining the density contrast of peaks, $\delta_{\rm
pk}\equiv n_{\rm pk}/\bar{n}_{\rm pk}-1$, we find
\begin{eqnarray}
\label{eq:Qr1}
 \langle Q_r\rangle(\bm{\theta}) = \frac1{f_{\rm sky}}
\int \frac{d^2\hat{\mathbf n}}{4\pi}M(\hat{\mathbf n})\langle \delta_{\rm
pk}(\hat{\mathbf{n}})Q_r(\hat{\mathbf{n}}+\hat{\bm{\theta}})\rangle,\\
\label{eq:Ur1}
\langle U_r\rangle(\bm{\theta}) = \frac1{f_{\rm sky}}
\int  \frac{d^2\hat{\mathbf n}}{4\pi}M(\hat{\mathbf n})\langle \delta_{\rm
pk}(\hat{\mathbf{n}})U_r(\hat{\mathbf{n}}+\hat{\bm{\theta}})\rangle,
\end{eqnarray}
where $f_{\rm sky}\equiv \int M(\hat{\mathbf n})d^2\hat{\mathbf
n}/(4\pi)$ is the fraction of sky outside of the mask, and we have used
$N_{\rm pk}=4\pi f_{\rm sky}\bar{n}_{\rm pk}$.
	
In Appendix~\ref{app:stacking}, we use the statistics of peaks of Gaussian
random fields to relate $\langle Q_r\rangle$ 
to the temperature-$E$-mode polarization cross power spectrum 
$C_l^{\rm TE}$, $\langle U_r\rangle$ to the temperature-$B$-mode
polarization cross power spectrum $C_l^{\rm TB}$, and the stacked
temperature profile, $\langle T\rangle$, to the temperature power
spectrum $C_l^{\rm TT}$. We find
\begin{eqnarray}
\label{eq:Qr}
 \langle Q_r\rangle(\theta)
&=& 
-\int\frac{ldl}{2\pi}
W_l^TW_l^P(\bar{b}_\nu+\bar{b}_\zeta l^2)C_l^{\rm TE}J_2(l\theta),\\
\label{eq:Ur}
 \langle U_r\rangle(\theta)
&=& 
-\int\frac{ldl}{2\pi}
W_l^TW_l^P(\bar{b}_\nu+\bar{b}_\zeta l^2)C_l^{\rm TB}J_2(l\theta),\\
\label{eq:T}
 \langle T\rangle(\theta)
&=& 
\int\frac{ldl}{2\pi}
(W_l^T)^2(\bar{b}_\nu+\bar{b}_\zeta l^2)C_l^{\rm TT}J_0(l\theta),
\end{eqnarray}
where $W_l^T$ and $W_l^P$ are the harmonic transform of window
functions, which are a 
combination of the experimental beam, pixel window, and any other additional
smoothing applied to the temperature and polarization data,
respectively, and $\bar{b}_\nu+\bar{b}_\zeta l^2$ is the
``scale-dependent bias'' of peaks 
found by \citet{desjacques:2008} averaged over peaks. See
Appendix~\ref{app:stacking} for details.

\subsubsection{Prediction  and Physical Interpretation}

\begin{figure*}[t]
\centering \noindent
\includegraphics[width=16cm]{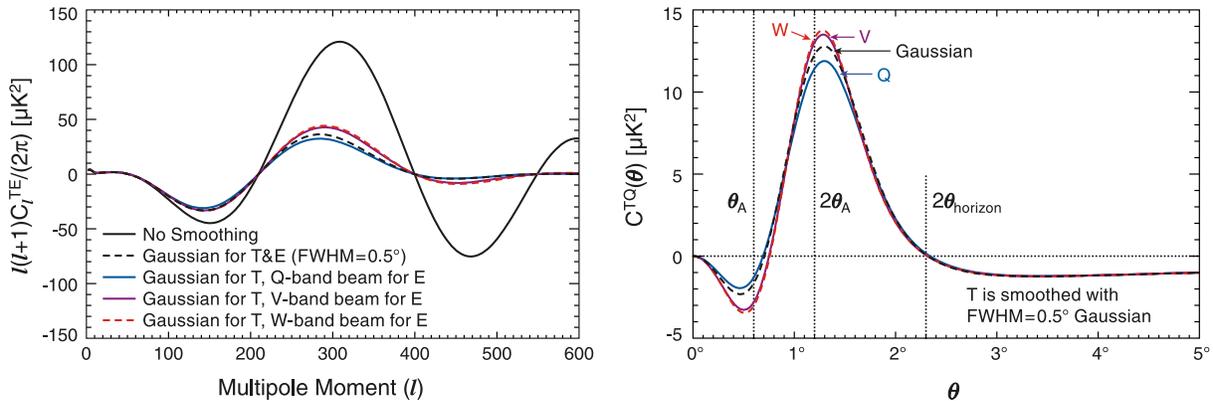}
\caption{%
Temperature-polarization cross correlation with various smoothing
 functions. 
(Left) The TE power spectrum with no smoothing is shown in the black solid
 line. For the other curves, the temperature is always smoothed with a 
 $0.5^\circ$ (FWHM) Gaussian, whereas the polarization is smoothed with
 either the same
 Gaussian (black dashed), Q-band beam (blue solid), V-band beam (purple
 solid), or W-band beam (red dashed). 
(Right) The corresponding spatial temperature-$Q_r$ correlation functions. The
 vertical dotted lines indicate (from left to right): the acoustic
 scale, $2\times$the acoustic scale, and $2\times$the horizon size,
 all evaluated at the decoupling epoch. 
} 
\label{fig:clte}
\end{figure*} 
\begin{figure*}[t]
\centering \noindent
\includegraphics[width=16cm]{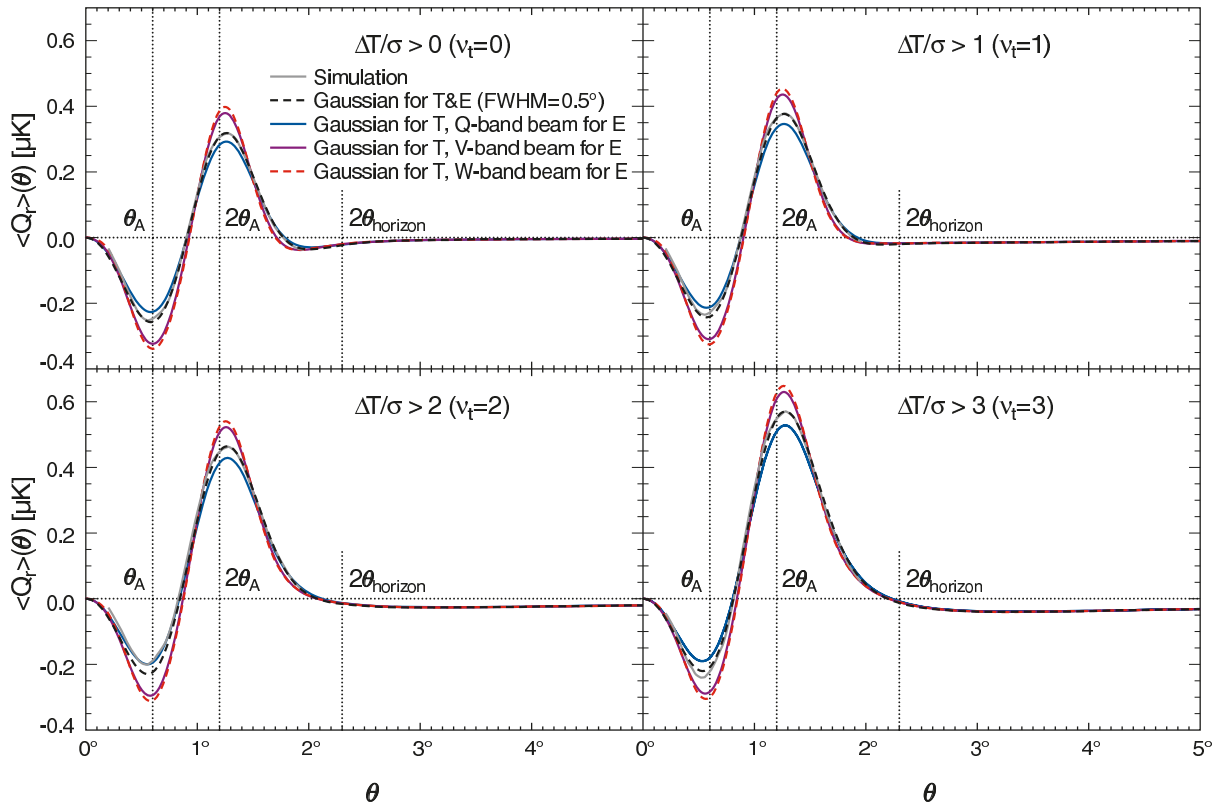}
\caption{%
Predicted temperature peak-polarization cross correlation, as measured by the
 stacked profile of the transformed Stokes $Q_r$, computed from
 equation~(\ref{eq:Qr}) for 
 various values of the threshold peak 
 heights.  The temperature is always smoothed with a $0.5^\circ$ (FWHM) Gaussian, whereas the polarization is smoothed with either the same
 Gaussian (black dashed), Q-band beam (blue solid), V-band beam (purple
 solid), or W-band beam (red dashed). 
(Top left) All temperature hot spots are stacked. 
(Top right) Spots greater than 1$\sigma$ are stacked.
(Bottom left) Spots greater than 2$\sigma$ are stacked.
(Bottom right) Spots greater than 3$\sigma$ are stacked.
The light gray lines show the average of the measurements from noiseless
 simulations with a Gaussian smoothing of $0.5^\circ$ FWHM. 
The agreement is excellent.		
} 
\label{fig:cpk}
\end{figure*} 
\begin{figure*}[t]
\centering \noindent
\includegraphics[width=16cm]{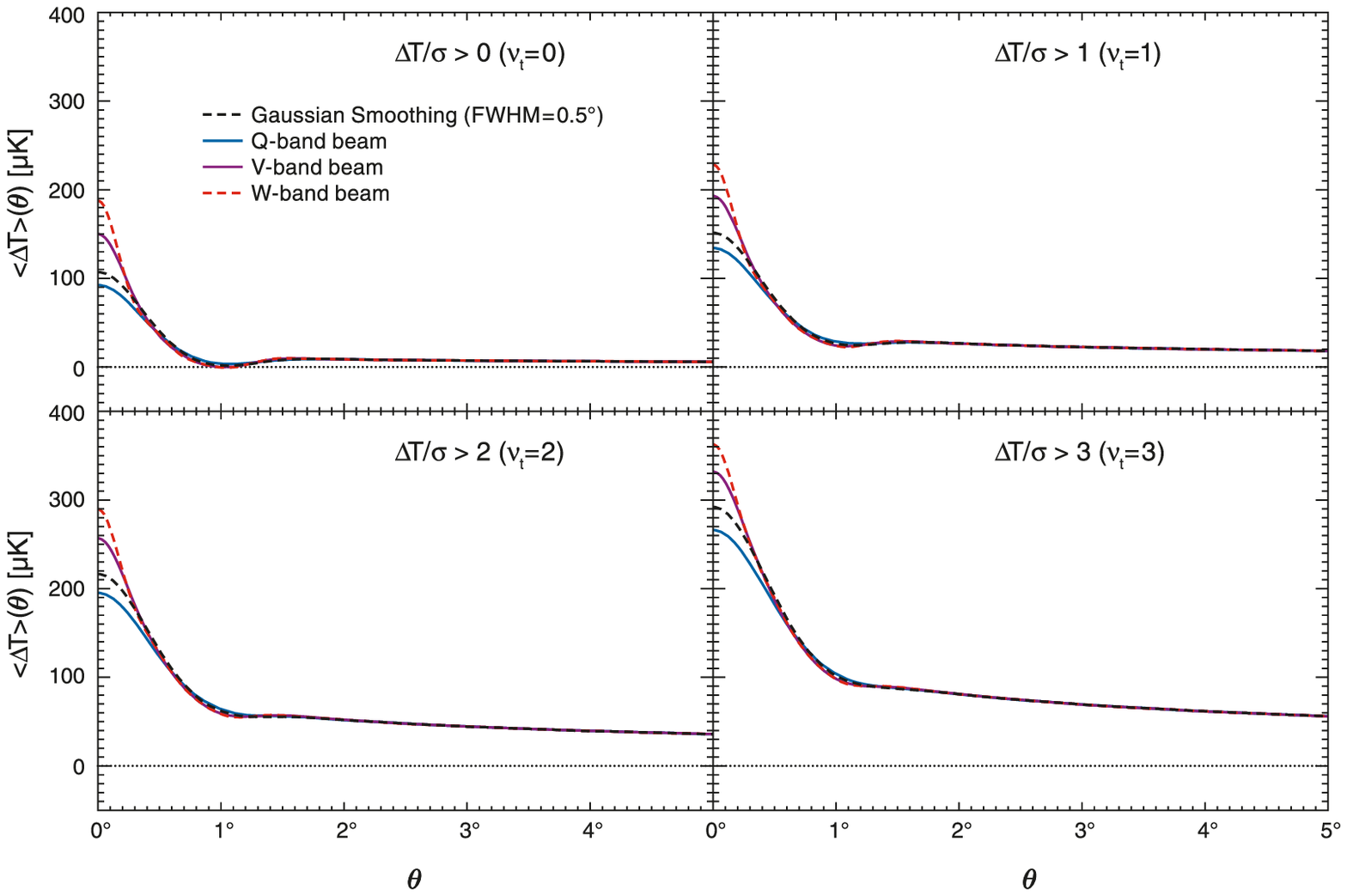}
\caption{%
Predicted temperature peak-temperature correlation, as measured by the
 stacked temperature profile, computed from equation~(\ref{eq:T}) for
 various values of the threshold peak 
 heights.  The choices of the smoothing functions and the threshold
 peak heights are the same as in Figure~\ref{fig:cpk}.
} 
\label{fig:cpt}
\end{figure*} 

What do  $\langle
	   Q_r\rangle(\theta)$ and $\langle U_r\rangle(\theta)$ look
	   like? 
The $Q_r$ map  is expected to be non-zero for a cosmological
  signal, while the $U_r$ map is expected to vanish in a
  parity-conserving universe unless some systematic error rotates
  the polarization plane uniformly.

To understand the shape of $Q_r$ as well as its physical
	   implications, let us begin by showing the smoothed $C_l^{\rm TE}$
	   spectra and the corresponding
	   temperature-$Q_r$ correlation functions, $C^{TQ_r}(\theta)$, 
	   in Figure~\ref{fig:clte}. (Note that $C^{TQ_r}$ and $C^{TU_r}$
	   can be computed from equations~(\ref{eq:Qr}) and
	   (\ref{eq:Ur}), respectively, with $b_\nu=1$ and $b_\zeta=0$.)
This shows three distinct effects
	   causing polarization of CMB \citep[see][for a pedagogical
	   review]{hu/white:1997}: 
\begin{itemize}
 \item [(a)] $\theta\gtrsim 2\theta_{\rm horizon}$, where $\theta_{\rm
       horizon}$ is the angular size of the radius of
       the horizon size at the decoupling epoch. Using the comoving
       horizon size of $r_{\rm horizon}=0.286$~Gpc and the comoving angular
       diameter distance to the decoupling epoch of $d_{\rm
       A}=14$~Gpc as derived from the \map\ data, we find $\theta_{\rm
       horizon}= 1.2^\circ$.  
As this scale is so much greater than the sound horizon size (see below), only
       gravity affects the physics. Suppose that there is a
       Newtonian gravitational potential, $\Phi_N$, at the center of a
       perturbation, 
       $\theta=0$. If it is overdense at the center, $\Phi_N<0$,
       and thus it is a cold spot according to the Sachs--Wolfe formula
       \citep{sachs/wolfe:1967}, $\Delta T/T=\Phi_N/3<0$.
       The photon fluid in this region will flow into the gravitational
       potential well, creating a converging flow. Such a flow creates
       the quadrupole temperature anisotropy around an electron at
       $\theta\ge 2\theta_{\rm horizon}$, producing polarization
       that is radial, i.e., $Q_r>0$. Since the temperature is negative,
       we obtain $\langle TQ_r\rangle<0$, i.e., anti-correlation
       \citep{coulson/crittenden/turok:1994}. On the
       other hand, if it is overdense at the center, then the photon
       fluid moves outward, producing polarization
       that is tangential, i.e., $Q_r<0$. Since the temperature is positive,
       we obtain $\langle TQ_r\rangle<0$, i.e., anti-correlation. The
       anti-correlation at $\theta\ge 2\theta_{\rm horizon}$ is a
       smoking-gun for the presence of super-horizon fluctuations at the
       decoupling epoch \citep{spergel/zaldarriaga:1997}, which has been
       confirmed by the \map\ data \citep{peiris/etal:2003}.
\item[(b)] $\theta\simeq 2\theta_A$, where $\theta_A$ is the angular
	   size of the radius of the {\it sound} horizon size at the
	   decoupling epoch. Using the comoving sound 
       horizon size of $r_s=0.147$~Gpc and $d_{\rm
       A}=14$~Gpc as derived from the \map\ data, we find
	   $\theta_A=0.6^\circ$.   Again, consider a potential 
	   well with $\Phi_N<0$ at the center. As the photon fluid
	   flows into the well, it compresses, increasing the temperature
	   of the photons.
	   Whether or not this increase can reverse the sign
	   of the temperature fluctuation (from negative to positive)
	   depends on whether the initial perturbation was adiabatic. If it
	   was adiabatic, then the temperature would reverse sign
	   at $\theta\lesssim 2\theta_{\rm horizon}$. Note that the
	   photon fluid is still flowing in, and thus the polarization
	   direction is radial, $Q_r>0$. However, now that the temperature is
	   positive, the correlation reverses sign: $\langle
	   TQ_r\rangle>0$. A similar argument (with the opposite sign)
	   can be used to show the same result, $\langle TQ_r\rangle>0$,
	   for $\Phi_N>0$ at the center. As an aside, the temperature
	   reverses sign on smaller  angular scales for isocurvature
	   fluctuations. 
\item[(c)] $\theta\simeq \theta_A$. Again, consider a potential
	   well with $\Phi_N<0$ at the center. At $\theta\lesssim
	   2\theta_A$, the pressure of the photon fluid is so
	   great that it can slow down the flow of the fluid. Eventually,
   	   at $\theta\sim  \theta_A$, the pressure becomes
	   large enough to reverse the direction of the flow (i.e., the
	   photon fluid expands). As a
	   result the polarization
	   direction becomes tangential, $Q_r<0$; however, as the
	   temperature is still 
	   positive,  the correlation reverses sign again: $\langle
	   TQ_r\rangle<0$. 
\end{itemize}
On even smaller scales, the correlation reverses sign again
\citep[see Figure~2 of][]{coulson/crittenden/turok:1994}
because the temperature gets too cold due to expansion. We do not see
this effect in Figure~\ref{fig:clte} because of the smoothing. 
Lastly,  there is no
correlation between $T$ and $Q_r$  at $\theta=0$ because of symmetry.

These features are essentially preserved in the peak-polarization
correlation as measured by the stacked polarization profiles. We show
them in Figure~\ref{fig:cpk} for various values of the threshold peak
heights. The important difference is that, thanks to the scale-dependent
bias $\propto l^2$, the small-scale trough at $\theta\simeq \theta_{\rm
A}$ is enhanced, making it easier to observe. On the other hand, the
large-scale anti-correlation is suppressed. 
We can therefore conclude that, with the \map\
data, we should be able to measure the compression phase at
$\theta\simeq 2\theta_A = 1.2^\circ$, as well as the reversal
phase at $\theta\simeq \theta_A = 0.6^\circ$.
We also show the profiles calculated from numerical simulations (gray
solid lines). The agreement with equation~(\ref{eq:Qr}) is excellent. 
We also show the predicted profiles of the stacked temperature data in
Figure~\ref{fig:cpt}. 

\subsection{Analysis Method}

\subsubsection{Temperature Data}
We use the foreground-reduced V$+$W temperature map 
 at the HEALPix resolution of $N_{\rm side}=512$ to find temperature
peaks. First, we smooth the foreground-reduced temperature maps in
6 differencing assemblies (DAs) (V1,
V2, W1, W2, W3, W4) to a common resolution of $0.5^\circ$ (FWHM) using
\begin{equation}
 \Delta T(\hat{\mathbf n}) = \sum_{lm}
  a_{lm}\frac{W_l^T}{b_l}Y_{lm}(\hat{\mathbf n}),
\end{equation}
where $b_l$ is the appropriate beam transfer function for each DA
\citep{jarosik/etal:prep}, and $W_l^T=p_l\exp[-l(l+1)\sigma_{\rm
FWHM}^2/(16\ln 2)]$ is the pixel window function for $N_{\rm
side}=512$, $p_l$,  times the spherical harmonic transform of a Gaussian with
$\sigma_{\rm FWHM}=0.5^\circ$.
We then coadd the foreground-reduced V- and W-band maps with
the inverse noise variance weighting, and remove the monopole from the region
outside of the mask (which is already negligibly small, $1.07\times
10^{-4}~\mu$K). For the mask, we combine the new 7-year {\it KQ85} mask,
{\it KQ85y7} \citep[defined in][also see
Section~\ref{sec:analysis_wmap}]{gold/etal:prep} and {\it P06} 
masks, leaving 68.7\% of the sky available for the analysis.

We find the locations of minima and
maxima using the software ``{\sf hotspot}'' in the HEALPix package
\citep{gorski/etal:2005}.
Over the full sky (without the
mask), we find 20953 maxima and 20974 minima. 
As the maxima and minima found by {\sf hotspot} still contain negative
and positive peaks, respectively, we further select the ``hot spots'' by
removing all negative peaks from maxima, and the ``cold spots'' by
removing all positive peaks from minima. This procedure corresponds to
setting the threshold peak height to $\nu_t=0$; thus, our prediction for
$\langle Q_r\rangle(\theta)$ is
the top-left panel of Figure~\ref{fig:cpk}.

Outside of the mask, we find 12387 hot spots and 12628 cold spots. The
r.m.s. temperature fluctuation is $\sigma_0=83.9~\mu{\rm K}$.
What does the theory predict? Using equation~(\ref{eq:npk}) with the
power spectrum $C_l^{\rm TT}=(C_l^{\rm TT,{\rm
signal}}p_l^2+N_l^{\rm TT}/b_l^2)\exp[-l(l+1)\sigma_{\rm
FWHM}^2/(8\ln 2)]$ where $N_l^{\rm TT}
=7.47\times 10^{-3}~\mu{\rm K}^2~{\rm sr}$ is the
noise bias of the V$+$W map before Gaussian smoothing and $C_l^{\rm TT,{\rm
signal}}$ is the 5-year best-fitting power-law $\Lambda$CDM temperature
power spectrum, we find $4\pi f_{\rm sky}\bar{n}_{\rm pk}=12330$ for
$\nu_t=0$ and $f_{\rm sky}=0.687$; thus, the number of observed hot and
cold spots is consistent with the predicted number.\footnote{Note that
the predicted number is $4\pi f_{\rm sky}\bar{n}_{\rm pk}=10549$ if we
ignore the noise bias; thus, even with a Gaussian smoothing, the
contribution from noise is not negligible.}

\subsubsection{Polarization  Data}

As for the polarization data, we use the {\it raw} (i.e., without
foreground cleaning) polarization maps in V and W bands. We have checked
that the cleaned maps give similar results with slightly larger
error bars, which is consistent with the excess noise introduced by the
template foreground cleaning procedure
\citep{page/etal:2007,gold/etal:2009,gold/etal:prep}. 
As we are focused on relatively small angular scales, $\theta\lesssim
2^\circ$, in this analysis, the results presented in this section would
not be affected by a potential systematic effect causing an excess power
in the W-band polarization data on large angular scales, $l\lesssim
10$. However, note that this excess power could just be a statistical
fluctuation \citep{jarosik/etal:prep}.
We form two sets of the data: (i) V, W, and V$+$W band maps smoothed to a common
resolution of $0.5^\circ$, and (ii) V, W, and V$+$W band maps without any
additional smoothing. The first set is used only for visualization, 
whereas the second set is used for the $\chi^2$ analysis.

We extract a square region of $5^\circ\times 5^\circ$ around each
temperature hot or cold spot. We then coadd the extracted $T$ images with
uniform weighting, and $Q$ and $U$ images with the inverse noise variance
weighting. 
We have eliminated the pixels masked by {\it KQ85y7} and {\it
P06} from each $5^\circ\times
5^\circ$ region when we coadd images, and thus
the resulting stacked image has the smallest noise at the center
(because the masked pixels usually appear near the edge of each
image). 
We also accumulate the inverse noise variance per pixel as we coadd $Q$
and $U$ maps. The
coadded inverse noise variance maps of $Q$ and $U$ will be used to
estimate the errors of the stacked images of $Q$ and $U$ per pixel,
which will then be used for the $\chi^2$ analysis.

We find that the stacked images of $Q$ and $U$ have constant
offsets, which is not surprising. Since these affect our determination
of polarization directions, we remove monopoles from the stacked images of $Q$
and $U$.  The size of each pixel in the stacked image is $0.2^\circ$,
and the number of pixels is $25^2=625$. 

Finally, we compute $Q_r$ and $U_r$ from the stacked images of Stokes
$Q$ and $U$ using equations~(\ref{eq:Qrdef}) and (\ref{eq:Urdef}),
respectively. 

\subsection{Results}
\label{sec:chi2}
\begin{deluxetable}{lccc}
\tablecolumns{4}
\small
\tablewidth{0pt}
\tablecaption{%
Statistics of the results from the stacked polarization analysis
}
\tablehead{ 
\colhead{Data Combination\tablenotemark{a}}
& \colhead{$\chi^2$\tablenotemark{b}}
& \colhead{best-fitting Amplitude\tablenotemark{c}}
& \colhead{$\Delta\chi^2$\tablenotemark{d}}
}
\startdata
Hot, $Q$, V$+$W & 661.9 & $0.57\pm 0.21$ & $-7.3$ \nl
Hot, $U$, V$+$W & 661.1 & $1.07\pm 0.21$ & $-24.7$ \nl
Hot, $Q_r$, V$+$W & 694.2 & $0.82\pm 0.15$ & $-29.2$ \nl
Hot, $U_r$, V$+$W & 629.2 & $-0.13\pm 0.15$ & $-0.18$ \nl
\hline
Cold, $Q$, V$+$W & 668.3 & $0.89\pm 0.21$ & $-18.2$ \nl
Cold, $U$, V$+$W & 682.7 & $0.86\pm 0.21$ & $-16.7$ \nl
Cold, $Q_r$, V$+$W & 682.2 & $0.90\pm 0.15$ & $-36.2$ \nl
Cold, $U_r$, V$+$W & 657.8 & $0.20\pm 0.15$ & $-0.46$ \nl
\hline
Hot, $Q$, V$-$W & 559.8 & & \nl
Hot, $U$, V$-$W & 629.8 & & \nl
Hot, $Q_r$, V$-$W & 662.2 & & \nl
Hot, $U_r$, V$-$W & 567.0 & & \nl
\hline
Cold, $Q$, V$-$W & 584.0 & & \nl
Cold, $U$, V$-$W & 668.2 & & \nl
Cold, $Q_r$, V$-$W & 616.0 & & \nl
Cold, $U_r$, V$-$W & 636.9 & & 
\enddata
\tablenotetext{a}{``Hot'' and
 ``Cold'' denote the stacking around temperature hot spots and cold
 spots, respectively.}
\tablenotetext{b}{Computed with respect to zero signal. The
 number of degrees of freedom (DOF) is $25^2=625$.}
\tablenotetext{c}{Best-fitting amplitudes for the
 corresponding theoretical predictions.  The
 quoted errors show the 68\% confidence level.
Note that, for $U_r$, we used the
 prediction for $Q_r$; thus, the fitted amplitude may be interpreted as
 $\sin(2\Delta\alpha)$, where $\Delta\alpha$ is the rotation of
 the polarization plane due to, e.g., violation of global parity symmetry.}
\tablenotetext{d}{Difference between the second column
 and $\chi^2$ after removing the model with the best-fitting amplitude
 given in the 3rd column.}
\label{tab:chi2}
\end{deluxetable}

\begin{figure*}[t]
\centering \noindent
\includegraphics[width=16cm]{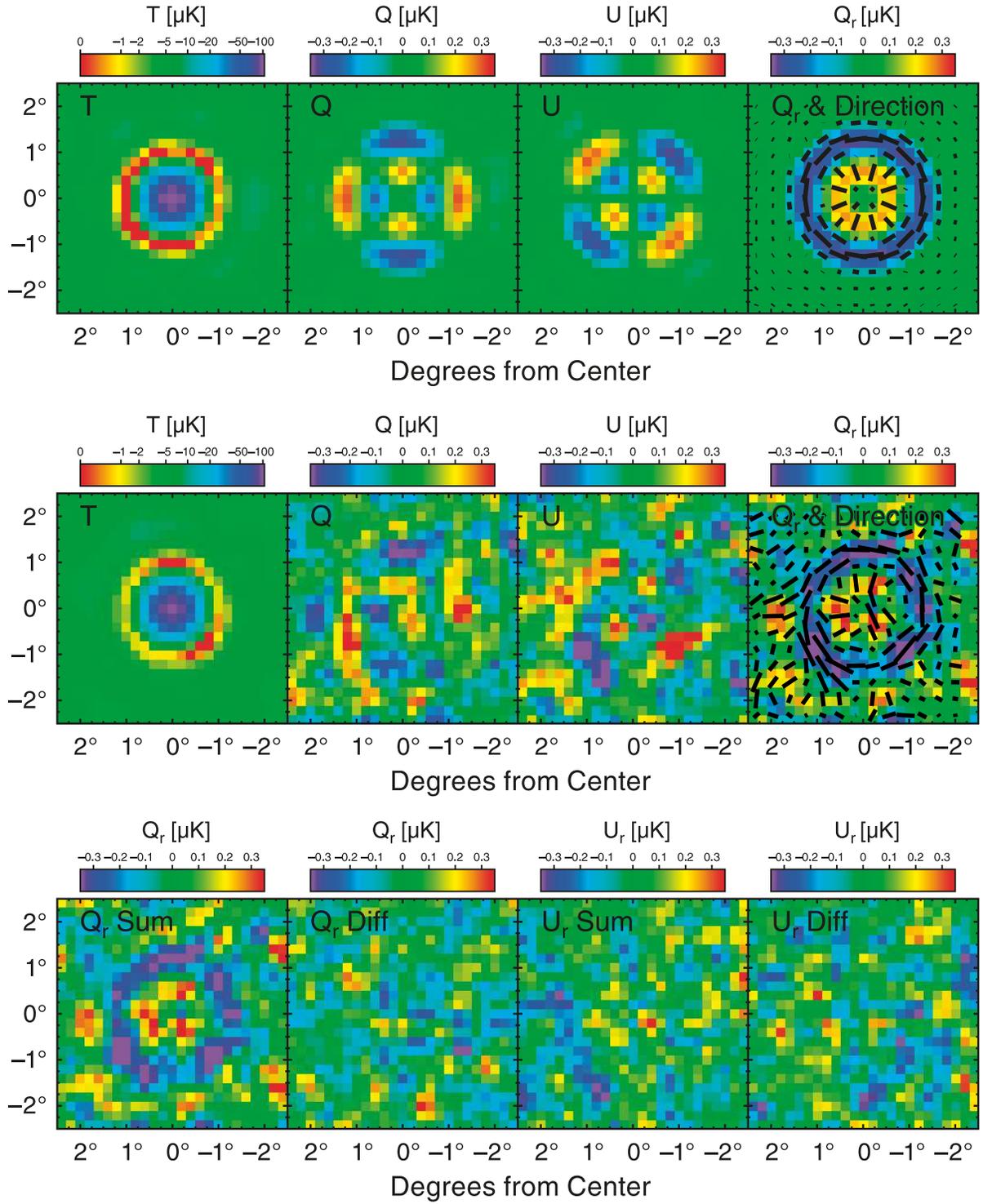}
\caption{%
Stacked images of temperature and polarization data around temperature
 {\it cold} 
 spots. Each panel shows a $5^\circ\times 5^\circ$ region with north up
 and east left. Both the temperature and polarization data have been
 smoothed  to a common resolution of $0.5^\circ$.
(Top) Simulated images with no instrumental noise. From left to right:
 the stacked temperature, 
 Stokes $Q$, Stokes $U$, and transformed Stokes $Q_r$ (see
 equation~(\ref{eq:Qrdef})) overlaid with the
 polarization directions.
(Middle) \map\ 7-year V$+$W data. In the observed map of $Q_r$, the
 compression phase at $1.2^\circ$ and the reversal phase at $0.6^\circ$
 are clearly visible. 
(Bottom) Null tests. From left to right: the stacked $Q_r$ from the sum
 map and from the difference map (V$-$W)/2, the stacked $U_r$ from the
 sum map and from the difference map. The latter three maps are all
 consistent with noise. Note that $U_r$, which probes the TB correlation
 (see equation~(\ref{eq:Ur})), is expected to vanish in a
 parity-conserving universe.
} 
\label{fig:minima}
\end{figure*}

\begin{figure*}[t]
\centering \noindent
\includegraphics[width=16cm]{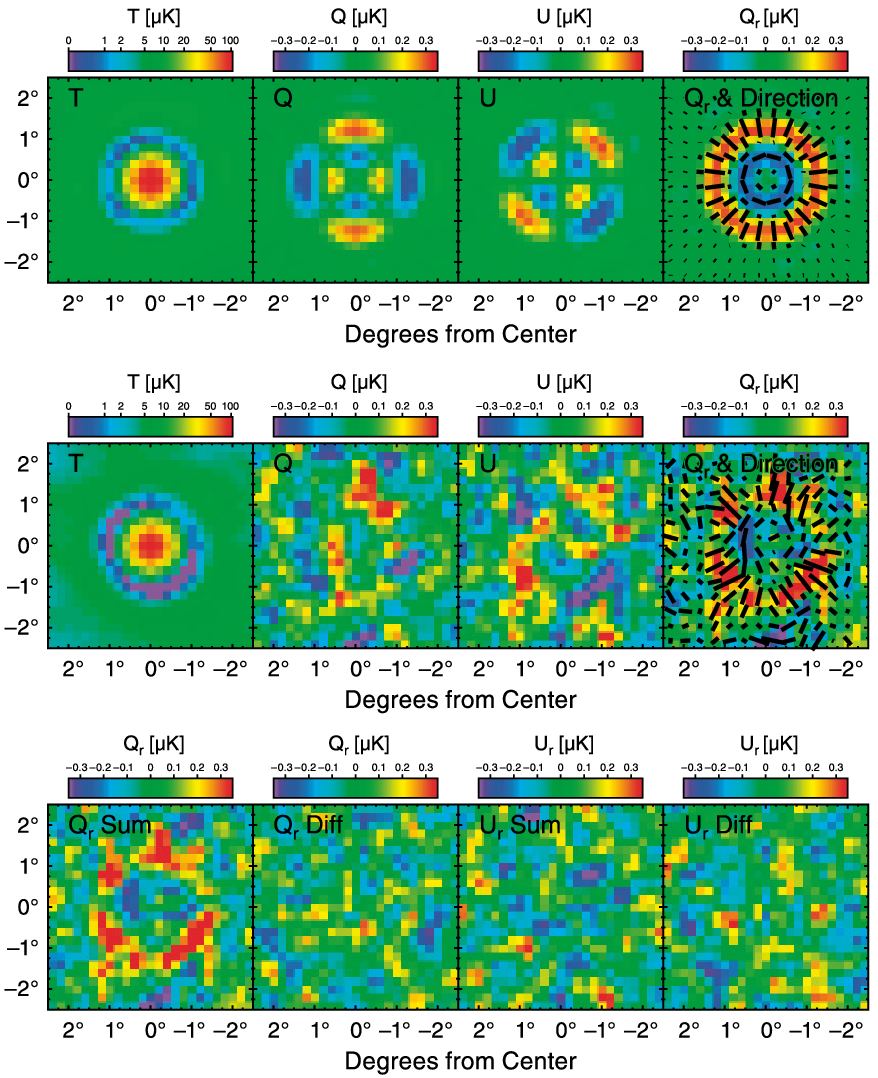}
\caption{%
Same as Figure~\ref{fig:minima} but for temperature {\it hot} spots. 
} 
\label{fig:maxima}
\end{figure*}

In Figure~\ref{fig:minima} and \ref{fig:maxima}, we show the stacked
images of $T$, $Q$, $U$, $Q_r$, and $U_r$ around temperature cold spots and hot
spots, respectively. 
The peak values of the stacked temperature profiles agree with the
predictions (see the dashed line in the top-left panel of
Figure~\ref{fig:cpt}). A dip in temperature (for hot spots; a bump for
cold spots) 
at $\theta\simeq 1^\circ$ is clearly visible in the data.
While the Stokes $Q$ and $U$ measured from the data exhibit the expected
features, they are still fairly noisy. The most striking images are the
stacked $Q_r$ (and $T$). The predicted features are clearly visible,
particularly the
compression phase at $1.2^\circ$ and the reversal phase at
$0.6^\circ$ in $Q_r$: the polarization directions around
temperature cold spots are radial at
$\theta\simeq 0.6^\circ$ and tangential at $\theta\simeq 1.2^\circ$,
and those around temperature hot spots show the opposite patterns, as
predicted. 

How significant are these features? Before performing the quantitative
$\chi^2$ analysis, 
we first compare $Q_r$ and $U_r$ using both the (V$+$W)/2 sum map (here,
V$+$W refers to the inverse noise variance weighted average) as well as
  the (V$-$W)/2 difference map (bottom panels of 
Figure~\ref{fig:minima} and \ref{fig:maxima}).
  The $Q_r$ map (which is expected to be non-zero for a cosmological
  signal) shows clear differences between the sum and difference
  maps, while the $U_r$ map  (which is expected to vanish in a
  parity-conserving universe unless some systematic error rotates
  the polarization plane uniformly) is consistent with zero in both the
  sum and difference maps.

Next, we perform the standard $\chi^2$ analysis. We summarize the
results in Table~\ref{tab:chi2}.
We report  the values of $\chi^2$ measured with respect to zero signal
in the second column, where the number of degrees of freedom (DOF) is
625. For each sum map combination, we fit the data to the predicted
signal to find the best-fitting amplitude. 

The largest improvement in
$\chi^2$ is observed for $Q_r$, as expected from the visual inspection
of Figure~\ref{fig:minima} and \ref{fig:maxima}: 
we find $0.82\pm 0.15$ and $0.90\pm 0.15$ for the stacking of $Q_r$
around hot and cold spots, respectively. The improvement in $\chi^2$ is
$\Delta\chi^2=-29.2$ and $-36.2$, respectively; thus, we detect the
expected polarization patterns around hot and cold spots at the level of 
5.4$\sigma$ and 6$\sigma$, respectively. The combined significance
exceeds 8$\sigma$. 

On the other hand, we do not find any evidence for
$U_r$. The $\chi^2$ values with respect to zero signal per DOF are
$629.2/625$ (hot spots) and $657.8/625$ (cold spots), and the
probabilities of finding larger values of $\chi^2$ are 44.5\% and 18\%,
respectively. But, can we learn anything about cosmology from this result? 
While the standard model predicts $C_l^{\rm TB}=0$ and hence $\langle
U_r\rangle=0$, models in which the global parity symmetry is violated
can create $C_l^{\rm TB}=\sin(2\Delta\alpha)C_l^{\rm TE}$
\citep{lue/wang/kamionkowski:1999,carroll:1998,feng/etal:2005}.
Therefore, we fit the measured $U_r$ to the predicted $Q_r$, finding 
a null result:  $\sin(2\Delta\alpha)=-0.13\pm 0.15$ and $0.20\pm 0.15$~(68\%~CL), or equivalently
$\Delta\alpha=-3.7^\circ\pm 4.3^\circ$ and $5.7^\circ\pm 4.3^\circ$~(68\%~CL)
for hot and cold spots, respectively.
Averaging these numbers, we obtain $\Delta\alpha=1.0^\circ\pm 3.0^\circ$~(68\%~CL),
which is consistent with (although not as stringent as) the limit we find
from the full analysis presented in Section~\ref{sec:TB}.
Finally, all the $\chi^2$ values measured from the difference maps are
consistent with a null signal. 

How do these results compare to the full analysis of the TE power
spectrum? By fitting the 7-year $C_l^{\rm TE}$ data to the same power
spectrum used above (5-year best-fitting power-law $\Lambda$CDM model from
$l=24$ to $800$, i.e., DOF$=$777), we find the best-fitting amplitude of $0.999\pm 0.048$ and 
$\Delta\chi^2=-434.5$, i.e., a 21$\sigma$ detection of the TE signal. 
This is reasonable, as we used only the V- and W-band data for the
stacking analysis, while we used also the Q-band data for measuring the
TE power spectrum; 
$\langle Q_r\rangle(\theta)$ is insensitive to
information on $\theta\gtrsim 2^\circ$ (see top left panel of
Figure~\ref{fig:cpk}); and the smoothing suppresses the power at
$l\gtrsim 400$ (see left panel of Figure~\ref{fig:clte}).
Nevertheless, there is probably a way to extract more information from
$\langle Q_r\rangle(\theta)$ by, for example, combining data at different
threshold peak heights and smoothing scales.

\subsection{Discussion}
\label{sec:polarizationdiscussion}
If the temperature fluctuations of the CMB obey Gaussian statistics and
global parity symmetry is respected on cosmological scales, the
temperature-$E$-mode polarization cross power spectrum, $C_l^{\rm TE}$,
contains all the information about the temperature-polarization
correlation. In this sense, the stacked polarization images do not add
any new information. 

The detection and measurement of the temperature-$E$ mode
polarization cross-correlation power spectrum, $C_l^{\rm TE}$
\citep{kovac/etal:2002,kogut/etal:2003,spergel/etal:2003},
can be regarded as equivalent to finding the
predicted polarization patterns around hot and cold spots. While we 
have shown that one can write the stacked
polarization profile around temperature spots in terms of an integral of
$C_l^{\rm TE}$,
the formal equivalence between this new method
and $C_l^{\rm TE}$ is valid only when temperature fluctuations obey Gaussian
statistics, as the stacked $Q$ and $U$ maps measure correlations between
temperature {\it peaks} and polarization.
So far there is no convincing evidence for non-Gaussianity in the
temperature fluctuations observed by \map\ \citep[][see
Section~\ref{sec:NG} for the 7-year limits on primordial
non-Gaussianity, and Bennett et al.~2010 for discussion on other non-Gaussian
features]{komatsu/etal:2003}.  

Nevertheless, they provide striking confirmation of
our understanding of the physics at the decoupling epoch in the form of
radial and tangential polarization patterns at two characteristic
angular scales that are important for the physics of acoustic
oscillation: the compression phase at $\theta=2\theta_A$ and the
reversal phase at $\theta=\theta_A$. 

Also, this analysis does not
require any analysis in harmonic space, nor decomposition to $E$ and $B$
modes. The analysis is so straightforward and intuitive that the 
method presented here would also be useful for null tests and systematic
error checks. The stacked image of $U_r$ should be particularly useful
for systematic error checks.

Any experiments that measure both temperature and polarization
should be able to produce the stacked images such as presented in
Figure~\ref{fig:minima} and \ref{fig:maxima}. 

\section{Summary of 7-year Parameter Estimation}
\label{sec:analysis}

\begin{deluxetable}{lccc}
\tablecolumns{4}
\small
\tablewidth{0pt}
\tablecaption{%
Polarization Data: Improvements from the 5-year data
}
\tablehead{
\colhead{$l$ Range}
& \colhead{Type}
& \colhead{7-year}
& \colhead{5-year}
}
\startdata
High $l$\tablenotemark{a}  
& TE & Detected at 21$\sigma$ & Detected at 13$\sigma$ \nl
& TB & $\Delta\alpha=-0.9^\circ\pm 1.4^\circ$
& $\Delta\alpha=-1.2^\circ\pm 2.2^\circ$ \nl
\hline
Low $l$\tablenotemark{b}  
& EE & $\tau=0.088\pm 0.015$ & $\tau=0.087\pm 0.017$ \nl
& BB & $r<2.1$ (95\%~CL) & $r<4.7$  (95\%~CL) \nl
& EE/BB & $r<1.6$  (95\%~CL) & $r<2.7$  (95\%~CL) \nl
& TB/EB & $\Delta\alpha=-3.8^\circ\pm 5.2^\circ$
& $\Delta\alpha=-7.5^\circ\pm 7.3^\circ$ \nl
\hline
All $l$ 
& TE/EE/BB
& $r<0.93$ (95\%~CL) & $r<1.6$ (95\%~CL) \nl
& TB/EB\tablenotemark{c}
& $\Delta\alpha=-1.1^\circ\pm 1.4^\circ$
& $\Delta\alpha=-1.7^\circ\pm 2.1^\circ$
\enddata
\tablenotetext{a}{$l\ge 24$. The Q-, V-, and W-band data are used for
 the 7-year analysis, whereas only the Q- and V-band data were used for
 the 5-year analysis.}
\tablenotetext{b}{$2\le l\le 23$. The Ka-, Q-, and V-band data are used
 for both the 7-year and 5-year  analyses.}
\tablenotetext{c}{The quoted errors are statistical only, and do not
 include the systematic error $\pm 1.5^\circ$ (see Section~\ref{sec:TB}).}
\label{tab:improvements}
\end{deluxetable}

\subsection{Improvements from the 5-year Analysis}
\label{sec:analysis_wmap}

{\bf Foreground Mask}. 
The 7-year temperature analysis masks, {\it KQ85y7} and {\it KQ75y7}, have
been slightly enlarged to mask the regions that have excess
foreground emission, particularly in the HII regions Gum and Ophiuchus,
identified in the difference between foreground-reduced maps at
different frequencies.  As a result, the new {\it KQ85y7} and {\it 
KQ75y7} masks eliminate an additional 
3.4\% and 1.0\% of the sky,
leaving 78.27\% and 70.61\% of the sky for the cosmological
analyses, respectively. See Section~2.1 of \citet{gold/etal:prep} for details. 
There is no change in the polarization {\it P06} mask \citep[see
Section~4.2 of][for definition of this mask]{page/etal:2007}, which
leaves 73.28\% of the sky.

{\bf Point Sources and the SZ Effect}.
We continue to marginalize over a contribution from unresolved point
sources, assuming that the antenna temperature of point
sources declines with frequency as $\nu^{-2.09}$ \citep[see equation~(5)
of][]{nolta/etal:2009}. 
The 5-year estimate of the power spectrum from
unresolved point sources 
in Q band in units of antenna temperature, $A_{\rm ps}$, 
was $10^3A_{\rm ps}=11\pm 1~\mu{\rm K}^2~{\rm sr}$
\citep{nolta/etal:2009}, and we used this value and the error bar to
marginalize over the power spectrum of residual point sources in the 7-year
parameter estimation.
The subsequent analysis showed that the 7-year estimate of the power
spectrum is 
$10^3A_{\rm ps}=9.0\pm 0.7~\mu{\rm K}^2~{\rm sr}$
\citep{larson/etal:prep},
which is somewhat lower than the
5-year value because more sources are resolved by \map\ and included in
the source mask. The difference in the diffuse mask (between {\it KQ85y5}
and {\it KQ85y7}) does not affect the value of $A_{\rm ps}$ very much:
we find $9.3$ instead of $9.0$ if we use the 5-year diffuse mask and the
7-year source mask.
The source power spectrum is sub-dominant in the total power. 
We have checked that the parameter results are insensitive to the
difference between the 5-year and 7-year residual source estimates.

We continue to marginalize over a contribution
from the SZ effect using the same template as for the
3- and 5-year analyses \citep{komatsu/seljak:2002}. We assume a uniform
prior on the amplitude of this template as $0< A_{\rm SZ}< 2$, which
is now justified by the latest limits from the 
SPT collaboration, $A_{\rm SZ}=0.37\pm 0.17$
\citep[68\%~CL;][]{lueker/etal:2010}, and the ACT collaboration, $A_{\rm
SZ}<1.63$ \citep[95\%~CL;][]{fowler/etal:2010}. 

{\bf High-$l$ Temperature and Polarization}.
We increase the multipole range of the power spectra used for the
cosmological parameter estimation from $2-1000$ to $2-1200$ for the TT
power spectrum, and from $2-450$ to $2-800$ for the TE power
spectrum. We use the 7-year V- and W-band maps \citep{jarosik/etal:prep}
to measure the high-$l$ TT power spectrum in $l=33-1200$. While we used
only Q- and V-band maps to measure the high-$l$ TE and TB power spectra
for the 5-year analysis \citep{nolta/etal:2009}, we also include
W-band maps in the 7-year high-$l$ polarization analysis. 

With these data, we now detect the high-$l$ TE power spectrum
at 21$\sigma$, compared to 13$\sigma$ for the 5-year
high-$l$ TE data. This is a consequence of adding two more years of data
and the W-band data. 
The TB data can be used to probe a rotation angle of
the polarization plane, $\Delta\alpha$, due to potential parity-violating
effects or systematic effects.  With the 7-year high-$l$ TB data we find
a limit $\Delta\alpha=-0.9^\circ\pm 1.4^\circ$ (68\%~CL). For comparison, the
limit from the 5-year high-$l$ TB power spectrum was
$\Delta\alpha=-1.2^\circ\pm
2.2^\circ$ \citep[68\%~CL;][]{komatsu/etal:2009}. 
See Section~\ref{sec:TB} for the 7-year limit on $\Delta\alpha$ from the
full analysis.

{\bf Low-$l$ Temperature and Polarization}. 
Except for using the 7-year maps and the new temperature {\it KQ85y7}
mask, there is no change in the analysis of
the low-$l$ temperature and polarization data:
we use the Internal Linear Combination (ILC) map
\citep{gold/etal:prep} to measure the low-$l$ TT power spectrum in
$l=2-32$, and calculate the likelihood using the Gibbs sampling and
Blackwell-Rao (BR) estimator
\citep{jewell/levin/anderson:2004,wandelt:2003,wandelt/larson/lakshminarayanan:2004,odwyer/etal:2004,eriksen/etal:2004,eriksen/etal:2007c,eriksen/etal:2007b,chu/etal:2005,larson/etal:2007}. For
the implementation of the BR estimator in the 5-year analysis, see
Section~2.1 of \citet{dunkley/etal:2009}. 
We use Ka-, Q-, and V-band maps for the low-$l$ polarization analysis
in $l=2-23$, and evaluate the likelihood directly in pixel space as
described in Appendix~D of \citet{page/etal:2007}.

To get a feel for improvements in the low-$l$ polarization data with two
additional years of integration, we note that the 7-year limits on the
optical depth, and the tensor-to-scalar ratio and rotation angle from
the low-$l$ polarization data {\it alone}, are $\tau=0.088\pm 0.015$
\citep[68\%~CL; 
see][]{larson/etal:prep}, $r<1.6$
(95\%~CL; see Section~\ref{sec:GW}), and $\Delta\alpha=-3.8^\circ\pm
5.2^\circ$ (68\%~CL; see Section~\ref{sec:TB}), respectively. The
corresponding 5-year limits were 
$\tau=0.087\pm 0.017$ \citep{dunkley/etal:2009}, $r<2.7$ (see
Section~\ref{sec:GW}), and 
$\Delta\alpha=-7.5^\circ\pm 7.3^\circ$ \citep{komatsu/etal:2009},
respectively. 

In Table~\ref{tab:improvements}, we summarize the improvements from the
5-year  data mentioned above.

\subsection{External data sets}
\label{sec:analysis_ex}

The \map\ data are statistically powerful enough to constrain 6
parameters of a flat $\Lambda$CDM model with a tilted spectrum. 
However, to constrain deviations from this minimal model, other CMB data
probing smaller angular scales and astrophysical data probing the
expansion rates, distances, and growth of structure are useful.

\subsubsection{Small-scale CMB Data}

The best limits on the primordial helium abundance, $Y_p$, 
are obtained when the \map\ data are combined with the power spectrum
data from other 
CMB experiments probing smaller angular scales, $l\gtrsim 1000$.

We use the temperature power spectra from  
the Arcminute Cosmology Bolometer Array Receiver
\citep[ACBAR;][]{reichardt/etal:2009} and 
 QUEST at DASI (QUaD)  \citep{brown/etal:2009} experiments. 
For the former, we use the temperature power spectrum binned in 16 band
powers in the multipole range $900<l<2000$. For the latter, we use the
temperature power spectrum binned in 13 band powers in
$900<l<2000$.

We marginalize over the
beam and calibration errors of each experiment: 
for ACBAR, the beam error is 2.6\% on a 5 arcmin (FWHM) Gaussian beam and
the calibration error is 2.05\% in temperature. 
For QUaD, the beam error combines a 2.5\% error on 5.2 and 3.8 arcmin
(FWHM) Gaussian beams at 100~GHz and 150~GHz, respectively, with an
additional term accounting for the sidelobe uncertainty \citep[see
Appendix A of][for details]{brown/etal:2009}. The calibration error is
3.4\% in temperature.

The ACBAR data are calibrated to the \map\ 5-year temperature data, and
the QUaD data are calibrated to the BOOMERanG data
\citep{masi/etal:2006} which are, in turn, calibrated to the \map\
1-year temperature data. (The QUaD team takes into account the change in
the calibration from the 1-year to the 5-year \map\ data.)
The calibration errors quoted above are much greater than the
calibration uncertainty of the \map\ 5-year data
\citep[0.2\%][]{hinshaw/etal:2007}. This is due to the noise of the
ACBAR, QUaD, and BOOMERanG data. In other words, the above calibration
errors are dominated by the statistical errors that are uncorrelated
with the \map\ data. We thus treat the \map, ACBAR, and QUaD data as
independent. 

\begin{figure*}[t]
\centering \noindent
\includegraphics[width=17cm]{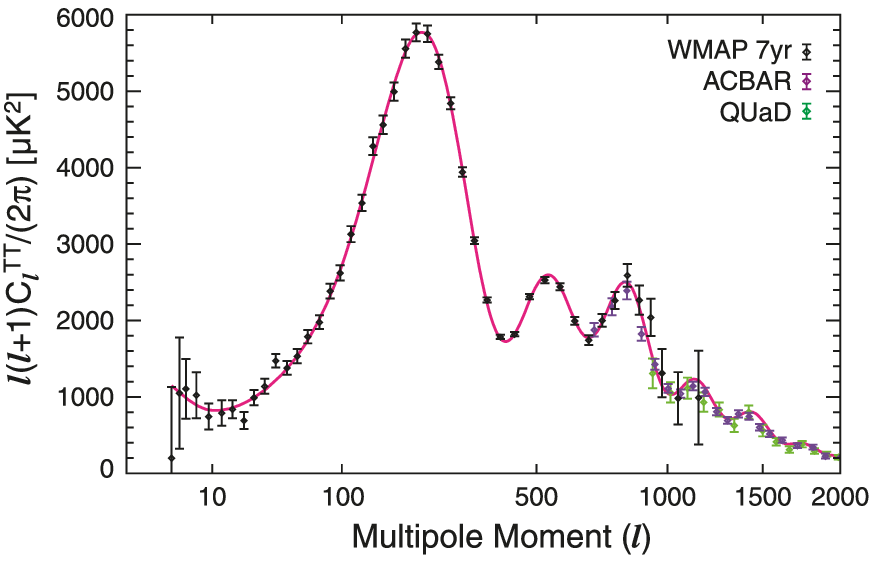}
\caption{%
The \map\ 7-year temperature power spectrum \citep{larson/etal:prep},
 along with the temperature power spectra from the ACBAR
 \citep{reichardt/etal:2009} and QUaD \citep{brown/etal:2009}
 experiments. We show the ACBAR and QUaD data only at $l\ge 690$, where
 the errors in the \map\ power spectrum are dominated by noise. We do not
 use the power 
 spectrum at $l>2000$ because of a potential contribution from the SZ
 effect and point sources.
 The solid line shows the best-fitting 6-parameter flat
 $\Lambda$CDM model to the \map\ data alone (see the 3rd column of
 Table~\ref{tab:summary} for the maximum likelihood parameters). 
} 
\label{fig:cl}
\end{figure*}

Figure~\ref{fig:cl} shows the \map\ 7-year temperature power spectrum
\citep{larson/etal:prep} as well as the temperature power spectra from
ACBAR and QUaD.

We do not use the other, previous small-scale CMB data, as their
statistical errors are much larger than those of ACBAR and QUaD, and
thus adding them would not improve the constraints on the cosmological
parameters significantly. The new power spectrum data from the SPT
\citep{lueker/etal:2010} and ACT \citep{fowler/etal:2010}
collaborations were not yet available at the time of our analysis.

\subsubsection{Hubble Constant and Angular Diameter Distances}
\label{sec:wmap+bao+h0}
There are two main astrophysical priors that we shall use in
this paper: the Hubble constant and the angular diameter distances out
to $z=0.2$ and $0.35$.
\begin{itemize}
 \item A Gaussian prior on the present-day Hubble constant,
$H_0=74.2\pm 3.6~{\rm km~s^{-1}~Mpc^{-1}}$
\citep[68\%~CL;][]{riess/etal:2009}. The quoted error includes both statistical
and systematic errors. 
This measurement of $H_0$ is obtained from the magnitude-redshift
       relation of 240 low-$z$ Type Ia supernovae at $z<0.1$. The
       absolute magnitudes of supernovae are calibrated using new observations
       from {\sl HST} of 240 Cepheid variables in six local Type Ia supernovae
       host galaxies and the maser galaxy NGC 4258. The systematic
       error is minimized by calibrating supernova luminosities directly using
       the geometric maser distance measurements. This is a significant
       improvement over the prior 
       that we adopted for the 5-year analysis, $H_0=72\pm 8~{\rm
       km~s^{-1}~Mpc^{-1}}$, which is from the Hubble Key Project final results
       \citep{freedman/etal:2001}. 
 \item Gaussian priors on the distance ratios, $r_s/D_V(z=0.2)=0.1905\pm
       0.0061$ and $r_s/D_V(z=0.35)=0.1097\pm
       0.0036$, measured from the {\sl Two-Degree Field Galaxy Redshift
       Survey} ({\sl 2dFGRS}) and 
       the {\sl Sloan Digital Sky Survey} Data
       Release 7 ({\sl SDSS} DR7) \citep{percival/etal:2009}. The inverse
       covariance matrix is given by equation~(5) of
       \citet{percival/etal:2009}. These priors are improvements from
       those we adopted for the 5-year analysis, 
       $r_s/D_V(z=0.2)=0.1980\pm
       0.0058$ and $r_s/D_V(z=0.35)=0.1094\pm
       0.0033$ \citep{percival/etal:2007c}.

       The above measurements can be translated into a measurement of
       $r_s/D_V(z)$ at a single, ``pivot''
       redshift: $r_s/D_V(z=0.275)=0.1390\pm 0.0037$
       \citep{percival/etal:2009}. \citet{kazin/etal:2010} used the
       two-point correlation function 
       of {\sl SDSS}-DR7 LRGs to measure $r_s/D_V(z)$ at $z=0.278$. They
       found $r_s/D_V(z=0.278)=0.1394\pm 0.0049$, which is an excellent
       agreement with the above measurement by
       \citet{percival/etal:2009} at a similar redshift. The excellent
       agreement between these two independent studies, which are based on
       very different methods, indicates that the systematic error in
       the derived values of $r_s/D_V(z)$ may be much smaller than the
       statistical error.

       Here, $r_s$ is the comoving sound
       horizon size at the baryon drag epoch $z_d$,
\begin{equation}
 r_s(z_d)
=\frac{c}{\sqrt{3}}\int_0^{1/(1+z_d)}
\frac{da}{a^2H(a)\sqrt{1+(3\Omega_b/4\Omega_\gamma)a}}.
\label{eq:rs}
\end{equation} 
For $z_d$, we use the fitting formula proposed by
       \citet{eisenstein/hu:1998}. The
        effective distance measure, $D_V(z)$
       \citep{eisenstein/etal:2005}, is  given by  
\begin{equation}
 D_V(z) \equiv \left[(1+z)^2D_A^2(z)\frac{cz}{H(z)}\right]^{1/3},
\end{equation}
where $D_A(z)$ is the proper (not comoving) angular diameter distance:
\begin{equation}
 D_A(z) = \frac{c}{H_0}\frac{f_k\left[H_0\sqrt{\left|\Omega_{
				k}\right|}\int_0^z\frac{dz'}{H(z')}\right]}{(1+z)\sqrt{\left|\Omega_{
 k}\right|}},
\label{eq:da}
\end{equation}
where $f_k[x]=\sin x$, $x$, and $\sinh x$ for $\Omega_{k}<0$
($k=1$; positively curved), $\Omega_{k}=0$ ($k=0$; flat), and
       $\Omega_{k}>0$ ($k=-1$; negatively curved), 
respectively. The Hubble expansion rate, which has contributions from
       baryons, cold dark matter, photons, massless and massive
       neutrinos, curvature, and dark energy, is given by
       equation~(\ref{eq:hubble}) in Section~\ref{sec:exactnu}.
\end{itemize}
The cosmological parameters determined by combining the \map\ data,
BAO, and $H_0$ will be called ``\map+BAO+$H_0$,'' and they constitute
our best estimates of the cosmological parameters, unless noted
otherwise. 

Note that, when redshift is much less than unity, the effective distance
approaches $D_V(z)\to cz/H_0$. Therefore, the effect of different
cosmological models on $D_V(z)$ do not appear until one goes to higher
redshifts. If redshift is very low, $D_V(z)$ is simply measuring the
Hubble constant.

\subsubsection{Power Spectrum of Luminous Red Galaxies}
\label{sec:lrg}
A combination of the \map\ data and the power
spectrum of Luminous Red Galaxies (LRGs) measured from the {\sl SDSS} DR7 is a
powerful probe of the total mass of neutrinos, $\sum m_\nu$, and the
effective number of neutrino species, $N_{\rm eff}$
\citep{reid/etal:2010,reid/etal:2010b}. We thus combine the LRG power
spectrum \citep{reid/etal:2010} with the \map\ 7-year data and the
Hubble constant \citep{riess/etal:2009} to update the constraints on
$\sum m_\nu$ and $N_{\rm eff}$ reported in \citet{reid/etal:2010}.
Note that BAO and the LRG power spectrum cannot be treated as independent
data sets because a part of the measurement of BAO used LRGs as well. 

\subsubsection{Luminosity Distances}
\label{sec:sn}
The luminosity distances out to high-$z$ Type Ia supernovae
have been the 
most powerful data for first discovering the existence of dark energy
\citep{riess/etal:1998,perlmutter/etal:1999} and then constraining the
properties of dark energy, such as the equation of state parameter,
$w$ \citep[see][for a recent review]{frieman/turner/huterer:2008}.
With more than 400 Type Ia supernovae discovered, the constraints
on the properties of dark energy inferred from Type Ia supernovae are now
limited by systematic errors rather than by statistical errors.

There is an indication that the constraints on dark energy parameters
are different when different methods are used to fit the light curves of
Type Ia supernovae \citep{hicken/etal:2009,kessler/etal:2009}. 
We also found that the parameters of the minimal 6-parameter
$\Lambda$CDM model derived from two 
compilations of \citet{kessler/etal:2009} are different: one compilation
uses the light curve 
fitter called {\sf SALT-II} \citep{guy/etal:2007} while the other uses
the light curve fitter called {\sf MLCS2K2}
\citep{jha/riess/kirshner:2007}.
For example, $\Omega_\Lambda$ derived from
\map+BAO+{\sf SALT-II} and \map+BAO+{\sf MLCS2K2}  are different by
nearly $2\sigma$, despite being derived from the same data sets
(but processed with two different light curve fitters).
If we allow the dark energy equation of state parameter, $w$, to vary,
we find that $w$
derived from
\map+BAO+{\sf SALT-II} and \map+BAO+{\sf MLCS2K2} are different by $\sim
2.5\sigma$. 

At the moment it
is not obvious how to estimate systematic errors and properly
incorporate them in the likelihood analysis, 
in order
to reconcile different methods and data sets. 

In this paper, we shall use one compilation of the supernova data called
the ``Constitution'' samples \citep{hicken/etal:2009}. The reason for this
choice over the others, such as the compilation by
\citet{kessler/etal:2009} that includes the latest data from the SDSS-II
supernova survey, is that the Constitution samples are an extension of the
``Union'' samples \citep{kowalski/etal:2008} that we used for the 5-year
analysis \citep[see Section~2.3 of][]{komatsu/etal:2009}. More
specifically, the Constitution samples are the Union samples plus the
latest samples of nearby Type Ia supernovae optical photometry from the Center
for Astrophysics (CfA) supernova group \citep[CfA3
sample;][]{hicken/etal:2009b}. Therefore, the parameter constraints from
a combination of the \map\ 7-year data, the latest BAO data described above
\citep{percival/etal:2009}, and the Constitution supernova data may be
directly compared to the ``\map+BAO+SN'' parameters given in Table~1 and
2 of \citet{komatsu/etal:2009}. This is a useful comparison, as it shows
how much the limits on parameters have improved by adding two more years
of data. 

However, given the scatter of results among different compilations of
the supernova data, we have decided to choose the ``\map+BAO+$H_0$'' 
(see Section~\ref{sec:wmap+bao+h0}) as
our best data combination to constrain the cosmological parameters,
except for dark energy parameters. For dark energy parameters,
we compare the results from \map+BAO+$H_0$ and \map+BAO+SN in
Section~\ref{sec:darkenergy}. 
Note that we always marginalize over the absolute magnitudes of
Type Ia supernovae with a uniform prior. 

\subsubsection{Time-delay Distance}
\label{sec:timedelay}
Can we measure angular diameter distances out to higher redshifts? 
Measurements of gravitational lensing time delays 
offer a way to determine absolute distance scales \citep{refsdal:1964}.
When a foreground galaxy lenses a background variable source (e.g., 
quasars) and
produces multiple images of the source, changes of the source
luminosity due to variability appear on multiple images at different
times. 

The time delay at a given image position $\bm{\theta}$ for a
given source position $\bm{\beta}$, $t(\bm{\theta},\bm{\beta})$,
depends on the angular diameter distances as 
\citep[see, e.g.,][for a review]{schneider/kochanek/wambsganss:2006}
\begin{equation}
 t(\bm{\theta},\bm{\beta}) = \frac{1+z_{\rm l}}{c}\frac{D_{\rm l}D_{\rm
  s}}{D_{\rm 
  ls}}\phi_{\rm F}(\bm{\theta},\bm{\beta}),  
\end{equation}
where $D_{\rm l}$, $D_{\rm s}$, and $D_{\rm ls}$ are the angular
diameter distances out to a  lens galaxy, to a source galaxy, and
between them, respectively, and $\phi_{\rm F}$ is the so-called Fermat
potential, which depends on the path length of light rays and
gravitational potential of the lens galaxy. 

The biggest challenge for this method is to control systematic errors in
our knowledge of $\phi_{\rm F}$, which requires a detailed modeling of
mass distribution of the lens. One can, in principle, minimize this
systematic error by finding a lens system where the mass distribution of
lens is relatively simple.

The lens system B1608+656 is not a simple system, with two lens galaxies
and dust extinction; however, it has one of the most precise
time-delay measurements of quadruple lenses. 
The lens redshift of this
system is relatively large, $z_{\rm l}=0.6304$
\citep{myers/etal:1995}. 
The source redshift is $z_{\rm s}=1.394$ \citep{fassnacht/etal:1996}.
This system has been used to determine $H_0$
to 10\% accuracy \citep{koopmans/etal:2003}. 

\citet{suyu/etal:2009} have obtained more data from the deep {\sl HST}
Advanced Camera 
for Surveys (ACS) observations of the 
asymmetric and spatially extended lensed images, and constrained 
the slope of mass distribution of the lens galaxies.
They also obtained ancillary data (for stellar dynamics and
lens environment studies) to control the systematics, particularly the
the so-called ``mass-sheet degeneracy,'' which the strong lensing data
alone cannot break. By doing so, they were able to reduce the error in
$H_0$ (including the systematic error) by a factor of two 
\citep{suyu/etal:2010}. They find a constraint on the ``time-delay
distance,'' $D_{\Delta t}$, as
\begin{equation}
 D_{\Delta t} \equiv (1+z_{\rm l})
\frac{D_{\rm l}D_{\rm s}}{D_{\rm
  ls}}
\simeq 5226\pm 206~{\rm Mpc},
\end{equation}
where the number is found from a Gaussian fit to the likelihood of
$D_{\Delta t}$\footnote{S.~H.~Suyu 2009, private communication.};
however, the actual shape of the likelihood is slightly 
non-Gaussian. We thus  use:
\begin{itemize}
 \item Likelihood of $D_{\Delta t}$ out to the lens system B1608+656 given
       by \citet{suyu/etal:2010},
\begin{equation}
 P(D_{\Delta t})
= \frac{\exp\left[-(\ln(x-\lambda)-\mu)^2/(2\sigma^2)\right]}
{\sqrt{2\pi}(x-\lambda)\sigma},
\end{equation}
where $x=D_{\Delta t}/(1~{\rm Mpc})$, $\lambda=4000$,
       $\mu=7.053$, and $\sigma=0.2282$. 
This likelihood includes systematic errors due to the 
       mass-sheet degeneracy, which dominates the total error budget
       \citep[see Section~6 of][for more details]{suyu/etal:2010}.
Note that this is the only lens system for which $D_{\Delta t}$ (rather
       than $H_0$) has been constrained.\footnote{
As the time-delay distance, $D_{\Delta t}$, is the angular diameter
       distance to the lens, 
$D_{\rm l}$, multiplied by the distance ratio, $D_{\rm s}/D_{\rm ls}$,
the sensitivity of $D_{\Delta t}$ to cosmological parameters is somewhat
limited compared to that of $D_{\rm l}$
       \citep{fukugita/futamase/kasai:1990}. On the other hand, if the 
density profile of the lens galaxy is approximately given by
$\rho\propto 1/r^2$,  the observed Einstein radius and velocity
dispersion of the lens galaxy can be used to infer the same distance
ratio, $D_{\rm s}/D_{\rm ls}$, and thus one can use this property to
constrain cosmological parameters as well
\citep{futamase/yoshida:2001,yamamoto/futamase:2001,yamamoto/etal:2001,ohyama/etal:2002,dobke/etal:2009},
up to uncertainties in the density profile
\citep{chiba/takahashi:2002}. By combining measurements of the
time-delay, Einstein ring, and velocity dispersion, one can in principle
measure $D_{\rm l}$ directly, thereby turning strong gravitational lens
systems into standard rulers \citep{paraficz/hjorth:2009}. While the accuracy
of the current data for B1608+656 does not permit us to determine
$D_{\rm l}$ precisely yet (S.~H.~Suyu and P.~J.~Marshall 2009,
private communication), there seems to be exciting future prospects for this
method. Future prospects of the time-delay method are also discussed in
\citet{oguri:2007,coe/moustakas:2009}. 
}
\end{itemize}

\subsection{Treating Massive Neutrinos in $H(a)$ Exactly}
\label{sec:exactnu}
When we evaluate the likelihood of external astrophysical data sets, we
often need to 
compute the Hubble expansion rate, $H(a)$. While we treated the effect
of massive neutrinos on $H(a)$ approximately for the 5-year analysis of
the external data sets presented in \citet{komatsu/etal:2009}, we treat
it exactly for the 7-year analysis, as described below.

The total energy density of massive neutrino species, $\rho_\nu$, is
given by (in natural units)
\begin{equation}
 \rho_\nu(a) = 2\int \frac{d^3p}{(2\pi)^3}\frac1{e^{p/T_\nu(a)}+1}
\sum_{i} \sqrt{p^2+m^2_{\nu,i}},
\end{equation}
where $m_{\nu,i}$ is the mass of each neutrino species. Using the comoving
momentum, $q\equiv pa$, and the present-day neutrino temperature,
$T_{\nu 0}=(4/11)^{1/3}T_{\rm cmb}=1.945$~K, we write
\begin{equation}
 \rho_\nu(a) = \frac1{a^4}\int \frac{q^2dq}{\pi^2}\frac1{e^{q/T_{\nu 0}}+1}
\sum_{i} \sqrt{q^2+m^2_{\nu,i}a^2}.
\label{eq:rhonu}
\end{equation}
Throughout this paper, we shall assume that 
all massive neutrino species have the equal mass $m_\nu$, i.e.,
$m_{\nu,i}=m_\nu$ for all $i$.\footnote{While the current cosmological
data are not yet sensitive to the mass of individual neutrino species,
that is, the mass hierarchy, this situation may change in the future,
with high-$z$ galaxy redshift surveys or weak lensing surveys
\citep{takada/komatsu/futamase:2006,slosar:2006,hannestad/wang:2007,kitching/etal:2008,abdalla/rawlings:2007}.} 

When neutrinos are relativistic, one may relate $\rho_\nu$
to the photon energy density, $\rho_\gamma$, as
\begin{equation}
 \rho_\nu(a)\to \frac78 \left(\frac{4}{11}\right)^{4/3} N_{\rm
  eff}\rho_\gamma(a)
\simeq 0.2271 N_{\rm eff}\rho_\gamma(a),
\label{eq:neff}
\end{equation}
where $N_{\rm eff}$ is the effective number of neutrino species.
Note that $N_{\rm eff}=3.04$ for the standard neutrino
species.\footnote{A recent estimate gives $N_{\rm eff}=3.046$ \citep{mangano/etal:2005}.}
This motivates our writing equation~(\ref{eq:rhonu}) as
\begin{equation}
\rho_\nu(a) =  0.2271{N_{\rm eff}}\rho_\gamma(a) 
f(m_{\nu}a/T_{\nu 0}),
\label{eq:rhonu2}
\end{equation}
where 
\begin{equation}
 f(y)\equiv \frac{120}{7\pi^4}
\int_0^\infty dx\frac{x^2\sqrt{x^2+y^2}}{e^x+1}.
\end{equation}
The limits of this function are $f(y)\to 1$ for $y\to 0$, and 
$f(y)\to \frac{180\zeta(3)}{7\pi^4}y$ for $y\to
\infty$, where $\zeta(3)\simeq 1.202$ is the Riemann zeta function. We
find that $f(y)$ can be 
approximated by the following fitting formula:\footnote{Also see
Section~5 of \citet{wright:2006}, where $\rho_\nu$ is normalized by the
density in the non-relativistic limit. Here, $\rho_\nu$ is normalized by the
density in the relativistic limit. Both results agree with the same precision.}
\begin{equation}
 f(y)\approx \left[1+(Ay)^p\right]^{1/p},
\end{equation}
where $A=\frac{180\zeta(3)}{7\pi^4}\simeq 0.3173$ and $p=1.83$. This
fitting formula is constructed such that it reproduces the asymptotic
limits in $y\to 0$ and $y\to\infty$ exactly. This fitting formula
underestimates $f(y)$ by 0.1\% at $y\simeq 2.5$, and overestimates by
0.35\% at $y\simeq 10$. The errors are smaller than these values at
other $y$'s.

Using this result, we write the Hubble expansion rate as
\begin{eqnarray}
\nonumber
 H(a) &=& H_0\left\{
\frac{\Omega_c+\Omega_b}{a^3}+\frac{\Omega_\gamma}{a^4}
\left[1+0.2271N_{\rm eff} f(m_{\nu}a/T_{\nu
 0})\right]\right.\\
& &\left.+
\frac{\Omega_k}{a^2} + \frac{\Omega_\Lambda}{a^{3(1+w_{\rm eff}(a))}}
\right\}^{1/2},
\label{eq:hubble}
\end{eqnarray}
where $\Omega_\gamma=2.469\times 10^{-5}h^{-2}$ for $T_{\rm cmb} =
2.725$~K. 
Using the massive neutrino density parameter, $\Omega_\nu h^2 = \sum
m_\nu/(94~{\rm eV})$, for the standard 3 neutrino species, we find
\begin{equation}
 \frac{m_\nu a}{T_{\nu 0}}
= \frac{187}{1+z}\left(\frac{\Omega_\nu h^2}{10^{-3}}\right).
\end{equation}
One can check that $(\Omega_\gamma/a^4) 0.2271N_{\rm eff} f(m_{\nu}a/T_{\nu
 0})\to \Omega_\nu/a^3$ for $a\to \infty$.
One may compare equation~(\ref{eq:hubble}), which is exact (if we
 compute $f(y)$ exactly), to
equation~(7) of \citet{komatsu/etal:2009}, which is approximate.

Throughout this paper, we shall use $\Omega_\Lambda$ to denote the dark energy
density parameter at present: $\Omega_\Lambda \equiv \Omega_{de}(z=0)$.
The function $w_{\rm eff}(a)$ in equation~(\ref{eq:hubble}) is the
effective equation of state of dark energy given by 
$w_{\rm eff}(a)\equiv \frac1{\ln a}\int_0^{\ln a} d\ln a'~w(a')$,
and $w(a)$ is the usual dark energy equation of state, i.e., the dark
energy pressure divided by the dark energy density:
$w(a)\equiv{P_{de}(a)}/{\rho_{de}(a)}$.
For vacuum energy (cosmological constant), $w$ does not depend on time,
and $w=-1$.

\section{Cosmological Parameters Update Except For Dark
 Energy}\label{sec:parameters} 

\begin{deluxetable*}{lccccc}
\tablecolumns{6}
\small
\tablewidth{0pt}
\tablecaption{Primordial tilt $n_s$, running index $dn_s/d\ln k$, and
 tensor-to-scalar ratio $r$}
\tablehead{
\colhead{Section}
& \colhead{Model} 
& \colhead{Parameter\tablenotemark{a}}
& \colhead{7-year \map\tablenotemark{b}}
& \colhead{\map+ACBAR+QUaD\tablenotemark{c}}
& \colhead{\map+BAO+$H_0$}
}
 \startdata
 Section~\ref{sec:GW} 
& Power-law\tablenotemark{d}
& $n_s$ 
& $0.967\pm 0.014$ 
& $0.966^{+0.014}_{-0.013}$ 
& $0.968\pm 0.012$ 
\nl
\hline
 Section~\ref{sec:running} 
& Running 
& $n_s$ 
& \ensuremath{1.027^{+ 0.050}_{- 0.051}}
\tablenotemark{e}
& \ensuremath{1.041^{+ 0.045}_{- 0.046}} 
& \ensuremath{1.008\pm 0.042}\tablenotemark{f}
\nl
& 
& $dn_s/d\ln k$ 
& \ensuremath{-0.034\pm 0.026} 
& \ensuremath{-0.041^{+ 0.022}_{- 0.023}}
& \ensuremath{-0.022\pm 0.020}
\nl
\hline
  Section~\ref{sec:GW} 
& Tensor 
& $n_s$ 
& \ensuremath{0.982^{+ 0.020}_{- 0.019}} 
& \ensuremath{0.979^{+ 0.018}_{- 0.019}}
& \ensuremath{0.973\pm 0.014} 
\nl
& 
& $r$ 
& \ensuremath{< 0.36\ \mbox{(95\% CL)}} 
& \ensuremath{< 0.33\ \mbox{(95\% CL)}} 
& \ensuremath{< 0.24\ \mbox{(95\% CL)}}
\nl 
\hline
Section~\ref{sec:running} 
& Running 
& $n_s$ 
& \ensuremath{1.076\pm 0.065} 
& %
& \ensuremath{1.070\pm 0.060}
\nl
&+Tensor
& $r$ 
& \ensuremath{< 0.49\ \mbox{(95\% CL)}} 
& N/A%
& \ensuremath{< 0.49\ \mbox{(95\% CL)}}
\nl
& 
& $dn_s/d\ln k$ 
& \ensuremath{-0.048\pm 0.029} 
& %
& \ensuremath{-0.042\pm 0.024}
\enddata
\tablenotetext{a}{Defined at $k_0=0.002~{\rm Mpc}^{-1}$.}
\tablenotetext{b}{\citet{larson/etal:prep}.}
\tablenotetext{c}{ACBAR \citep{reichardt/etal:2009}; QUaD
 \citep{brown/etal:2009}.} 
\tablenotetext{d}{The parameters in this row are based on {\sf RECFAST}
 version 1.5 (see Appendix~\ref{app:comparison}), while the parameters
 in all the other rows are based on {\sf RECFAST}
 version 1.4.2.}
\tablenotetext{e}{At the pivot point for \map\ only, where $n_s$ and
 $dn_s/d\ln k$ are uncorrelated, $n_s(k_{\rm pivot})=0.964\pm 0.014$. 
The ``pivot wavenumber'' may be defined in two ways: (i)
$k_{\rm pivot}=0.0805~{\rm
 Mpc}^{-1}$ from $n_s(k_{\rm pivot})=n_s(k_0)+\frac12(dn_s/d\ln
 k)\ln(k_{\rm pivot}/k_0)$,
or (ii)
$k_{\rm pivot}=0.0125~{\rm Mpc}^{-1}$
from $\left.d\ln\Delta^2_{\cal R}/d\ln k\right|_{k=k_{\rm
 pivot}}=n_s(k_0)-1+(dn_s/d\ln k)\ln(k_{\rm pivot}/k_0)$.
}
\tablenotetext{f}{At the pivot point for \map+BAO+$H_0$,  where $n_s$ and $dn_s/d\ln k$ are uncorrelated,
 $n_s(k_{\rm pivot})=0.964\pm 0.013$.
The ``pivot wavenumber'' may be defined in two ways: (i)
$k_{\rm pivot}=0.106~{\rm
 Mpc}^{-1}$ from $n_s(k_{\rm pivot})=n_s(k_0)+\frac12(dn_s/d\ln
 k)\ln(k_{\rm pivot}/k_0)$,
or (ii)
$k_{\rm pivot}=0.0155~{\rm Mpc}^{-1}$
from $\left.d\ln\Delta^2_{\cal R}/d\ln k\right|_{k=k_{\rm
 pivot}}=n_s(k_0)-1+(dn_s/d\ln k)\ln(k_{\rm pivot}/k_0)$.
}
\label{tab:ns}
\end{deluxetable*}

\subsection{Primordial Spectral Index and Gravitational Waves}
\label{sec:GW}
\begin{figure}[t]
\centering \noindent
\includegraphics[width=8cm]{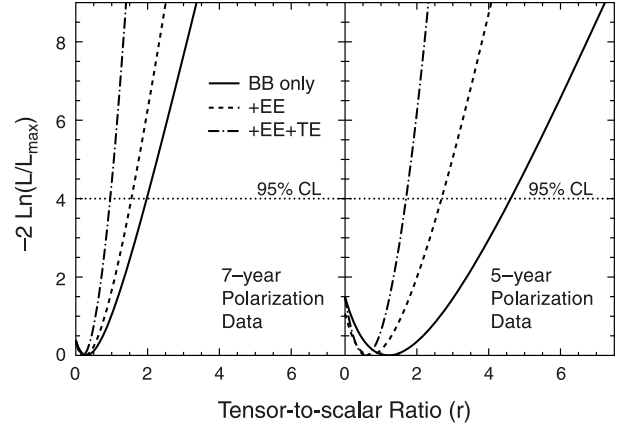}
\caption{%
Limits on the tensor-to-scalar ratio, $r$, from the polarization data
 (BB, EE and TE) alone. All the other cosmological parameters, including
 the optical depth, are fixed at the 5-year best-fit $\Lambda$CDM model
 \citep{dunkley/etal:2009}. 
 The vertical axis shows $-2\ln(L/L_{\rm max})$,
 where $L$ is the likelihood and $L_{\rm max}$ is the maximum
 value. This quantity may be interpreted as the standard $\chi^2$, as
 the likelihood is approximately a Gaussian near the maximum; thus,
 $-2\ln(L/L_{\rm max})=4$ corresponds to the 95.4\%~CL limit.
The solid, dashed and dot-dashed lines show the likelihood as a function
 of $r$ from the BB-only, BB$+$EE, and BB$+$EE$+$TE data. 
(Left) The 7-year polarization data. We find $r<2.1$, 1.6, and 0.93
 (95.4\%~CL) from the BB-only, BB$+$EE, and BB$+$EE$+$TE data,
 respectively.
(Right) The 5-year polarization data. 
We find $r<4.7$, 2.7, and 1.6
 (95.4\%~CL) from the BB-only, BB$+$EE, and BB$+$EE$+$TE data,
 respectively.
} 
\label{fig:tens}
\end{figure}

The 7-year \map\ data combined with BAO and $H_0$ exclude the
scale-invariant spectrum by 99.5\%~CL, if we ignore tensor
modes (gravitational waves). 

For a power-law power spectrum of primordial curvature
perturbations ${\cal R}_k$, i.e., 
\begin{equation}
 \Delta^2_{\cal R}(k) = \frac{k^3\langle |{\cal R}_k|^2\rangle}{2\pi^2}
= \Delta^2_{\cal R}(k_0)\left(\frac{k}{k_0}\right)^{n_s-1},
\label{eq:powerlaw}
\end{equation}
where $k_0=0.002~{\rm Mpc}^{-1}$, we find 
$$
n_s=0.968\pm 0.012~(68\%~{\rm CL}).
$$
For comparison, the \map\ data-only limit is 
$n_s=0.967\pm 0.014$
\citep{larson/etal:prep}, and 
the \map\ plus the small-scale CMB experiments ACBAR
\citep{reichardt/etal:2009} and QUaD \citep{brown/etal:2009} is
$n_s=0.966^{+0.014}_{-0.013}$.
As explained in Section~3.1.2 of
\citet{komatsu/etal:2009}, the small-scale CMB data do not reduce the
error bar in $n_s$ very much because of relatively large statistical errors, beam
errors, and calibration errors. 

How about tensor modes? While the B-mode polarization is a smoking-gun
for tensor modes
\citep{seljak/zaldarriaga:1997,kamionkowski/kosowsky/stebbins:1997}, the
\map\ data mainly constrain the amplitude of tensor modes by the
low-$l$ temperature power spectrum \citep[see Section~3.2.3
of][]{komatsu/etal:2009}. Nevertheless, it is still useful to see how
much constraint one can obtain from the 7-year polarization data. 

We first fix the cosmological parameters at the 5-year \map\ best-fit
values of a power-law $\Lambda$CDM model. We then calculate the tensor
mode contributions to the B-mode, E-mode, and TE power spectra as a
function of one parameter: the amplitude, in the form of the
tensor-to-scalar ratio, $r$, defined as
\begin{equation}
 r \equiv \frac{\Delta_h^2(k_0)}{\Delta_{\cal R}^2(k_0)},
\end{equation}
where $\Delta_h^2(k)$ is the power spectrum of tensor metric perturbations,
$h_k$, given by
\begin{equation}
 \Delta_h^2(k) = \frac{4k^3\langle |h_k|^2\rangle}{2\pi^2}
= \Delta^2_{h}(k_0)\left(\frac{k}{k_0}\right)^{n_t}.
\end{equation}

In Figure~\ref{fig:tens}, we show the limits on $r$ from the B-mode
power spectrum only ($r<2.1$, 95\%~CL), from the B- and E-mode
power spectra combined ($r<1.6$), and from the B-mode, E-mode, and TE
power spectra combined ($r<0.93$). These limits are significantly better
than those from the 
5-year data ($r<4.7$, 2.7, and 1.6, respectively), because of the
smaller noise and shifts in the best-fitting values.
For comparison, the B-mode power spectrum from the BICEP 2-year data
gives $r<0.73$ \citep[95\%~CL;][]{chiang/etal:2010}.

If we add the temperature power spectrum, but still fix all the other
 cosmological parameters including $n_s$, then we find $r<0.15$~(95\%~CL)
 from both 5-year and 7-year data; however, due to a strong correlation
 between $n_s$ and $r$, this would be an underestimate of the error. 
 For a 7-parameter model (a flat $\Lambda$CDM model with a tilted
 spectrum, tensor modes, and $n_t=-r/8$), we find 
\ensuremath{r < 0.36\ \mbox{(95\% CL)}} from the
 \map\ data alone \citep{larson/etal:prep}, 
\ensuremath{r < 0.33\ \mbox{(95\% CL)}} 
from \map\  plus ACBAR and QUaD, 
$$
\ensuremath{r < 0.24\ \mbox{(95\% CL)}} 
$$
from \map+BAO+$H_0$, and 
\ensuremath{r < 0.20\ \mbox{(95\% CL)}} 
from
 \map+BAO+SN, where ``SN'' is the Constitution samples compiled by
 \citet{hicken/etal:2009} (see Section~\ref{sec:sn}). 

We give a summary
 of these numbers in Table~\ref{tab:ns}.

\subsection{Running Spectral Index}
\label{sec:running}

Let us relax the assumption that the power spectrum is a pure power-law,
and add a ``running index,'' $dn_s/d\ln k$ as \citep{kosowsky/turner:1995}
\begin{equation}
 \Delta^2_{\cal R}(k) 
= \Delta^2_{\cal
R}(k_0)\left(\frac{k}{k_0}\right)^{n_s(k_0)-1+\frac12\ln(k/k_0)dn_s/d\ln
k}.
\end{equation}
Ignoring  tensor modes again, we find 
$$
\ensuremath{dn_s/d\ln{k} = -0.022\pm 0.020}~(68\%~{\rm CL}),
$$
from \map+BAO+$H_0$. 
For comparison, the \map\ data-only limit is 
\ensuremath{dn_s/d\ln{k} = -0.034\pm 0.026}
\citep{larson/etal:prep}, and the \map+ACBAR+QUaD limit is 
\ensuremath{dn_s/d\ln{k} = -0.041^{+ 0.022}_{- 0.023}}.

None of these data combinations require $dn_s/d\ln k$: improvements in a
goodness-of-fit relative to a power-law model
(equation~(\ref{eq:powerlaw})) are $\Delta\chi^2=-2\ln (L_{\rm running}/L_{\rm
power-law})=-1.2$, $-2.6$, and $-0.72$ for the \map-only,
\map+ACBAR+QUaD, and \map+BAO+$H_0$, respectively.
See Table~\ref{tab:ns} for the case where both $r$ and $dn_s/d\ln k$ are
allowed to vary. 

A simple power-law primordial power spectrum without tensor modes
continues to be an excellent  fit to the data. 
While we have not done a non-parametric study of the shape of the power
spectrum, recent studies after the 5-year data release continue to show that
there is no convincing deviation from a simple power-law spectrum
\citep{peiris/verde:2010,ichiki/nagata/yokoyama:2010,hamann/shafieloo/souradeep:2010}. 

\subsection{Spatial Curvature}
\label{sec:OK}

While the \map\ data alone cannot constrain the spatial curvature
parameter of the observable universe, $\Omega_k$, very well, combining
the \map\ data with other distance indicators such as $H_0$, BAO, or supernovae
can constrain $\Omega_k$ \citep[e.g.,][]{spergel/etal:2007}.
  
Assuming a $\Lambda$CDM model ($w=-1$), we find
$$
\ensuremath{-0.0133<\Omega_k<0.0084\ \mbox{(95\% CL)}},
$$
from \map+BAO+$H_0$.\footnote{The
68\%~CL limit is 
 \ensuremath{\Omega_k = -0.0023^{+ 0.0054}_{- 0.0056}}.
}
However, the limit weakens significantly if dark energy is allowed to be
dynamical, $w\ne -1$, as this data combination, \map+BAO+$H_0$, cannot constrain
$w$ very well. We need additional information from Type Ia supernovae to
constrain $w$ and 
$\Omega_k$ simultaneously \citep[see Section~5.3
of][]{komatsu/etal:2009}. We shall explore this possibility in
Section~\ref{sec:darkenergy}.

\subsection{Non-Adiabaticity: Implications for Axions}
\label{sec:AD}

Non-adiabatic fluctuations are a powerful probe of the origin of matter
and the physics of inflation. Following  Section~3.6 of
\citet{komatsu/etal:2009}, we focus on
two physically motivated models for non-adiabatic fluctuations:
axion-type
\citep{seckel/turner:1985,linde:1985,linde:1991,turner/wilczek:1991} and
curvaton-type
\citep{linde/mukhanov:1997,lyth/wands:2003,moroi/takahashi:2001,moroi/takahashi:2002,bartolo/liddle:2002}. 

For both cases, we consider non-adiabatic fluctuations between photons
and cold dark matter: 
\begin{equation}
 {\cal S}=\frac{\delta\rho_c}{\rho_c}-\frac{3\delta\rho_\gamma}{4\rho_\gamma},
\end{equation}
and report  limits on the ratio of the power spectrum of ${\cal S}$
and that of the curvature perturbation ${\cal R}$
\citep[e.g.,][]{bean/dunkley/pierpaoli:2006}:
\begin{equation}
 \frac{\alpha(k_0)}{1-\alpha(k_0)}
= \frac{P_{\cal S}(k_0)}{P_{\cal R}(k_0)},
\end{equation}
where $k_0=0.002~{\rm Mpc}^{-1}$. We denote the limits on axion-type and
curvaton-type by $\alpha_0$ and $\alpha_{-1}$, respectively.\footnote{
The limits on $\alpha$ can also be converted into the numbers showing
``how much the adiabatic relation (${\cal S}=0$) can be violated,''
$\delta_{\rm adi}$, which can be calculated from
\begin{equation}
 \delta_{\rm
  adi}=\frac{\delta\rho_c/\rho_c-{3\delta\rho_\gamma}/(4\rho_\gamma)}{\frac12[\delta\rho_c/\rho_c+{3\delta\rho_\gamma}/(4\rho_\gamma)]}
\approx \frac{\sqrt{\alpha}}{3},
\end{equation}
for $\alpha\ll 1$ \citep{komatsu/etal:2009}.}

We find no evidence for non-adiabatic fluctuations. The \map\ data-only
limits are 
\ensuremath{\alpha_{0} < 0.13\ \mbox{(95\% CL)}}
and
\ensuremath{\alpha_{-1} < 0.011\ \mbox{(95\% CL)}}
\citep[95\%~CL;][]{larson/etal:prep}. 
With \map+BAO+$H_0$, we find 
$$
\ensuremath{\alpha_{0} < 0.077\ \mbox{(95\% CL)}}~
\mbox{and}~
\ensuremath{\alpha_{-1} < 0.0047\ \mbox{(95\% CL)}},
$$ 
 while with \map+BAO+SN, we find
\ensuremath{\alpha_{0} < 0.064\ \mbox{(95\% CL)}}
and
\ensuremath{\alpha_{-1} < 0.0037\ \mbox{(95\% CL)}}.

The limit on $\alpha_0$ has an important implication for axion dark
matter. In particular, a limit on $\alpha_0$ is related to a limit on
the tensor-to-scalar ratio, $r$
\citep{kain:2006,beltran/garcia-bellido/lesgourgues:2007,sikivie:2008,kawasaki/sekiguchi:2008}. The
explicit formula is given by equation~(48) of \citet{komatsu/etal:2009}
as\footnote{This formula assumes that the axion field began to oscillate
before the QCD phase transition. The formula in the other limit will be
given later. We shall assume that the energy density of the universe was
dominated by 
radiation when the axion field began to oscillate; however, this may not always
be true \citep{kawasaki/moroi/yanagida:1996,kawasaki/takahashi:2005}
when there was a significant amount of entropy production after the QCD
phase transition, i.e., $\gamma\ll 1$.}
\begin{eqnarray}
 r &=& \frac{4.7\times 10^{-12}}{\theta_a^{10/7}}\left(\frac{\Omega_ch^2}{\gamma}\right)^{12/7}
\left(\frac{\Omega_c}{\Omega_a}\right)^{2/7}
\frac{\alpha_{0}}{1-\alpha_{0}},
\label{eq:rlimitfromaxion}
\end{eqnarray}
where 
$\Omega_a\le \Omega_c$ is the axion density parameter, 
$\theta_a$ is the phase of the Peccei-Quinn field
within our observable universe, and
$\gamma\le 1$ is a ``dilution factor'' representing the amount by
which the axion density parameter, $\Omega_ah^2$, would have been
diluted due to a potential late-time entropy production by, e.g., decay
of some (unspecified) heavy particles, between 200~MeV and the epoch of
nucleosynthesis, 1~MeV.

 Where does this formula come from? Within the
context of the ``misalignment'' scenario of axion dark
matter\footnote{We make the following assumptions: the Peccei-Quinn
symmetry was broken during inflation but before the fluctuations we observe
today left the horizon, and was not restored before
or after the end of inflation (reheating). That the Peccei-Quinn
symmetry was not restored 
before reheating requires
the expansion rate during inflation not to exceed the axion decay
constant, $H_{\rm inf}<f_a$ \citep{lyth/stewart:1992}.
That the Peccei-Quinn symmetry was not restored
after reheating requires the reheating temperature after inflation not to exceed
$f_a$.}, there are 
two observables one can use to place limits on the axion properties: the
dark matter density and $\alpha_0$. They are given by
\citep[e.g.,][and references therein]{kawasaki/sekiguchi:2008}
\begin{eqnarray}
\label{eq:alpha0axion}
 \frac{\alpha_0(k)}{1-\alpha_0(k)}
&=& \frac{\Omega^2_a}{\Omega^2_c}\frac{8\epsilon}{\theta^2_a(f_a/M_{\rm
pl})^2},\\
\Omega_ah^2 &=& 1.0\times 10^{-3}\gamma\theta_a^2\left(\frac{f_a}{10^{10}~{\rm
					 GeV}}\right)^{7/6},
\label{eq:oah2}
\end{eqnarray}
where $f_a$ is the axion decay constant, and
$\epsilon=-\dot{H}_{\rm inf}/H^2_{\rm inf}$ is the so-called slow-roll
parameter (where $H_{\rm inf}$
is the Hubble expansion rate during inflation).
For single-field inflation models, $\epsilon$ is
related to $r$ as $r=16\epsilon$.
By eliminating the axion decay constant, one obtains
equation~(\ref{eq:rlimitfromaxion}). 

In deriving the above formula for $\Omega_ah^2$
(equation~(\ref{eq:oah2})), we have assumed that the axion field began
to oscillate before the QCD phase transition.\footnote{Specifically, the
temperature at which the axion field began to oscillate, $T_1$, can be
calculated from the condition $3H(T_1)=m_a(T_1)$, where
$m_a(T)\approx 0.1m_{a0}(0.2~{\rm GeV}/T)^{4}$ is the mass of axions before
the QCD phase transition, $T\gtrsim 0.2~{\rm GeV}$, and $m_{a0}=13~{\rm
MeV}(1~{\rm GeV}/f_a)$ is the mass of axions at the zero
temperature. Here, we have used the pion decay constant of
$F_\pi=184~{\rm MeV}$ to calculate $m_{a0}$, following equation~(3.4.16) of
\citet{weinberg:COS}. The Hubble expansion rate during radiation era is
given by $M_{pl}^2H^2(T)=(\pi^2/90)g_*T^4$, where $M_{pl}=2.4\times
10^{18}$~GeV is the reduced Planck mass and $g_*$ is the number of
relativistic degrees of freedom. Before the QCD phase transition,
$g_*=61.75$. After the QCD phase transition but before the
electron-positron annihilation, $g_*=10.75$.}
This is true when $f_a<{\cal O}(10^{-2})M_{pl}$; however, when
$f_a>{\cal O}(10^{-2})M_{pl}$, the axions are so light that the axion
field would not start oscillating after the QCD phase
transition.\footnote{This dividing point, $f_a={\cal
O}(10^{-2})M_{pl}$, can be found from the condition $T_1=0.2~{\rm GeV}$
and $3H(T_1)=m_a(T_1)$. See \citet{hertzberg/tegmark/wilczek:2008} for
more accurate numerical estimate. Note that
\citet{hertzberg/tegmark/wilczek:2008} used $F_\pi=93~{\rm MeV}$ for the
pion decay constant when calculating the axion mass at the zero temperature.}
In this limit, the formula for $\Omega_ah^2$ is given by
\begin{equation}
 \Omega_ah^2=1.6\times 10^5\gamma\theta_a^2\left(\frac{f_a}{10^{17}~{\rm
					    GeV}}\right)^{3/2}.
\label{eq:oah22}
\end{equation}
By eliminating $f_a$ from equation~(\ref{eq:alpha0axion}) and
(\ref{eq:oah22}), we obtain another formula for $r$:
\begin{equation}
 \label{eq:rlimitfromaxion2}
r=\frac{4.0\times 10^{-10}}{\theta_a^{2/3}}
\left(\frac{\Omega_ch^2}{\gamma}\right)^{4/3}
\left(\frac{\Omega_c}{\Omega_a}\right)^{2/3}
\frac{\alpha_{0}}{1-\alpha_{0}}.
\end{equation}

Equation~(\ref{eq:rlimitfromaxion}) and (\ref{eq:rlimitfromaxion2}),
combined with our limits on $\Omega_ch^2$ 
and $\alpha_0$, implies that the axion dark matter
scenario in which axions account for most of the observed amount of dark
matter, $\Omega_a\sim \Omega_c$, must satisfy 
\begin{eqnarray}
 r&<&\frac{7.6\times 10^{-15}}{\theta_a^{10/7}\gamma^{12/7}}
\quad \mbox{for $f_a<{\cal O}(10^{-2})M_{pl}$},\\
 r&<&\frac{1.5\times 10^{-12}}{\theta_a^{2/3}\gamma^{4/3}}
\quad \mbox{for $f_a>{\cal O}(10^{-2})M_{pl}$}.
\end{eqnarray}
Alternatively, one can express this constraint as
\begin{eqnarray}
\nonumber
 \theta_a\gamma^{6/5}&<&3.3\times
  10^{-9}\left(\frac{10^{-2}}{r}\right)^{7/10}
 \mbox{for $f_a<{\cal O}(10^{-2})M_{pl}$},\\
\nonumber
 \theta_a\gamma^{2}&<&1.8\times
  10^{-15}\left(\frac{10^{-2}}{r}\right)^{3/2}
 \mbox{for $f_a>{\cal O}(10^{-2})M_{pl}$}.
\end{eqnarray}
Therefore, a future detection of tensor modes at the level of
$r=10^{-2}$ would imply a fine-tuning of $\theta_a$ or $\gamma$ or both
of these parameters \citep{komatsu/etal:2009}.
If such fine-tunings are not permitted, axions cannot account for the
observed abundance of dark matter (in the misalignment scenario that we
have considered here).

Depending on one's interest, one may wish to eliminate the phase,
leaving the axion decay constant in the formula
\citep[see equation~(B7) of][]{komatsu/etal:2009}:
\begin{equation}
  r = (1.6\times
  10^{-12})\left(\frac{\Omega_ch^2}{\gamma}\right)
\left(\frac{\Omega_c}{\Omega_a}\right)\left(\frac{f_a}{10^{12}~{\rm
					 GeV}}\right)^{5/6}
\frac{\alpha_0}{1-\alpha_0},
\end{equation}
for $f<{\cal O}(10^{-2})M_{pl}$. This formula gives
\begin{equation}
 f_a>1.8\times 10^{26}~{\rm
  GeV}~\gamma^{6/5}\left(\frac{r}{10^{-2}}\right)^{6/5},
\end{equation}
which is inconsistent with the condition $f_a<{\cal O}(10^{-2})M_{pl}$ (unless
$r$ is extremely small). The formula that is valid for $f>{\cal
O}(10^{-2})M_{pl}$ is 
\begin{equation}
r= (2.2\times
  10^{-8})\left(\frac{\Omega_ch^2}{\gamma}\right)
\left(\frac{\Omega_c}{\Omega_a}\right)\left(\frac{f_a}{10^{17}~{\rm
					 GeV}}\right)^{1/2}
\frac{\alpha_0}{1-\alpha_0},
\end{equation}
which gives
\begin{equation}
 f_a>3.2\times 10^{32}~{\rm
  GeV}~\gamma^{2}\left(\frac{r}{10^{-2}}\right)^{2}. 
\end{equation}
Requiring $f_a<M_{pl}=2.4\times 10^{18}$~GeV, we obtain
\begin{equation}
 r < \frac{8.7\times 10^{-10}}{\gamma}.
\end{equation}
Thus, a future detection of tensor modes at the level of $r=10^{-2}$
implies a significant amount of entropy production, $\gamma\ll 1$, or a
super-Planckian axion decay constant, $f_a\gg M_{pl}$, or both.
Also see \citet{hertzberg/tegmark/wilczek:2008,mack:prep,mack/steinhardt:prep}
for similar studies.

For the implications of $\alpha_{-1}$ for curvaton dark matter, see
Section~3.6.4 of \citet{komatsu/etal:2009}.

\subsection{Parity Violation}
\label{sec:TB}

While the TB and EB correlations vanish in a parity-conserving universe,
they may not vanish when global parity symmetry is broken on
cosmological scales \citep{lue/wang/kamionkowski:1999,carroll:1998}.
In pixel space, they would show up as a non-vanishing $\langle
U_r\rangle$. As we showed already in Section~\ref{sec:chi2}, 
the \map\
7-year $\langle
U_r\rangle$ data are consistent with noise. What can we learn from this? 

It is now a routine work of CMB experiments to deliver the TB and EB
data, and constrain a rotation angle of the polarization plane 
due to a parity-violating effect (or a rotation due to
some systematic error). Specifically, a 
rotation of the polarization plane by an angle $\Delta\alpha$ gives the
following 5 
transformations:
%
%
\begin{eqnarray}
 C_l^{\rm TE,obs} &=& C_l^{\rm TE}\cos(2\Delta\alpha),\\
 C_l^{\rm TB,obs} &=& C_l^{\rm TE}\sin(2\Delta\alpha),\\
 C_l^{\rm EE,obs} &=& C_l^{\rm EE}\cos^2(2\Delta\alpha),\\
 C_l^{\rm BB,obs} &=& C_l^{\rm EE}\sin^2(2\Delta\alpha),\\
 C_l^{\rm EB,obs} &=& \frac12C_l^{\rm EE}\sin(4\Delta\alpha),
\end{eqnarray}
where $C_l$'s on the right hand side 
are the primordial power spectra in the absence
of rotation, while $C_l^{\rm obs}$'s on the left hand side are what
we would observe in the presence of rotation. 

Note that these equations are not exact but valid only when 
the primordial $B$-mode polarization is negligible compared to the
$E$-mode polarization, i.e., $C_l^{BB}\ll C_l^{EE}$. For the full
expression including $C_l^{BB}$, see 
 \citet{lue/wang/kamionkowski:1999} and \citet{feng/etal:2005}.

Roughly speaking, when the polarization data are still dominated by
noise rather than by the cosmic signal, the uncertainty in
$\Delta\alpha$ is given by a half of the inverse of the signal-to-noise
ratio of TE or EE, i.e., 
\begin{eqnarray}
{\rm Err}[\Delta\alpha^{\rm TB}]
\nonumber
&\simeq&
\frac1{2(S/N)^{\rm TE}},\\
\nonumber
{\rm Err}[\Delta\alpha^{\rm EB}]
&\simeq&
\frac1{2(S/N)^{\rm EE}}.
\end{eqnarray}
(Note that we use the full likelihood code to find the best-fitting values
and error bars. These equations should only be used to provide
an intuitive feel of how the errors scale with signal-to-noise.)
As we mentioned in the last paragraph of Section~\ref{sec:chi2}, with
the 7-year polarization data we
detect the TE power spectrum at $21\sigma$ from
$l=24$ to $800$. We thus expect
${\rm Err}[\Delta\alpha^{\rm TB}]\simeq 1/42\simeq 0.024~{\rm radian}\simeq
1.4^\circ$, which is significantly better than the 5-year value,
$2.2^\circ$ \citep{komatsu/etal:2009}. On the other hand, we detect the
EE power spectrum at $l\ge 24$ only at a few $\sigma$ level, and thus
${\rm Err}[\Delta\alpha^{\rm EB}]\gg {\rm Err}[\Delta\alpha^{\rm TB}]$,
implying that we may ignore the high-$l$ EB data.

The magnitude of polarization rotation angle, $\Delta\alpha$, depends on
the path length over which photons experienced a parity-violating
interaction. As pointed out by \citet{liu/lee/ng:2006}, this leads to
the polarization angle that depends on $l$. We can divide this
$l$-dependence in two regimes: (i) $l\lesssim 20$: the polarization
signal was generated during reionization \citep{zaldarriaga:1997}. We
are sensitive only to the polarization rotation between the reionization
epoch and present epoch. (ii) $l\gtrsim 20$: the polarization signal was generated at the decoupling epoch.  We are sensitive to the polarization rotation between the decoupling epoch and present epoch; thus, we have the largest path length in this case.

Using the high-$l$ TB data from $l=24$ to $800$, we find
$\Delta\alpha=-0.9^\circ\pm 1.4^\circ$, which is a significant improvement
over the 5-year high-$l$ result, $\Delta\alpha=-1.2^\circ\pm 2.2^\circ$
\citep{komatsu/etal:2009}.  

Let us turn our attention to lower multipoles, $l\le 23$. Here, with the
7-year polarization data, 
the EE power spectrum is detected at $5.1\sigma$, whereas the TE power
spectrum is only marginally seen ($1.9\sigma$). 
(The overall significance level of detection of the E-model polarization
at $l\le 23$, including EE and TE, is $5.5\sigma$.) We 
therefore use both the TB and EB data at $l\le 23$.
We find $\Delta\alpha=-3.8^\circ\pm 5.2^\circ$, which is also a good
improvement over the 5-year low-$l$ value, $\Delta\alpha=-7.5^\circ\pm
7.3^\circ$.  

Combining the low-$l$ TB/EB and high-$l$ TB data, we find  
$\Delta\alpha=-1.1^\circ\pm 1.4^\circ$
(the 5-year combined limit was $\Delta\alpha=-1.7^\circ\pm 2.1^\circ$),
where the quoted error is purely statistical; however, the \map\ instrument
can measure the polarization angle to within $\pm 
1.5^\circ$ of the design orientation
\citep{page/etal:2003,page/etal:2007}. We thus add 
$1.5^\circ$ as an estimate of a potential systematic error. Our final
7-year limit is  
$$
 \Delta\alpha= -1.1^\circ \pm 1.4^\circ~\mbox{({\rm stat.})} \pm 1.5^\circ~\mbox{({\rm
syst.})}~\mbox{(68\%~CL)}, 
$$
or $-5.0^\circ<\Delta\alpha<2.8^\circ$ (95\%~CL), for which we have
added the statistical and systematic errors in quadrature
(which may be an under-estimate of the total error).
The statistical error and systematic error are now comparable.

Several research groups have obtained limits on $\Delta\alpha$ from
various data sets
\citep{feng/etal:2006,kostelecky/mewes:2007,cabella/natoli/silk:2007,xia/etal:2008,xia/etal:2008b,wu/etal:2009,gubitosi/etal:2009}. 
Recently, the BOOMERanG collaboration \citep{pagano/etal:2009}
revisited a limit on $\Delta\alpha$ from their
2003 flight (B2K), taking into account the effect of systematic errors
rotating the polarization angle by $-0.9^\circ\pm 0.7^\circ$. By
removing this, they find $\Delta\alpha=-4.3^\circ\pm 4.1^\circ$ (68\% CL).
The QUaD collaboration used their final data set to
find $\Delta\alpha=
0.64^\circ \pm 0.50^\circ~\mbox{({\rm stat.})} \pm 0.50^\circ~\mbox{({\rm
syst.})}$ \citep[68\%~CL;][]{brown/etal:2009}.
\citet{xia/li/zhang:2010} used the BICEP 2-year data
\citep{chiang/etal:2010} to find $\Delta\alpha=-2.6^\circ\pm 1.0^\circ$
(68\%~CL statistical); however, a systematic error of $\pm 0.7^\circ$ needs
to be added to this error budget \citep[see ``Polarization orientation
uncertainty'' in Table~3 of][]{takahashi/etal:2010}. 
Therefore, basically the systematic errors in recent 
measurements of $\Delta\alpha$ from \map\ 7-year, QUaD final, and BICEP
2-year data are comparable to the statistical errors. 

Adding the
statistical and systematic errors in quadrature and averaging over
\map, QUaD and BICEP with the inverse variance weighting, we find 
$\Delta\alpha=-0.25^\circ\pm 0.58^\circ$ (68\%~CL), or 
$-1.41^\circ<\Delta\alpha<0.91^\circ$ (95\%~CL).
We therefore conclude that the microwave background data are comfortably
consistent with a parity-conserving universe.
See, e.g., \citet{kostelecky/mewes:2008,arvanitaki/etal:prep} and
references therein for 
implications of this result for 
potential violations of Lorentz invariance and CPT symmetry.

\subsection{Neutrino Mass}
\label{sec:massnu}

Following Section~6.1 of \citet{komatsu/etal:2009} (also see references
therein), we constrain the 
total mass of neutrinos, $\sum m_\nu=94~{\rm eV}(\Omega_\nu h^2)$,
mainly from the 7-year \map\ data combined with the distance
information. A new component in the analysis is 
the exact treatment of massive neutrinos when calculating the likelihood
of the BAO data, 
as described in Section~\ref{sec:exactnu} \citep[also see][]{wright:2006}.

For a flat $\Lambda$CDM model, i.e., $w=-1$ and $\Omega_k=0$, the
\map-only limit is 
\ensuremath{\sum m_\nu < 1.3\ \mbox{eV}\ \mbox{(95\% CL)}}, while the \map+BAO+$H_0$ limit is 
$$
\ensuremath{\sum m_\nu < 0.58\ \mbox{eV}\ \mbox{(95\% CL)}}~\quad (\mbox{for}~w=-1).
$$
The latter is the best upper limit on
$\sum m_\nu$ without information on the growth of structure, which is
achieved by a better measurement of the early Integrated Sachs-Wolfe
(ISW) effect through the third acoustic peak of the 7-year temperature
power spectrum \citep{larson/etal:prep}, as well as by a better
determination of $H_0$ from \citet{riess/etal:2009}. For explanations
of this effect, see \citet{ichikawa/fukugita/kawasaki:2005} or
Section~6.1.3 of \citet{komatsu/etal:2009}.

\citet{sekiguchi/etal:2010} combined the 5-year version of
\map+BAO+$H_0$ with the small-scale CMB data to find 
$\sum m_\nu<0.66$~eV~(95\%~CL). 
Therefore, the improvement from this value to our
7-year limit, $\sum m_\nu<0.58~{\rm eV}$, indeed comes from a better
determination of the amplitude of the third acoustic peak in the 7-year
temperature data. 

The limit improves when information on the growth of structure is
added. For example, with \map+$H_0$ and the power spectrum of LRGs
\citep[][see Section~\ref{sec:lrg}]{reid/etal:2010}
combined, we find
\ensuremath{\sum m_\nu < 0.44\ \mbox{eV}\ \mbox{(95\% CL)}} for
$w=-1$.

The \map+BAO+$H_0$ limit on the neutrino mass  weakens significantly to 
\ensuremath{\sum m_\nu < 1.3\ \mbox{eV}\ \mbox{(95\% CL)}}
for $w\ne -1$
because we do not use information of Type Ia supernovae here to constrain
$w$. This is driven by $w$ being too negative: there is an
anti-correlation between $w$ and $\sum m_\nu$ \citep{hannestad:2005b}.
The best-fitting value of
$w$ in this case is 
\ensuremath{w = -1.44\pm 0.27}~(68\%~CL).\footnote{
That the neutrino mass and $w$ are anti-correlated implies 
that the neutrino mass limit would improve if we impose a prior on $w$ as $w\ge
-1$.}
For \map+LRG+$H_0$, we find $\sum m_\nu<0.71~{\rm
eV}$~(95\%~CL) for $w\ne -1$.
When the Constitution supernova data are included (\map+BAO+SN), we find $\sum
m_\nu<0.71$\footnote{The 7-year \map+BAO+SN limit for $w=-1$ is slightly
weaker than the 5-year 
\map+BAO+SN limit, 0.67~eV.  The 5-year limit was derived using an
approximate treatment 
of the effect of massive neutrinos on $r_s/D_V(z)$.
The 7-year limit
we quote here, which uses the exact treatment of massive neutrinos
(Section~\ref{sec:exactnu}),  is more reliable.
} and 0.91~eV (95\%~CL) for $w=-1$ and $w\ne -1$,
respectively. 

Recent studies after the 5-year data release combined the \map\
5-year data with information on the growth of structure to find
various improved limits. \citet{vikhlinin/etal:2009} added the abundance
of X-ray-selected clusters of galaxies, which were found in the {\sl 
ROSAT} All Sky 
Survey and followed up by the {\sl Chandra} X-ray Observatory \citep[their
cluster catalog is described in][]{vikhlinin/etal:2009b}, to 
the \map\ 5-year data, the BAO measurement from
\citet{eisenstein/etal:2005}, and the Type Ia supernova data from
\citet{davis/etal:2007}, to find
$\sum 
m_\nu<0.33$~eV (95\%~CL) for $w\ne -1$. 
\citet{mantz/etal:2010b} added a different cluster catalog, also derived
from the {\sl ROSAT} All Sky 
Survey and followed up by the {\sl Chandra} X-ray Observatory
\citep[their
cluster catalog is described in][]{mantz/etal:2010c}, and the measurement
of the gas mass fraction of relaxed clusters \citep{allen/etal:2008}
to 
the \map\ 5-year data, the BAO measurement from
\citet{percival/etal:2007c}, and the ``Union'' Type Ia supernova samples from
\citet{kowalski/etal:2008} \citep[all of which constitute the 5-year
``\map+BAO+SN'' set in][]{komatsu/etal:2009}, to find
$\sum 
m_\nu<0.33$ and 0.43~eV (95\%~CL) for $w=-1$ and $w\ne -1$, respectively.

\citet{reid/etal:2010b} added a prior on the
amplitude of matter density fluctuations,
$\sigma_8(\Omega_m/0.25)^{0.41}=0.832\pm
0.033$ \citep[68\%~CL;][]{rozo/etal:2010}, which was derived from the
abundance of 
optically-selected clusters of galaxies called the ``maxBCG cluster
catalog'' \citep{koester/etal:2007}, to the 5-year \map+BAO+SN, and
found $\sum m_\nu<0.35$ and 0.52~eV 
(95\%~CL) for $w=-1$ and $w\ne -1$,
respectively. \citet{thomas/abdalla/lahav:2010} added the angular power
spectra of photometrically selected samples of LRGs called ``MegaZ'' to
the 5-year \map+BAO+SN, and found $\sum m_\nu<0.325$~eV (95\%~CL) for $w=-1$.
\citet{wang/etal:2005} pointed out that the limit on $\sum
m_\nu$ from galaxy clusters would improve significantly by not only
using the abundance but also the power spectrum of clusters.

In order to exploit the full information contained in the growth of
structure, it is essential to understand the effects of massive
neutrinos on the {\it non-linear} growth. All of the work to date (including
\map+LRG+$H_0$ presented above) included the effects of massive neutrinos
on the {\it linear} growth, while ignoring their non-linear effects. The
widely-used phenomenological calculation of the non-linear
matter power spectrum called the  {\sf HALOFIT} \citep{smith/etal:2003} has not
been calibrated for models with massive neutrinos. Consistent
treatments of massive neutrinos in the non-linear structure formation
using cosmological perturbation theory
\citep{saito/takada/taruya:2008,saito/takada/taruya:2009,wong:2008,lesgourgues/etal:2009,shoji/komatsu:2009}
and numerical simulations \citep{brandbyge/etal:2008,brandbyge/hannestad:2009} have just begun to be
explored. More work along these lines would be necessary to exploit
the information on the growth structure to constrain the mass of
neutrinos.

\subsection{Relativistic Species}
\label{sec:neff}
\begin{deluxetable*}{lccccc}
\tablecolumns{6}
\small
\tablewidth{0pt}
\tablecaption{%
Improvements in $N_{\rm eff}$: 7-year versus 5-year
}
\tablehead{
\colhead{Parameter}
&\colhead{Year}
&\colhead{\map\ only}
&\colhead{\map+BAO+SN+{\sl HST}}
&\colhead{\map+BAO+$H_0$}
&\colhead{\map+LRG+$H_0$}
}
\startdata
$z_{\rm eq}$
& 5-year
& $3141^{+154}_{-157}$
& $3240^{+99}_{-97}$
& 
& 
\nl
& 7-year
& \ensuremath{3145^{+ 140}_{- 139}}
& 
& \ensuremath{3209^{+ 85}_{- 89}}
& \ensuremath{3240\pm 90}
\nl
\hline
$\Omega_mh^2$
& 5-year
& $0.178^{+0.044}_{-0.041}$
& $0.160\pm 0.025$ 
& 
& 
\nl
& 7-year
& \ensuremath{0.184^{+ 0.041}_{- 0.038}}
& 
& \ensuremath{0.157\pm 0.016}
& \ensuremath{0.157^{+ 0.013}_{- 0.014}}
\nl
\hline
$N_{\rm eff}$
& 5-year 
& $>2.3$ (95\%~CL)
& $4.4\pm 1.5$
& 
& 
\nl
& 7-year
& \ensuremath{> 2.7\ \mbox{(95\% CL)}} 
& 
& \ensuremath{4.34^{+ 0.86}_{- 0.88}} 
& \ensuremath{4.25^{+ 0.76}_{- 0.80}} 
\enddata
\label{tab:neff}
\end{deluxetable*}

\begin{figure*}[t]
\centering \noindent
\includegraphics[width=17cm]{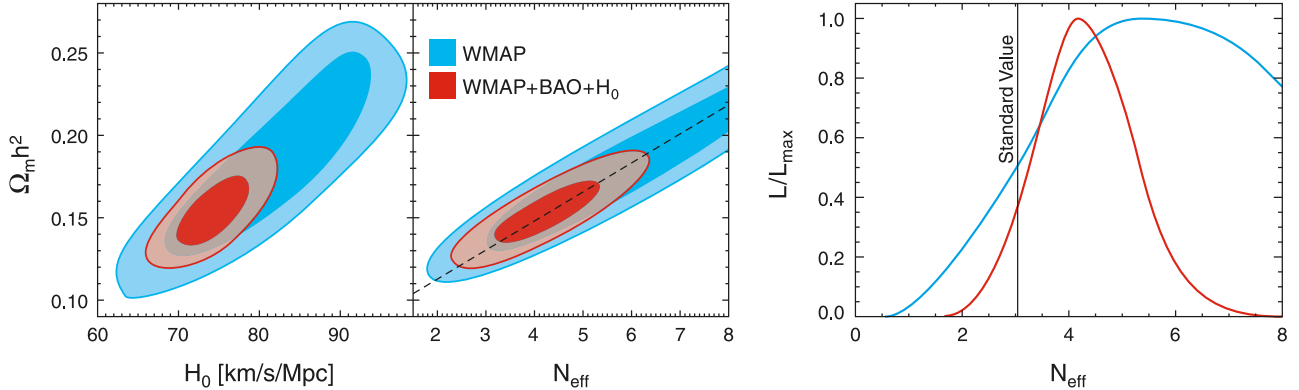}
\caption{%
Constraint on the effective number of neutrino species, $N_{\rm
 eff}$. (Left) Joint two-dimensional marginalized 
 distribution (68\% and 95\% CL), showing how a better determination of
 $H_0$ improves a limit on $\Omega_mh^2$. 
(Middle) A correlation between $N_{\rm eff}$ and $\Omega_mh^2$. The
 dashed line shows the line of correlation given by
 equation~(\ref{eq:neffformula}). A better determination of
 $H_0$ improves a limit on $\Omega_mh^2$ which, in turn, improves 
a limit on $N_{\rm eff}$. (Right)
One-dimensional marginalized distribution of $N_{\rm eff}$
from \map-only and \map+BAO+$H_0$.  The 68\% 
interval from \map+BAO+$H_0$, 
\ensuremath{N_{\rm eff} = 4.34^{+ 0.86}_{- 0.88}},
is consistent with the standard value, 3.04, which is shown by the
 vertical line. 
} 
\label{fig:omh2_vs_h}
\end{figure*}

How many relativistic species are there in the universe after the
matter-radiation equality epoch?
We parametrize the relativistic degrees of freedom using the
effective number of neutrino species, $N_{\rm eff}$, given in
equation~(\ref{eq:neff}). This quantity can be written in terms of the
matter density, $\Omega_mh^2$, and the redshift of matter-radiation
equality, $z_{\rm eq}$, as \citep[see equation~(84)
of][]{komatsu/etal:2009}
\begin{equation}
 N_{\rm eff} = 3.04 + 
7.44 \left(\frac{\Omega_mh^2}{0.1308}\frac{3139}{1+z_{\rm eq}}-1\right).
\label{eq:neffformula}
\end{equation}
(Here, $\Omega_mh^2=0.1308$ and $z_{\rm eq}=3138$ are the 5-year maximum
likelihood values from the simplest $\Lambda$CDM model.)
This formula suggests that the variation in $N_{\rm eff}$ is
given by
\begin{equation}
 \frac{\delta N_{\rm eff}}{N_{\rm eff}}
\simeq 2.45 
\frac{\delta(\Omega_mh^2)}{\Omega_mh^2}
- 2.45\frac{\delta z_{\rm eq}}{1+z_{\rm eq}}.
\end{equation}
The equality redshift is one of the direct observables from the
temperature power spectrum. The \map\ data constrain $z_{\rm eq}$ mainly
from the ratio of the first peak to the third peak. As the 7-year
temperature power spectrum has a better determination of the amplitude
of the third peak
\citep{larson/etal:prep}, we expect a better limit on $z_{\rm eq}$
compared to the 5-year one.
For models where $N_{\rm eff}$ is different
from 3.04, we find 
\ensuremath{z_{\rm eq} = 3145^{+ 140}_{- 139}}
(68\%~CL) from the
\map\ data only\footnote{For models with $N_{\rm eff}=3.04$, we find
\ensuremath{z_{\rm eq} = 3196^{+ 134}_{- 133}}
(68\%~CL).},  
which is better than the 5-year limit by more than 10\%
(see Table~\ref{tab:neff}).

However, the fractional error in $\Omega_mh^2$ is much larger, and thus
we need to determine $\Omega_mh^2$ using external data. 
The BAO data provide one constraint. We also find that $\Omega_mh^2$ and
$H_0$ are strongly correlated in the models with $N_{\rm eff}\ne 3.04$
(see Figure~\ref{fig:omh2_vs_h}). Therefore, an improved measurement of
$H_0$ from \citet{riess/etal:2009} would help reduce the error in
$\Omega_mh^2$, thereby reducing the error in $N_{\rm eff}$.
The limit on $\Omega_mh^2$ from the 7-year \map+BAO+$H_0$ combination is
better than the 5-year ``\map+BAO+SN+{\sl HST}'' limit by 36\%. 

We find that the \map+BAO+$H_0$ limit on $N_{\rm eff}$ is 
$$
\ensuremath{N_{\rm eff} = 4.34^{+ 0.86}_{- 0.88}}~\mbox{(68\%~CL)},
$$
while the \map+LRG+$H_0$ limit is 
\ensuremath{N_{\rm eff} = 4.25^{+ 0.76}_{- 0.80}}~(68\%~CL),
which are significantly better than the 
5-year \map+BAO+SN+{\sl HST} limit,  $N_{\rm
eff}=4.4\pm 1.5$~(68\%~CL).

\citet{reid/etal:2010b} added the maxBCG prior, $\sigma_8(\Omega_m/0.25)^{0.41}=0.832\pm
0.033$ \citep[68\%~CL;][]{rozo/etal:2010}, to the 5-year
\map+BAO+SN+{\sl HST}, and found $N_{\rm eff}=3.5\pm 0.9$ 
(68\%~CL).  They also added the above prior to the 5-year
version of \map+LRG+$H_0$, finding $N_{\rm eff}=3.77\pm 0.67$ (68\%~CL).

The constraint on $N_{\rm eff}$ can also be
interpreted as an upper bound on the energy density in primordial
gravitational waves with frequencies $> 10^{-15}$~Hz.
Many cosmological mechanisms for the generation of stochastic
gravitational waves exist, such as certain inflationary models,
electroweak phase transitions, and cosmic strings. At low frequencies
($10^{-17} - 10^{-16}$~Hz), the background is constrained by the limit on tensor fluctuations  described in Section~\ref{sec:GW}.  Constraints
at higher frequencies come from pulsar timing measurements at
$\sim 10^{-8}$~Hz \citep{jenet/etal:2006}, recent data from the
Laser Interferometer Gravitational Wave Observatory (LIGO) at 100~Hz
\cite[with limits of $\Omega_{{\rm gw}} <6.9 \times
10^{-6}$][]{abbott/etal:2009}, 
and at frequencies $>10^{-10}$~Hz from measurements of light-element abundances.
A large gravitational wave energy density at nucleosynthesis would
alter the predicted abundances, and observations imply an upper bound of $\Omega_{{\rm gw}} h^2 < 7.8 \times 10^{-6}$ \citep{cyburt/etal:2005}.

The CMB provides a limit that reaches down to $10^{-15}$~Hz,
corresponding to the comoving horizon at recombination.
The gravitational wave background within the horizon behaves as
free-streaming massless 
particles, so affects the CMB and matter power spectra in 
the same way as massless neutrinos \citep{smith/pierpaoli/kamionkowski:2006}.
The density contributed by $N_{{\rm gw}}$  massless neutrino species is
$\Omega_{{\rm gw}} h^2 = 5.6 \times 10^{-6} N_{{\rm gw}}$. Constraints
have been found using the WMAP 3-year data 
combined with additional cosmological probes by \citet{smith/pierpaoli/kamionkowski:2006},
for both adiabatic and homogeneous initial conditions for the tensor
perturbations. 
With the current WMAP+BAO+$H_0$ data combination, we define $N_{\rm{gw}}
=  N_{\rm{eff}}-3.04$, and find limits of
$$
N_{{\rm gw}}< 2.85, \quad \Omega_{{\rm gw}}h^2 < 1.60 \times
10^{-5}~(95\% {\rm CL})
$$
for adiabatic initial conditions, imposing an $N_{\rm{eff}} \ge 3.04$ prior. Adiabatic conditions might be expected
if the gravitational waves were generated by the appearance of cusps in cosmic strings \citep{damour/vilenkin:2000,damour/vilenkin:2001,siemens/etal:2006}. For the WMAP+LRG+$H_0$ data, we find $N_{{\rm gw}}<2.64$, or $\Omega_{{\rm gw}}h^2 < 1.48 \times 10^{-5}$ at 95\% CL. Given a particular string model, these bounds can be used to constrain the cosmic string tension \citep[e.g.,][]{siemens/mandic/creighton:2007,copeland/kibble:2009}.

\subsection{Primordial Helium Abundance}
\label{sec:helium}

\begin{figure*}[t]
\centering \noindent
\includegraphics[width=17cm]{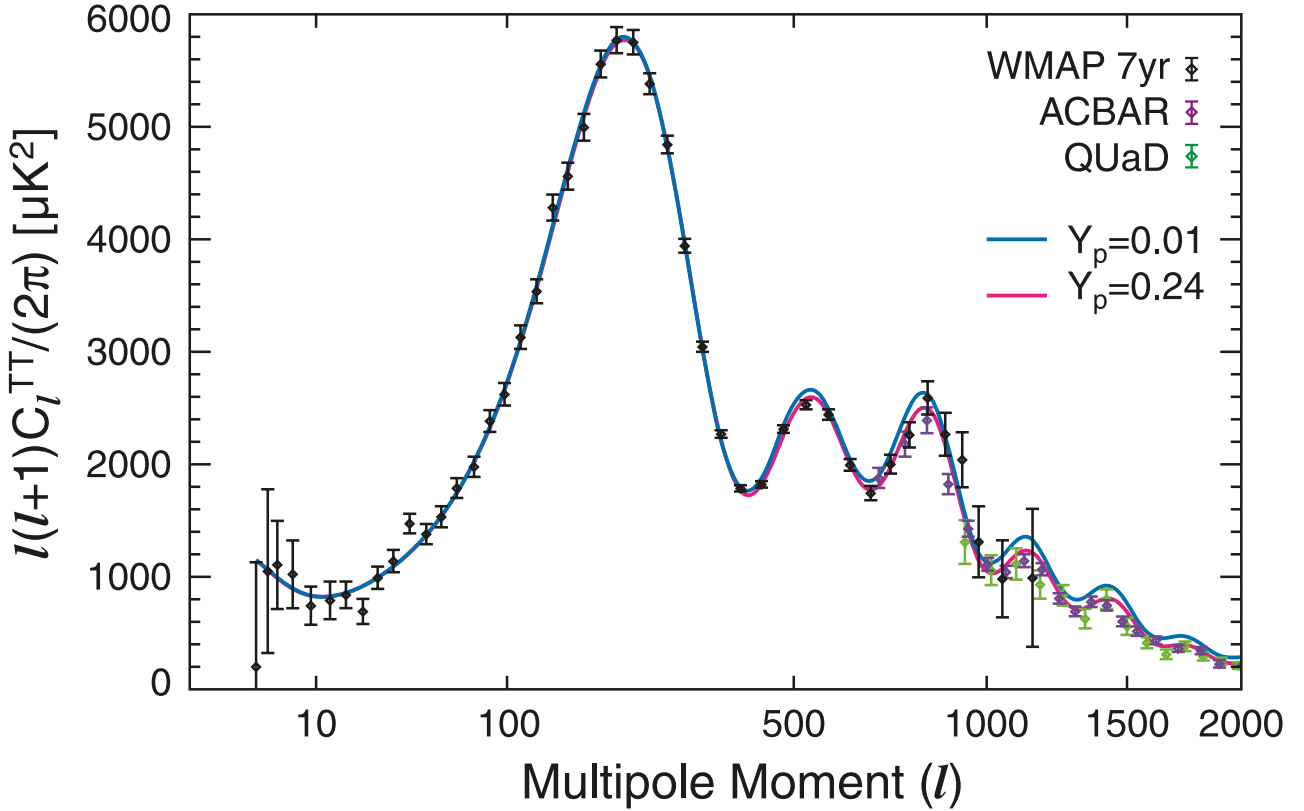}
\caption{%
Primordial helium abundance and the temperature power spectrum. The data
 points are the same as those in Figure~\ref{fig:cl}. The lower (pink)
 solid line 
 (which is the same as the solid line in Figure~\ref{fig:cl})
 shows the power spectrum with the nominal helium abundance, $Y_p=0.24$,
 while the upper (blue) solid line shows that with a tiny helium abundance,
 $Y_p=0.01$. The larger the helium abundance is, the smaller the number
 density of electrons during recombination becomes, which enhances the
 Silk damping of the power spectrum on small angular scales, $l\gtrsim 500$.
} 
\label{fig:zerohelium}
\end{figure*}
\begin{figure*}[t]
\centering \noindent
\includegraphics[width=17cm]{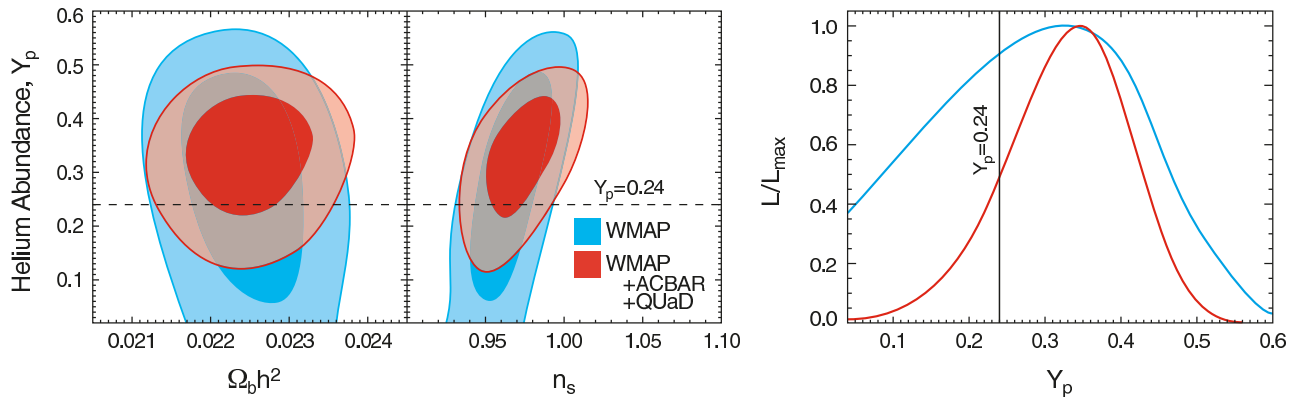}
\caption{%
Constraint on the primordial helium abundance, $Y_p$. 
 (Left) Joint two-dimensional marginalized
 distribution (68\% and 95\% CL), showing that $Y_p$ and $\Omega_bh^2$  
 are essentially uncorrelated. 
(Middle) A slight correlation exists between $Y_p$ and $n_s$: an
 enhanced Silk damping produced by a larger $Y_p$ can be partially
 canceled by a larger $n_s$. (Right)
One-dimensional marginalized distribution of $Y_p$
from \map-only and \map+ACBAR+QUaD.  The 68\% 
interval from \map+ACBAR+QUaD, 
$Y_p=0.326\pm 0.075$ is
 consistent with the nominal value, 0.24, which is shown by the
 vertical line.
} 
\label{fig:obh2_vs_yhe}
\end{figure*}

A change in the primordial helium abundance affects the shape of the
temperature power spectrum \citep{hu/etal:1995}. The most dominant
effect is a suppression of the power spectrum at $l\gtrsim
500$ due to an enhanced Silk damping effect.

For a given mass density of baryons (protons and helium nuclei), the
number density of electrons, $n_e$, can be related to the primordial
helium abundance. When both hydrogen and helium were ionized,
$n_e=(1-Y_p/2)\rho_b/m_p$. However, most of the helium recombines by $z\sim
1800$ \citep[see][and references therein]{switzer/hirata:2008}, much
earlier than the photon 
decoupling epoch, $z=1090$. 
As a result, the number density of free
electrons at around the decoupling epoch is given by
$n_e=(1-Y_p)\rho_b/m_p\propto 
(1-Y_p)\Omega_bh^2$ \citep{hu/etal:1995}.
The larger $Y_p$ is, the smaller $n_e$ becomes. If
the number of electrons is reduced, photons can free-stream longer (the
mean free path of photons, $1/(\sigma_Tn_e)$, gets larger), wiping out
more temperature anisotropy. Therefore, a larger $Y_p$ results in a
greater suppression of power on small angular scales.

\citet{ichikawa/sekiguchi/takahashi:2008} \citep[also
see][]{ichikawa/takahashi:2006} show that a 100\% change in $Y_p$
changes the heights of the second, third, and forth peaks by $\approx 1$\%,
3\%, and 3\%, respectively. Therefore, one expects that a
combination of the \map\ data and small-scale CMB experiments such as
ACBAR \citep{reichardt/etal:2009} and QUaD \citep{brown/etal:2009} would
be a powerful probe of the primordial helium abundance. 

In Figure~\ref{fig:zerohelium}, we compare the \map, ACBAR, and QUaD
data with the temperature power spectrum with the nominal value of
the primordial helium abundance, $Y_p=0.24$ (pink line), and that with
a tiny amount 
of helium, $Y_p=0.01$ (blue line). There is too much power in the case of
$Y_p=0.01$, making it possible to {\it detect} the
primordial helium effect using the CMB data alone.

However, one must be careful about a potential degeneracy between the
effect of helium and those of the other cosmological parameters. First,
as the number density of electrons is given by $n_e=(1-Y_p)n_b\propto
(1-Y_p)\Omega_bh^2$, $Y_p$ and $\Omega_bh^2$ may be correlated. Second,
a scale-dependent suppression of power such as this may be correlated
with the effect of tilt, $n_s$ \citep{trotta/hansen:2004}.

In the left panel of Figure~\ref{fig:obh2_vs_yhe}, we show that
$\Omega_bh^2$ and $Y_p$ are essentially uncorrelated: the baryon density
is determined by the first-to-second peak ratio relative to the
first-to-third peak ratio, which is now well measured by the \map\
data. Therefore, the current \map\ data allow $\Omega_bh^2$ to be
measured regardless of $Y_p$.
	
In the middle panel of Figure~\ref{fig:obh2_vs_yhe}, we show that there
is a slight positive correlation between $n_s$ and $Y_p$: an
 enhanced Silk damping produced by a larger $Y_p$ can be partially
 canceled by a larger $n_s$  \citep{trotta/hansen:2004}.

We find a 95\%~CL upper limit of $Y_p<0.51$ from the \map\ data
alone. When we add the ACBAR and QUaD data, {\it we find a significant
detection of the effect of primordial helium by more than $3\sigma$}
(see the right panel of Figure~\ref{fig:obh2_vs_yhe}),
$$
Y_p=0.326\pm 0.075~\mbox{(68\%~CL)}.
$$
The 95\%~CL limit is  $0.16<Y_p<0.46$. The 99\% CL lower limit is $Y_p>0.11$.
This value is broadly consistent with the helium abundances estimated from 
observations of low-metallicity extragalactic ionized (HII) regions, 
$Y_p\simeq 0.24-0.25$
\citep{gruenwald/steigman/viegas:2002,izotov/thuan:2004,olive/skillman:2004,fukugita/kawasaki:2006,peimbert/luridiana/peimbert:2007}. 
See \citet{steigman:2007} for a review. 

We can improve this limit by imposing an {\it upper limit} on $Y_p$ from
these astrophysical measurements. As the helium is created by nuclear
fusion in stars, the helium abundances measured from stars \cite[e.g.,
Sun; see][for a recent review]{asplund/etal:2009}
and HII regions are, in general, larger than the primordial abundance.
On the other hand, as we have just shown, the CMB data provide a {\it
lower limit} on $Y_p$. Even with a very conservative hard prior,
$Y_p<0.3$, we find $0.23<Y_p<0.3$~(68\%~CL)\footnote{The upper
limit is set by the hard prior. The 68\% lower limit, $Y_{p,{\rm
min}}=0.23$, is found such that the integral of the posterior
likelihood of $Y_p$ in $Y_{p,{\rm min}}\le Y_{p}<0.3$ is 68\% of the
integral in $0\le Y_{p}<0.3$.
Similarly, the 95\%~CL lower limit is $Y_p>0.14$ and the 99\% CL lower
limit is $Y_p>0.065$.}. 
Therefore, a combination
of the CMB 
and the solar constraints on $Y_p$ offers a new way for testing
the predictions of theory of the big bang nucleosynthesis (BBN). For
example, the BBN predicts that the helium abundance is related to the
baryon-to-photon ratio, $\eta$, and the number of additional neutrino
species (or any other additional relativistic degrees of freedom) during
the BBN epoch, $\Delta N_\nu\equiv N_\nu-3$, as \citep[see equation~(11) of][]{steigman:prep}
\begin{equation}
 Y_p = 0.2485 + 0.0016[(\eta_{10} - 6) + 100(S - 1)],
\end{equation}
where $S\equiv \sqrt{1+(7/43)\Delta N_\nu}\simeq 1+0.081\Delta N_\nu$ and $\eta_{10}\equiv
 10^{10}\eta=273.9(\Omega_bh^2)=6.19\pm 0.15$~(68\%~CL; \map+BAO+$H_0$).
\citep[See][for more discussion on this method.]{simha/steigman:2008}
For $\Delta N_\nu=1$, the helium abundance changes by $\Delta
 Y_p=0.013$, which is much smaller than our error bar, but is
 comparable to the expected error bar from {\sl Planck}
 \citep{ichikawa/sekiguchi/takahashi:2008}. 

There have been several attempts to measure $Y_p$ from the CMB data
\citep{trotta/hansen:2004,huey/cyburt/wandelt:2004,ichikawa/takahashi:2006,ichikawa/sekiguchi/takahashi:2008,dunkley/etal:2009}. The
previous best-limit is 
$Y_p=0.25_{-0.07(-0.17)}^{+0.10(+0.15)}$ at 68\%~CL (95\%~CL), which was
obtained by \citet{ichikawa/sekiguchi/takahashi:2008}
from the \map\ 5-year data combined with ACBAR
\citep{reichardt/etal:2009}, BOOMERanG
\citep{jones/etal:2006,piacentini/etal:2006,montroy/etal:2006}, and
Cosmic Background Imager \citep[CBI;][]{sievers/etal:2007}. Note that
the likelihood function of $Y_p$ is non-Gaussian, with a tail extending
to $Y_p=0$; 
thus, the level of significance of detection was less than $3\sigma$.

\section{Constraints on Properties of Dark Energy}
\label{sec:darkenergy} 

In this section, we provide limits on the properties of dark energy,
characterized by the equation of state parameter, $w$. 
We first focus on constant (time independent) equation of state in a
flat universe (Section~\ref{sec:w_flat}) and a curved universe
(Section~\ref{sec:w_curve}). We then constrain a time-dependent $w$
given by $w(a)=w_0+w_a(1-a)$, where $a=1/(1+z)$ is the scale factor, in Section~\ref{sec:w0wa}.
Next, we provide the 7-year ``\map\ normalization
prior'' in Section~\ref{sec:normalization_prior}, which is useful for
constraining $w$ (as well as the mass of neutrinos) from the growth of
cosmic density fluctuations. \citep[See, e.g.,][for an application of the
5-year normalization prior to the X-ray cluster abundance
data.]{vikhlinin/etal:2009} In Section~\ref{sec:distance_prior}, we
provide the 7-year ``\map\ distance prior,'' which is useful for
constraining a variety of time-dependent $w$ models for which the Markov
Chain Monte Carlo exploration of the parameter space may not be
available.
\citep[See, e.g.,][for applications of the
5-year distance prior.]{li/etal:2008,wang:2008,wang:2009,vikhlinin/etal:2009}

We give a summary of our limits on dark energy parameters in 
Table~\ref{tab:darkenergy}.

\subsection{Constant Equation of State: Flat Universe}
\label{sec:w_flat}
In a flat universe, $\Omega_k=0$, an accurate determination of $H_0$
helps improve a limit on a constant equation of state, $w$
\citep{spergel/etal:2003,hu:2005b}. 
Using \map+BAO+$H_0$, we find
$$
\ensuremath{w = -1.10\pm 0.14}~\mbox{(68\%~CL)},
$$ 
which improves to
\ensuremath{w = -1.08\pm 0.13}~(68\%~CL) if we add the
time-delay distance out to the lens system B1608+656 \citep[][see
Section~\ref{sec:timedelay}]{suyu/etal:2010}. 
{\it These limits are
independent of high-$z$ Type Ia supernova data.}

The high-$z$ supernova data provide the most stringent limit on
$w$. Using \map+BAO+SN, we find 
\ensuremath{w = -0.980\pm 0.053}~(68\%~CL). The error
does not include systematic errors in supernovae, which are comparable to the
statistical error \citep{kessler/etal:2009,hicken/etal:2009}; thus, the
error in $w$ from \map+BAO+SN is about a half of that from 
\map+BAO+$H_0$ or \map+BAO+$H_0$+$D_{\Delta t}$.

The cluster abundance data are sensitive to $w$ via the comoving volume
element, angular diameter distance, and growth of matter density
fluctuations \citep{haiman/mohr/holder:2001}.
By combining the cluster abundance data and the 5-year \map\ data,
\citet{vikhlinin/etal:2009} found 
$w=-1.08\pm 0.15~\mbox{(stat)}\pm 0.025~\mbox{(syst)}$ (68\%~CL) for a flat universe.
By adding BAO of 
\citet{eisenstein/etal:2005} and the supernova data of
\citet{davis/etal:2007}, they found 
$w=-0.991\pm 0.045~\mbox{(stat)}\pm 0.039~\mbox{(syst)}$
(68\%~CL). 
These results using the cluster abundance data \citep[also
see][]{mantz/etal:2010} agree well
with our corresponding \map+BAO+$H_0$ and \map+BAO+SN limits.

\subsection{Constant Equation of State: Curved Universe}
\label{sec:w_curve}
\begin{figure}[t]
\centering \noindent
\includegraphics[width=8cm]{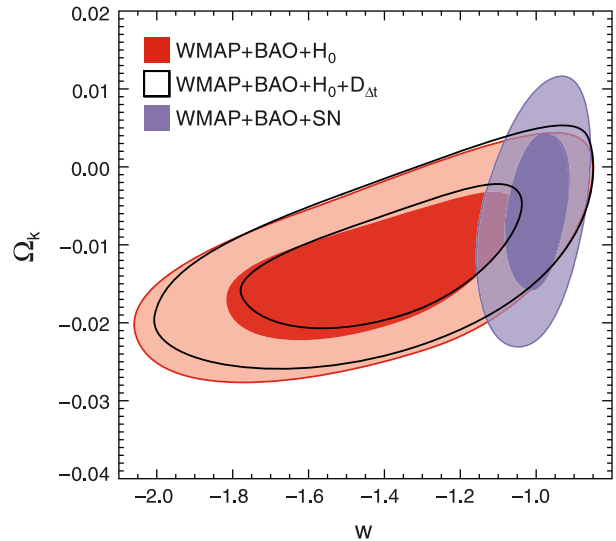}
\caption{%
 Joint two-dimensional marginalized constraint on the time-independent
 (constant) dark energy equation of state, $w$, and the curvature parameter,
 $\Omega_{k}$. The contours show the 68\% and 95\% CL from
 \map+BAO+$H_0$ (red), \map+BAO+$H_0$+$D_{\Delta t}$ (black), and
 \map+BAO+SN (purple). 
}
\label{fig:wok}
\end{figure}

When $\Omega_k\ne 0$, limits on $w$ significantly weaken, with a tail
extending to large negative values of $w$, unless supernova data are added. 

In
Figure~\ref{fig:wok}, we show that \map+BAO+$H_0$ allows for $w\lesssim
-2$, which can be excluded by adding information on the time-delay distance. 
In both cases, the spatial curvature is well
constrained: we find  $\Omega_k=-0.0125^{+0.0064}_{-0.0067}$ from \map+BAO+$H_0$,
and  $-0.0111^{+0.0060}_{-0.0063}$~(68\%~CL) from \map+BAO+$H_0$+$D_{\Delta
t}$, whose errors are comparable to that of the \map+BAO+$H_0$ limit on
$\Omega_k$ with $w=-1$,
\ensuremath{\Omega_k = -0.0023^{+ 0.0054}_{- 0.0056}}~(68\%~CL; see
Section~\ref{sec:OK}).

However, $w$ is poorly constrained: 
we find $w=-1.44\pm 0.27$ from \map+BAO+$H_0$, and
$-1.40\pm 0.25$~(68\%~CL) from \map+BAO+$H_0$+$D_{\Delta t}$.

Among the data combinations that do not use the information on the
growth of structure, 
the most powerful combination  for constraining $\Omega_k$ and $w$
simultaneously is a combination of the \map\ data, BAO
(or $D_{\Delta t}$), and supernovae, as \map+BAO (or $D_{\Delta t}$) primarily
constrains $\Omega_k$, and \map+SN primarily constrains $w$. 
With \map+BAO+SN, we find 
$w=-0.999^{+0.057}_{-0.056}$ and $\Omega_k=-0.0057^{+0.0066}_{-0.0068}$
(68\%~CL). 
Note that the error in the curvature is essentially the same as
that from \map+BAO+$H_0$, while the error in $w$ is $\sim 4$ times
smaller.

\citet{vikhlinin/etal:2009} combined their cluster abundance data with the
5-year \map+BAO+SN to find $w=-1.03\pm 0.06$ (68\%~CL) for a curved
universe.
\citet{reid/etal:2010} combined their LRG power spectrum with the 5-year
\map\ data and the Union supernova data to find 
$w=-0.99\pm 0.11$ and $\Omega_k=-0.0109\pm 0.0088$~(68\%~CL).
These results are in good agreement with our 7-year \map+BAO+SN limit.

\subsection{Time-dependent Equation of State}
\label{sec:w0wa}

\begin{figure}[t]
\centering \noindent 
\includegraphics[width=8cm]{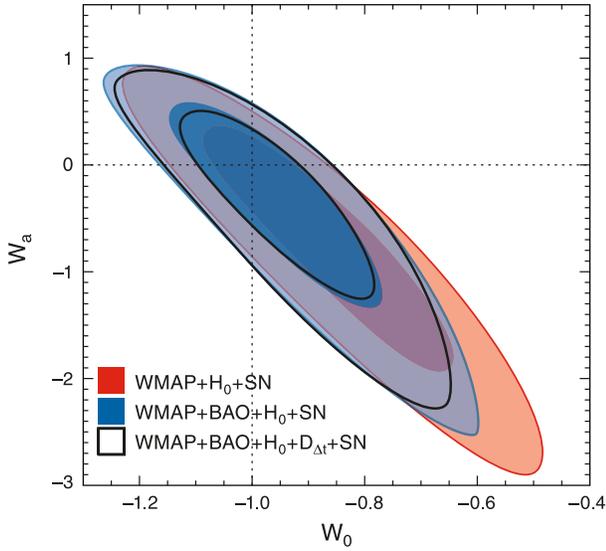}
\caption{%
Joint two-dimensional marginalized constraint on the 
linear evolution model of dark energy
 equation of state,  $w(a)=w_0+w_a(1-a)$. 
The contours show the 68\% and 95\% CL from \map+$H_0$+SN (red), 
 \map+BAO+$H_0$+SN (blue), and \map+BAO+$H_0$+$D_{\Delta t}$+SN (black),
 for a flat universe. 
}
\label{fig:w0wa}
\end{figure}

As for a time-dependent equation of state, we shall find constraints
on the present-day value of the equation of state and its derivative using
a linear form,
$w(a)=w_0+w_a(1-a)$ \citep{chevallier/polarski:2001,linder:2003}. 
We assume a flat universe, $\Omega_k=0$. \citep[For recent limits on
$w(a)$ with $\Omega_k\ne 0$, see][and references therein.]{wang:2009}
While we have constrained this model using the \map\ distance prior in
the 5-year analysis \citep[see Section~5.4.2 of][]{komatsu/etal:2009},
in the 7-year analysis
we shall present the full Markov Chain Monte Carlo exploration of this model.

For a time-dependent equation of state, one must be careful about the
treatment of perturbations in dark energy when $w$ crosses $-1$. We use the
``parametrized post-Friedmann'' (PPF) approach, implemented in the {\sf
CAMB} code following
\citet{fang/hu/lewis:2008}.\footnote{\citet{zhao/etal:2005} used a
multi-scalar-field model to treat $w$ crossing $-1$. The constraints on $w_0$
and $w_a$ have been obtained using this model and the previous years of
\map\ data \citep{xia/etal:2006,xia/etal:2008c,zhao/etal:2007}.}

In Figure~\ref{fig:w0wa}, we show the 7-year constraints on $w_0$ and
$w_a$ from \map+$H_0$+SN (red), 
 \map+BAO+$H_0$+SN (blue), and \map+BAO+$H_0$+$D_{\Delta t}$+SN
 (black). The angular diameter distances measured from BAO and
 $D_{\Delta t}$ help exclude models with large negative values of $w_a$.
 We find that the current data are consistent
with a cosmological constant, even when $w$ is allowed to depend on
time. However, a large range of values of $(w_0,w_a)$ are still allowed
by the data: we find 
$$
w_0=-0.93\pm 0.13~\mbox{and}~w_a=-0.41^{+0.72}_{-0.71}~\mbox{(68\%~CL)},
$$
from \map+BAO+$H_0$+SN. When the time-delay distance information is
added, the limits improve to 
$w_0=-0.93\pm 0.12$ and $w_a=-0.38^{+0.66}_{-0.65}$
(68\%~CL).
	
\citet{vikhlinin/etal:2009} combined their cluster abundance data with the
5-year \map+BAO+SN to find a limit on a linear combination of the parameters,
$w_a+3.64(1+w_0)=0.05\pm 0.17$ (68\%~CL).
Our data combination is sensitive to a different linear combination: we find
$w_a+5.14(1+w_0)=-0.05\pm 0.32$ (68\%~CL) for the 7-year \map+BAO+$H_0$+SN
combination.

The current data are consistent with a flat universe dominated by a
cosmological constant.

\subsection{\map\ Normalization Prior}
\label{sec:normalization_prior}
The growth of cosmological density fluctuations is a powerful probe of
dark energy, modified gravity, and massive neutrinos. 
The \map\ data provide a useful normalization of the cosmological
perturbation at the decoupling epoch, $z=1090$. By comparing this
normalization with the amplitude of matter density fluctuations in a low
redshift universe, one may distinguish between dark energy and modified gravity
\citep{ishak/upadhye/spergel:2006,koyama/maartens:2006,amarzguioui/etal:2006,dore/etal:2007,linder/cahn:2007,upadhye:2007,zhang/etal:2007,yamamoto/etal:2007,chiba/takahashi:2007,bean/etal:2007b,hu/sawicki:2007,song/hu/sawicki:2007,starobinsky:2007,daniel/etal:2008,jain/zhang:2008,bertschinger/zukin:2008,amin/wagoner/blandford:2008,hu:2008}
and determine the mass of neutrinos 
\citep{hu/eisenstein/tegmark:1998,lesgourgues/pastor:2006}.

In Section~5.5 of \citet{komatsu/etal:2009}, we provided a ``\map\ normalization
prior,'' which is a constraint on the power spectrum of curvature
perturbation, $\Delta^2_{\cal R}$.
\citet{vikhlinin/etal:2009} combined this with the number density of
clusters of galaxies to constrain the dark energy equation of state,
$w$, and the amplitude of matter density fluctuations, $\sigma_8$.

The matter density fluctuation in Fourier space, $\delta_{m,{\mathbf
k}}$, is related to ${\cal R}_{\mathbf k}$ as 
$\delta_{m,{\mathbf k}}(z) = \frac{2k^3}{5H_0^2\Omega_m}{\cal R}_{\mathbf k}T(k)D(k,z)$,
where $D(k,z)$ and $T(k)$ are the linear growth rate and the matter
transfer function normalized such that 
$T(k)\rightarrow 1$ as $k\rightarrow 0$,  and
$(1+z)D(k,z)\rightarrow 1$ 
 as $k\rightarrow 0$ during the matter era,  respectively. 
Ignoring the mass of neutrinos and modifications to gravity, one can
obtain the growth rate by solving a single differential equation
\citep{wang/steinhardt:1998,linder/jenkins:2003}.\footnote{See, e.g.,
equation~(79) of \citet{komatsu/etal:2009}. Note that there is a typo in
that equation: $w_{\rm eff}(a)$ needs to be replaced by $w(a)$.}

%

The 7-year normalization prior is
$$
 \Delta_{\cal R}^2(k_{WMAP}) = (2.208\pm 0.078)\times 10^{-9}~\mbox{(68\%~CL)},
$$
where $k_{WMAP}=0.027~{\rm Mpc}^{-1}$. For comparison, the 5-year
normalization prior was 
$\Delta_{\cal R}^2(0.02~{\rm Mpc}^{-1}) = (2.21\pm 0.09)\times 10^{-9}$.
This normalization prior is valid for models with $\Omega_k\neq 0$,
$w\neq -1$, or 
$m_\nu> 0$. However, these normalizations cannot be used 
for the models that have the tensor modes, $r>0$, or the running index,
$dn_s/d\ln k\neq 0$.

\subsection{\map\ Distance Prior}
\label{sec:distance_prior}
\begin{deluxetable}{lccc}
\tablecolumns{4}
\small
\tablewidth{0pt}
\tablecaption{%
\map\ distance priors obtained from the \map\ 7-year fit to models with spatial
 curvature and dark energy. The correlation coefficients are:
$r_{l_A,R}=0.1956$, 
$r_{l_A,z_*}=0.4595$, and
$r_{R,z_*}=0.7357$.
}
\tablehead{ 
& \colhead{7-year ML\tablenotemark{a}}
& \colhead{7-year Mean\tablenotemark{b}} 
& \colhead{Error, $\sigma$} 
}
\startdata
$l_A$ & 302.09 & 302.69 & 0.76 \nl
$R$   & 1.725 & 1.726  & 0.018 \nl
$z_*$ & 1091.3 & 1091.36 & 0.91
\enddata
\tablenotetext{a}{Maximum likelihood values (recommended).}
\tablenotetext{b}{Mean of the likelihood.}
\label{tab:wmap_prior}
\end{deluxetable}

\begin{deluxetable}{lccc}
\tablecolumns{4}
\small
\tablewidth{0pt}
\tablecaption{Inverse covariance matrix for the \map\ distance priors}
\tablehead{ & \colhead{$l_A$} & \colhead{$R$} & \colhead{$z_*$}}
\startdata
$l_A$ & $2.305$ &   $29.698$ &  $-1.333$ \nl
$R$   &         & $6825.270$ & $-113.180$ \nl
$z_*$      &         &            & $3.414$
\enddata
\label{tab:wmap_prior_cov}
\end{deluxetable}

The temperature power spectrum of CMB is sensitive to the physics at the
decoupling epoch, $z=1090$, as well as the physics between now and the
decoupling epoch. The former primarily affects the amplitude of acoustic
peaks, i.e., the ratios of the peak heights and the Silk damping. The
latter changes the {\it locations} of peaks via the angular diameter
distance out to the decoupling epoch. One can quantify this by (i) the
``acoustic scale'', $l_A$, 
\begin{equation}
 l_A=(1+z_*)\frac{\pi D_A(z_*)}{r_s(z_*)},
\end{equation}
where $z_*$ is the redshift of decoupling, for which we use the fitting
formula of \citet{hu/sugiyama:1996}, as well as by (ii) the ``shift
parameter,'' $R$ \citep{bond/efstathiou/tegmark:1997},
\begin{equation}
 R =\frac{\sqrt{\Omega_mH_0^2}}{c}(1+z_*)D_A(z_*). 
\end{equation}
These two parameters capture most of the constraining power of the \map\
data for dark energy properties
\citep{wang/mukherjee:2007,wright:2007,elgaroy/multamaki:2007,corasaniti/melchiorri:2008},
with one important difference. The distance prior does not capture the
information on the growth of structure probed by the late-time ISW
effect. As a result, the dark energy constraints 
derived from the distance prior are similar to, but weaker than, those
derived from the full analysis \citep{komatsu/etal:2009,li/etal:2008}.

We must understand the limitation of this method. Namely, the distance
prior is applicable only when the model in question is based on:
\begin{itemize}
 \item[1.] The standard Friedmann-Lemaitre-Robertson-Walker universe with
       matter, radiation, dark energy, as well as spatial curvature,
 \item[2.] Neutrinos with the effective
       number of neutrinos equal to 3.04, and the minimal mass
       ($m_\nu\sim 0.05$~eV), and
 \item[3.] Nearly power-law primordial power spectrum of curvature
       perturbations, $\left|dn_s/d\ln k\right|\ll 0.01$, negligible
	   primordial gravitational waves relative to the curvature
	   perturbations, $r\ll 0.1$, and negligible entropy fluctuations relative to the curvature perturbations, $\alpha\ll 0.1$. 
\end{itemize}

In Table~\ref{tab:wmap_prior} and \ref{tab:wmap_prior_cov}, we provide
the 7-year distance prior. The errors in $l_A$, $R$, and $z_*$ have
improved from the 5-year values by 12\%, 5\%, and 2\%, respectively. To
compute the likelihood, use 
\begin{equation}
 -2\ln L = \sum_{ij}(x_i-d_i)(C^{-1})_{ij}(x_j-d_j),
\end{equation}
where $x_i=(l_A,R,z_*)$ is the values predicted by a model in question, 
$d_i=(l_A^{WMAP},R^{WMAP},z_*^{WMAP})$ is the data given in
Table~\ref{tab:wmap_prior}, and $C^{-1}_{ij}$ is the inverse covariance
matrix given in Table~\ref{tab:wmap_prior_cov}.
Also see Section~5.4.1 of \citet{komatsu/etal:2009} for more information.

\section{Primordial non-Gaussianity}
\label{sec:NG}
\subsection{Motivation and Background}
During the period of cosmic inflation
\citep{starobinsky:1979,starobinsky:1982,guth:1981,sato:1981,linde:1982,albrecht/steinhardt:1982},
quantum fluctuations were generated and became the seeds for the cosmic
structures that we observe today
\citep{mukhanov/chibisov:1981,hawking:1982,starobinsky:1982,guth/pi:1982,bardeen/steinhardt/turner:1983}. \citep[Also
see][for
reviews.]{linde:1990,mukhanov/feldman/brandenberger:1992,liddle/lyth:CIALSS,liddle/lyth:PDP,bassett/tsujikawa/wands:2006,linde:2008}

Inflation predicts that the statistical distribution of primordial
fluctuations is nearly a Gaussian distribution with random phases. 
Measuring deviations from a Gaussian distribution, i.e., {\it non-Gaussian}
correlations in primordial fluctuations, is a 
powerful test of inflation, as how precisely the distribution is
(non-)Gaussian depends on the detailed physics of inflation
\citep[see][for reviews]{bartolo/etal:2004,komatsu/etal:astro2010}. 

In this paper, we constrain the amplitude of non-Gaussian correlations
using the angular bispectrum of CMB temperature anisotropy, the harmonic
transform of the 3-point correlation function \citep[see][for a
review]{komatsu:prep}. The observed angular bispectrum is related to the
3-dimensional bispectrum of primordial curvature perturbations,
$\langle\zeta_{{\mathbf k}_1}\zeta_{{\mathbf k}_2}\zeta_{{\mathbf
k}_3}\rangle =(2\pi)^3\delta^D({\mathbf k}_1+{\mathbf k}_2+{\mathbf
k}_3)B_\zeta(k_1,k_2,k_3)$. In the linear order, the primordial
curvature perturbation is related to Bardeen's curvature perturbation
\citep{bardeen:1980} in
the matter-dominated era, $\Phi$, by $\zeta=\frac53\Phi$
\citep[e.g.,][]{kodama/sasaki:1984}. The CMB 
temperature anisotropy in the Sachs--Wolfe limit
\citep{sachs/wolfe:1967} is given by $\Delta
T/T=-\frac13\Phi=-\frac15\zeta$.
We write the bispectrum of $\Phi$ as
\begin{equation}
\langle\Phi_{{\mathbf k}_1}\Phi_{{\mathbf k}_2}\Phi_{{\mathbf
k}_3}\rangle =(2\pi)^3\delta^D({\mathbf k}_1+{\mathbf k}_2+{\mathbf
k}_3)F(k_1,k_2,k_3).
\end{equation}

We shall explore 3 different shapes of the primordial bispectrum: ``local,''
``equilateral,'' and ``orthogonal.'' They are defined as follows:
\begin{itemize}
 \item [1.] {\bf Local form}. The local form bispectrum is given
       by \citep{gangui/etal:1994,verde/etal:2000,komatsu/spergel:2001}
\begin{eqnarray}
\nonumber
& &	F_{\rm local}(k_1,k_2,k_3) \\
\nonumber
 &=& 2\fnlKS
	 [P_\Phi(k_1)P_\Phi(k_2)+P_\Phi(k_2)P_\Phi(k_3)\\
\nonumber
& &+P_\Phi(k_3)P_\Phi(k_1)]\\
&=& 2A^2\fnlKS\left[\frac1{k^{4-n_s}_1k^{4-n_s}_2}+(2~\mbox{perm}.)\right],
\label{eq:Flocal}
\end{eqnarray}
where $P_\Phi=A/k^{4-n_s}$ is the power spectrum of $\Phi$ with a
       normalization factor $A$. This form is
       called the local form, as this bispectrum can arise from the
       curvature perturbation in the form of
       $\Phi=\Phi_L+\fnlKS\Phi^2_L$, where both sides are evaluated at
       the same location in space ($\Phi_L$ is a linear Gaussian
       fluctuation).\footnote{However, $\Phi=\Phi_L+\fnlKS\Phi^2_L$ is not the
       only way to produce this type of bispectrum. One can also produce
       this form from multi-scalar field inflation models where scalar
       field fluctuations are nearly scale invariant
       \citep{lyth/rodriguez:2005}; multi-scalar models called
       ``curvaton'' scenarios 
       \citep{linde/mukhanov:1997,lyth/ungarelli/wands:2003};
       multi-field models in which one field modulates the decay rate of
       inflaton field
       \citep{dvali/gruzinov/zaldarriaga:2004b,dvali/gruzinov/zaldarriaga:2004a,zaldarriaga:2004};
       multi-field models in which a violent production of particles and
       non-linear reheating,  
       called ``preheating,'' occur due to parametric resonances
       \citep{enqvist/etal:2005,jokinen/mazumdar:2006,chambers/rajantie:2007,bond/etal:2009};
       models in which the universe contracts first and then bounces
       \citep[see][for a review]{lehners:2008}.} 
       Equation~(\ref{eq:Flocal}) peaks at the so-called
       ``squeezed'' triangle for which $k_3\ll k_2\approx k_1$
       \citep{babich/creminelli/zaldarriaga:2004}. In this limit, we
       obtain \begin{equation}\label{eq:1}
 F_{\rm local}(k_1,k_1,k_3\to 0)=4\fnlKS P_\Phi(k_1)P_\Phi(k_3).
\end{equation}
How large is $\fnlKS$ from inflation?  
The earlier calculations showed that $\fnlKS$ from single-field slow-roll
inflation would be of order the slow-roll parameter, $\epsilon\sim
10^{-2}$
\citep{salopek/bond:1990,falk/rangarajan/srendnicki:1993,gangui/etal:1994}. 
More recently, \citet{maldacena:2003} and \citet{acquaviva/etal:2003}
       found that the coefficient of $P_\Phi(k_1)P_\Phi(k_3)$ from the simplest
single-field slow-roll inflation with the canonical kinetic term
in the squeezed limit is given by 
\begin{equation}
 F_{\rm local}(k_1,k_1,k_3\to 0)=\frac53(1-n_s) P_\Phi(k_1)P_\Phi(k_3).
\label{eq:singleprediction}
\end{equation}
Comparing this result with the form predicted by the $\fnlKS$ model,
one obtains $\fnlKS=(5/12)(1-n_s)$, which gives $\fnlKS=0.015$ for
       $n_s=0.963$. 
 \item [2.] {\bf Equilateral form}. The equilateral form bispectrum is given
       by \citep{creminelli/etal:2006}
\begin{eqnarray}
\nonumber
& &	F_{\rm equil}(k_1,k_2,k_3)= 6A^2\fnleq\\
\nonumber
& \times& \left\{
-\frac1{k^{4-n_s}_1k^{4-n_s}_2}-\frac1{k^{4-n_s}_2k^{4-n_s}_3}
-\frac1{k^{4-n_s}_3k^{4-n_s}_1}\right.\\
\nonumber
& &
-\frac2{(k_1k_2k_3)^{2(4-n_s)/3}}
+\left[\frac1{k^{(4-n_s)/3}_1k^{2(4-n_s)/3}_2k^{4-n_s}_3}\right.\\
& &\left.\left. +\mbox{(5 perm.)}\right]\right\}.
\label{eq:Fequil}
\end{eqnarray}
This function approximates the bispectrum forms that arise from a class of
       inflation models in which scalar fields have 
       non-canonical kinetic terms. One example is the so-called
       Dirac-Born-Infeld (DBI) inflation
       \citep{silverstein/tong:2004,alishahiha/silverstein/tong:2004},
       which gives $\fnleq\propto -1/c_s^2$ in the limit of $c_s\ll
       1$, where $c_s$ is the effective sound speed at which scalar
       field fluctuations propagate. There are various other models that
       can produce $\fnleq$ 
       \citep{arkani-hamed/etal:2004,seery/lidsey:2005,chen/etal:2007,cheung/etal:2008,li/wang/wang:2008}.
       The local and equilateral forms are nearly orthogonal to each
       other, which means that both can be measured nearly
       independently. 
 \item [3.] {\bf Orthogonal form}. The orthogonal form, which is
       constructed such that it is nearly orthogonal to both the local
       and equilateral forms, is given by
       \citep{senatore/smith/zaldarriaga:2010}
\begin{eqnarray}
\nonumber
& &	F_{\rm orthog}(k_1,k_2,k_3)= 6A^2\fnlor\\
\nonumber
& \times& \left\{
-\frac3{k^{4-n_s}_1k^{4-n_s}_2}-\frac3{k^{4-n_s}_2k^{4-n_s}_3}
-\frac3{k^{4-n_s}_3k^{4-n_s}_1}\right.\\
\nonumber
& &
-\frac8{(k_1k_2k_3)^{2(4-n_s)/3}}
+\left[\frac3{k^{(4-n_s)/3}_1k^{2(4-n_s)/3}_2k^{4-n_s}_3}\right.\\
& &\left.\left. +\mbox{(5 perm.)}\right]\right\}.
\label{eq:Forthog}
\end{eqnarray}
This form approximates the forms that arise from a linear combination of
       higher-derivative scalar-field interaction terms, each of which
       yields forms similar to the equilateral shape. 
\citet{senatore/smith/zaldarriaga:2010} found that, using the ``effective
       field theory of inflation'' approach \citep{cheung/etal:2008},
       a certain linear combination of similarly equilateral shapes can
       yield a distinct shape which is orthogonal to both the local and
       equilateral forms.
\end{itemize}
Note that these are not the most general forms one can write down, and
there are other forms which would probe different aspects of the physics
of inflation
\citep{moss/chun:2007,moss/graham:2007,chen/etal:2007,holman/tolley:2008,chen/wang:2010,chen/wang:2010b}.  

Of these forms, the local form bispectrum has special significance. 
\citet{creminelli/zaldarriaga:2004} showed that 
not only models with the canonical kinetic term, but
{\it all}
single-inflation models predict the
bispectrum in the squeezed limit given by
Eq.~(\ref{eq:singleprediction}), regardless of the form of potential,
kinetic term,
slow-roll, or initial vacuum state
\citep[also see][]{seery/lidsey:2005,chen/etal:2007,cheung/etal:2008}.
This means that a convincing detection of $\fnlKS$ would rule out {\it
all} single-field inflation models. 

\subsection{Analysis Method and Results}
\begin{deluxetable}{lccccc}
\tablecolumns{6}
\small
\tablewidth{0pt}
\tablecaption{%
Estimates\tablenotemark{a} and the corresponding 68\% intervals of the
 primordial non-Gaussianity parameters ($\fnlKS$, $\fnleq$, $\fnlor$)
 and the point source bispectrum amplitude, $\bsrc$ (in units of
 $10^{-5}~\mu{\rm K}^3~{\rm sr}^2$), from the \map\ 7-year temperature maps
}
\tablehead{ 
\colhead{Band}
& \colhead{Foreground\tablenotemark{b}} 
& \colhead{$\fnlKS$}
& \colhead{$\fnleq$}
& \colhead{$\fnlor$}
& \colhead{$\bsrc$}
}
\startdata
V+W & Raw  &  $59\pm 21$  &  $33\pm 140$  &  $-199 \pm 104$  & N/A \nl
V+W & Clean & $42\pm 21$  &  $29\pm 140$  &  $-198 \pm 104$  & N/A \nl
V+W & Marg.\tablenotemark{c} & $32\pm 21$  &  $26\pm 140$  &  $-202\pm
 104$ & $-0.08\pm 0.12$\nl
V   & Marg. & $43\pm 24$  &  $64\pm 150$  &  $-98 \pm 115$ & $0.32\pm
 0.23$ \nl
W   & Marg. &  $39\pm 24$  &  $36\pm 154$  &  $-257\pm 117$  & $-0.13\pm
 0.19$
\enddata
\tablenotetext{a}{The values quoted for ``V+W'' and ``Marg.'' are our
 best estimates from the \map\ 7-year data. In all cases, the
 full-resolution temperature maps at HEALPix $N_{\rm side}=1024$ are used.}
\tablenotetext{b}{In all cases, the {\it KQ75y7} mask is used.}
\tablenotetext{c}{``Marg.'' means that the foreground templates
 (synchrotron, free-free, and dust) have been marginalized over. When
 the foreground templates are marginalized over, the raw and clean maps
 yield the same $f_{\rm NL}$ values.}
\label{tab:fnl}
\end{deluxetable}

The first limit on $\fnlKS$ was obtained from the {\sl COBE} 4-year data
\citep{bennett/etal:1996} by \citet{komatsu/etal:2002}, using the
angular bispectrum. The limit was
improved by an order of magnitude when the \map\ first year data were
used to constrain $\fnlKS$ \citep{komatsu/etal:2003}. Since then the
limits have improved steadily as \map\ collect more years of data and
the bispectrum method for estimating $\fnlKS$ has improved
\citep{komatsu/spergel/wandelt:2005,creminelli/etal:2006,creminelli/etal:2007,spergel/etal:2007,yadav/wandelt:2008,komatsu/etal:2009,smith/senatore/zaldarriaga:2009}.\footnote{For
references to other methods for estimating $\fnlKS$, which do not use the
bispectrum directly, see Section~3.5 of
\citet{komatsu/etal:2009}. Recently, the ``skewness power spectrum'' has
been proposed as a new way to measure $\fnlKS$ and other non-Gaussian
components such as the secondary anisotropies and point sources
\citep{munshi/heavens:2010,smidt/etal:2009,munshi/etal:prep,calabrese/etal:2010}. In
the limit that noise is 
uniform, their estimator is equivalent to that of
\citet{komatsu/spergel/wandelt:2005}, which also allows for simultaneous
estimations of multiple sources of non-Gaussianity \citep[see Appendix A of][]{komatsu/etal:2009}.
The skewness power spectrum method provides a 
means to visualize the shape of various bispectra as a function of multipoles.} 

In this paper, we shall adopt the optimal estimator \citep[developed
by][]{babich:2005,creminelli/etal:2006,creminelli/etal:2007,smith/zaldarriaga:prep,yadav/etal:2008},
which builds on and significantly improves the original bispectrum
estimator proposed by \citet{komatsu/spergel/wandelt:2005}, especially
when the spatial distribution of instrumental noise is not uniform.
For details of the method, see Appendix~A of
\citet{smith/senatore/zaldarriaga:2009} for $\fnlKS$, and 
Section~4.1 of \citet{senatore/smith/zaldarriaga:2010} for $\fnleq$ and
$\fnlor$. To construct the optimal estimators, we need to specify the
cosmological parameters. We use the 5-year $\Lambda$CDM parameters from
\map+BAO+SN, for which $n_s=0.96$. 

We also constrain the bispectrum due
to residual (unresolved) point sources, $\bsrc$. The optimal estimator
for $\bsrc$ is constructed by replacing $a_{lm}/C_l$ in equation~(A24)
of \citet{komatsu/etal:2009} with $(C^{-1}a)_{lm}$, and using
their equations~(A17) and (A5). The $C^{-1}$ matrix is computed by the
multigrid-based algorithm of \citet{smith/zahn/dore:2007}.
 
We use the V- and W-band maps at the HEALPix resolution $N_{\rm
side}=1024$. As the optimal estimator weights the data optimally at all
multipoles, we no longer need to choose the maximum multipole used in
the analysis, i.e., we use all the data. We use both the
raw maps (before cleaning foreground) and foreground-reduced (clean)
maps to quantify the  foreground contamination of $f_{\rm NL}$ parameters.
For all cases,
we find the best limits on $f_{\rm NL}$ parameters by combining the V-
and W-band maps, and marginalizing
over the synchrotron, 
free-free, and dust foreground templates \citep{gold/etal:prep}.
As for the mask, we always use the {\it KQ75y7} mask \citep{gold/etal:prep}.

In Table~\ref{tab:fnl}, we summarize our results:
\begin{itemize}
 \item [1.] {\bf Local form results}. The 7-year best estimate of $\fnlKS$ is
$$\fnlKS=32\pm 21~\mbox{(68\%~CL)}.$$
The 95\% limit is $-10<\fnlKS<74$. When the raw maps are used, we find 
$\fnlKS=59\pm 21$~(68\%~CL). When the clean maps are used, but
       foreground templates are not marginalized over, we find 
$\fnlKS=42\pm 21$~(68\%~CL). These results (in particular the clean-map
       versus the foreground marginalized) indicate that the foreground
       emission makes a difference at the level of $\Delta\fnlKS~\sim
       10$.\footnote{The effect of the foreground marginalization
       depends on an estimator. Using the 
       needlet bispectrum, Cabella et al. \cite{cabella/etal:prep} found
       $\fnlKS=35\pm 42$ and $38\pm 47$~(68\%~CL) with and without the
       foreground 
       marginalization, respectively.} We find that the V+W result is lower than the V-band or W-band
       results. This is possible, as the V+W result contains
       contributions from the 
       cross-correlations of V and W such as $\langle {\rm VVW}\rangle$ and
       $\langle {\rm VWW}\rangle$. 
\item [2.] {\bf Equilateral form results}. The 7-year best estimate of $\fnleq$ is
$$\fnleq=26\pm 140~\mbox{(68\%~CL)}.$$
The 95\% limit is $-214<\fnleq<266$. For $\fnleq$, the foreground
      marginalization does not shift the central values very much,
      $\Delta\fnleq=-3$. This makes sense, as the equilateral bispectrum
      does not couple small-scale modes to very large-scale modes
      $l\lesssim 10$, which are sensitive to the
      foreground emission. On the other hand, the local form bispectrum
      is dominated by the squeezed triangles, which do couple large and
      small scales modes.
\item [3.] {\bf Orthogonal form results}. The 7-year best estimate of $\fnlor$ is
$$\fnlor=-202\pm 104~\mbox{(68\%~CL)}.$$
The 95\% limit is $-410<\fnlor<6$. The foreground marginalization has
little effect, $\Delta\fnlor=-4$.
\end{itemize}

As for the point-source bispectrum, we do not detect $\bsrc$ in V, W, or
V+W. In \citet{komatsu/etal:2009}, we estimated that the residual
sources could bias $\fnlKS$ by a small positive amount, and applied corrections
using Monte Carlo simulations. In this paper, we do not attempt to make
such corrections, but we note that sources could give $\Delta \fnlKS\sim
2$ (note that the simulations used by \citet{komatsu/etal:2009} likely
overestimated the effect of sources by a factor of two). As the
estimator has changed from that used by \citet{komatsu/etal:2009},
extrapolating the previous results is not trivial. Source
corrections to $\fnleq$ and $\fnlor$ could be larger
\citep{komatsu/etal:2009}, but we have not estimated the magnitude
of the effect for the 7-year data.

We used the linear perturbation theory to calculate the angular
bispectrum of primordial non-Gaussianity \citep{komatsu/spergel:2001}. 
Second-order effects
\citep{pyne/carroll:1996,mollerach/matarrese:1997,bartolo/matarrese/riotto:2006,bartolo/matarrese/riotto:2007,pitrou:2009,pitrou:2009b}
are expected to give $\fnlKS\sim 1$
\citep{nitta/etal:2009,senatore/tassev/zaldarriaga:2009,senatore/tassev/zaldarriaga:2009b,khatri/wandelt:2009,khatri/wandelt:prep,boubekeur/etal:2009,pitrou/uzan/bernardeau:2008}  and are
negligible given the noise level of the \map\ 7-year data.

Among various sources of secondary non-Gaussianities which might
contaminate measurements of primordial non-Gaussianity (in
particular $\fnlKS$), a 
coupling between 
the ISW effect and the weak gravitational lensing is the most dominant
source of confusion for $\fnlKS$
\citep{goldberg/spergel:1999,verde/spergel:2002,smith/zaldarriaga:prep,serra/cooray:2008,hanson/etal:2009,mangilli/verde:2009}. While
this contribution is expected to be detectable and bias the measurement
of $\fnlKS$ for {\sl Planck}, it is expected to be negligible for \map:
using the method of \citet{hanson/etal:2009}, we estimate that the
expected signal-to-noise ratio of this term in the \map\ 7-year data is
about $0.8$. We also estimate that this term can give $\fnlKS$ a
potential positive bias of $\Delta\fnlKS\sim 2.7$.
\citet{calabrese/etal:2010} used the skewness power spectrum method of \citet{munshi/etal:prep} to 
search for this term in the \map\ 5-year data and found a null result.
If we subtract $\Delta\fnlKS$ estimated above (for the residual source
and the ISW-lensing coupling) from the measured value, $\Delta\fnlKS$
becomes more consistent with zero.

From these results, we conclude that the \map\ 7-year data are
consistent with Gaussian primordial fluctuations to within 95\%~CL.
When combined with the limit on $\fnlKS$ from {\sl SDSS},
 $-29<\fnlKS<70$ \citep[95\%~CL][]{slosar/etal:2008}, we find $-5<\fnlKS<59$~(95\%~CL).

\section{Sunyaev--Zel'dovich Effect}
\label{sec:SZ}
We review the basics of the SZ effect in Section~\ref{sec:szintro}.
In Section~\ref{sec:coma}, we shall test our optimal estimator for
extracting the SZ signal from the \map\ data using the brightest SZ
source on the sky: the Coma cluster. We also present an improved
measurement of the SZ effect toward the Coma cluster ($3.6\sigma$). 

The most significant result from 
Section~\ref{sec:nearby} is the discovery of the thermal/dynamical
effect of clusters on the SZ effect.
We shall present the measurements of the SZ
effects toward nearby ($z\le 0.09$) galaxy clusters in Vikhlinin et al.'s sample
\citep{vikhlinin/etal:2009b}, which were used to infer the cosmological
parameters \citep{vikhlinin/etal:2009}. We then compare the measured SZ
flux to the expected flux from the X-ray data on the individual
clusters, finding a good agreement. 
Significance of detection (from merely 11 clusters, excluding Coma) is
$6.5\sigma$. By dividing the sample into cooling-flow and
non-cooling-flow clusters (or relaxed and non-relaxed clusters), we find
a significant difference in the SZ effect between these sub-samples.

In Section~\ref{sec:szstacking}, we shall report a significant
($\sim 8\sigma$) statistical detection of the SZ effect at hundreds of
positions of the known
clusters. We then compare the
measured SZ flux to theoretical models as well as to an
X-ray-calibrated empirical model, and discuss implications of our
measurement, especially a recent measurement of the
lower-than-theoretically-expected SZ power spectrum by the SPT collaboration.

Note that the analyses presented in Section~\ref{sec:nearby} and
\ref{sec:szstacking} are similar but different in one important aspect:
the former uses a handful (29) of clusters with well-measured {\it
Chandra} X-ray data, while the latter uses hundreds of clusters without
detailed X-ray data. Therefore, while the latter results have smaller
statistical errors (and much larger systematic errors), the former
results have much smaller systematic errors (and larger statistical errors). 

\subsection{Motivation and Background}
\label{sec:szintro}
When CMB photons encounter hot electrons in clusters of galaxies, the
temperature of CMB changes due to the inverse Compton scattering by
these electrons. This effect, known as the thermal SZ effect
\citep{zeldovich/sunyaev:1969,sunyaev/zeldovich:1972}, is a 
source of significant additional (secondary) anisotropies in the
microwave sky \citep[see][for reviews]{rephaeli:1995,birkinshaw:1999,carlstrom/holder/reese:2002}.

The temperature change due to the SZ effect in units of thermodynamic
temperature, $\Delta T_{\rm SZ}$, depends on frequency, $\nu$, and is
given by (for a spherically symmetric distribution of gas): 
\begin{equation}
\frac{\Delta T_{\rm SZ}(\theta)}{T_{\rm cmb}}
= g_\nu\frac{\sigma_T}{m_ec^2}\int_{-l_{\rm out}}^{l_{\rm out}}
dl~P_e\left(\sqrt{l^2+\theta^2D_A^2}\right),
\label{eq:sz}
\end{equation}
where $\theta$ is the angular distance from the center of a cluster of
galaxies on the sky, $D_A$ the proper (not comoving) 
angular diameter distance to the cluster
center, $l$ the radial coordinates from the cluster center along the line of
sight, $P_e(r)$ the {\it electron} pressure profile, $\sigma_T$ the
Thomson cross 
section, $m_e$ the 
electron mass, $c$ the speed of light, and $g_\nu$ the spectral function
given by 
\begin{equation}
 g_\nu\equiv x\coth\left(\frac{x}2\right)-4,
\end{equation}
where $x\equiv h\nu/(k_BT_{\rm cmb})\simeq \nu/(56.78~{\rm GHz})$ for
$T_{\rm cmb}=2.725$~K.
In the Rayleigh-Jeans limit, $\nu\to 0$, one finds $g_\nu\to -2$. 
At the \map\ frequencies, $g_\nu=-1.97$, $-1.94$, $-1.91$, $-1.81$, and
$-1.56$ at 23, 33, 41, 61, and 94~GHz, respectively. 
The integration boundary, $l_{\rm out}$, will be given later.

The thermal SZ effect (when relativistic corrections are ignored)
vanishes at $\simeq 217$~GHz. One then finds $g_\nu>0$ at higher
frequencies; thus, the thermal SZ effect produces a temperature
decrement at $\nu<217$~GHz, vanishes at 217~GHz, and produces a
temperature increment at $\nu>217$~GHz.

The angular power spectrum of temperature anisotropy caused by the SZ
effect is sensitive to both the gas distribution in clusters
\citep{atrio-barandela/mucket:1999,komatsu/kitayama:1999} and the
amplitude of matter density fluctuations, i.e., $\sigma_8$
\citep{komatsu/kitayama:1999,komatsu/seljak:2002,bond/etal:2005}. 
While we have not detected the SZ {\it power spectrum} in the \map\ data, we
have detected the SZ signal from the Coma cluster (Abell 1656)
in the 1-year \citep{bennett/etal:2003c} and 3-year
\citep{hinshaw/etal:2007} data.

We have also made a {\it statistical} detection of the SZ effect by
cross-correlating the \map\ data with the locations of known clusters in
the X-ray Brightest Abell-type Cluster
\citep[XBAC;][]{ebeling/etal:1996} catalog 
 \citep{bennett/etal:2003c,hinshaw/etal:2007}. 
In addition, there have been a number of statistical detections of the
SZ effect reported by many groups using various methods
\citep{fosalba/gaztanaga/castander:2003,hernandez/martin:2004,hernandez/etal:prep,myers/etal:2004,afshordi/lin/sanderson:2005,lieu/mittaz/zhang:2006,bielby/shanks:2007,afshordi/etal:2007,atrio-barandela/etal:2008,kashlinsky/etal:2008,diego/partridge:2009,melin/etal:prep}. 

\subsection{Coma Cluster}
\label{sec:coma}

\begin{figure}[t]
\centering \noindent
\includegraphics[width=8cm]{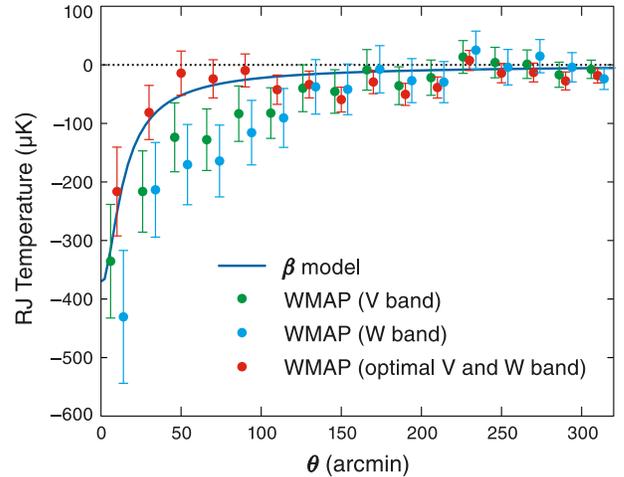}
\caption{%
Angular radial profile of the SZ effect toward the Coma cluster, in
 units of the Rayleigh-Jeans (RJ) temperature ($\mu$K). 
 While the V- (green) and W-band (blue) measurements are contaminated by the CMB
 fluctuations around Coma, our optimal estimator can separate the SZ
 effect and CMB when the V- and W-band measurements are combined (red). 
 The solid line shows the best-fitting spherical $\beta$ 
 model with the core radius of $\theta_c=10.5$~arcmin and
 $\beta=0.75$. The best-fitting central temperature decrement (fit to a
 $\beta$ model) is $T_{\rm
 SZ,RJ}(0)=-377\pm 105~\mu$K. Note that 10~arcmin corresponds to the
 physical distance of 0.195~$h^{-1}$~Mpc at the location of Coma. 
 The radius within which the mean overdensity is 500 times the critical density
 of the universe, $r_{500}$, corresponds to about 50~arcmin.
} 
\label{fig:coma}
\end{figure}

The Coma cluster (Abell 1656) is a nearby ($z=0.0231$) massive cluster
located near the north Galactic pole ($l$, $b$)=($56.75^\circ$,
$88.05^\circ$). The angular diameter distance to Coma, calculated from
$z=0.0231$ and $(\Omega_m,\Omega_\Lambda)=(0.277,0.723)$, is
$D_A=67~h^{-1}$~Mpc; thus, 10~arcmin on the sky corresponds to the physical
distance of $0.195~h^{-1}$~Mpc at the redshift of Coma.

To extract the SZ signal from the \map\ temperature map, we use the
optimal method described in Appendix~\ref{app:sz}: we write down the
likelihood function that contains CMB, noise, and the SZ effect, and
marginalize it over CMB. From the resulting likelihood function for the SZ
effect, which is given by equation~(\ref{eq:sz_p_d}), we find the
optimal estimator for the SZ effect in a given 
angular bin $\alpha$, $\hat{p}_\alpha$, as
\begin{equation}
 \hat{p}_\alpha
= F^{-1}_{\alpha\beta}(t_\beta)_{\nu' p'}
[N_{\rm pix}+\tilde{C}]^{-1}_{\nu'p',\nu p}
d_{\nu p},
\label{eq:phat}
\end{equation}
where the repeated symbols are summed. Here,
$d_{\nu p}$ is the measured temperature at a pixel $p$ in a frequency
band $\nu$, $(t_{\alpha})_{\nu p}$ is a map of an annulus 
corresponding to a given angular bin $\alpha$, which has been convolved
with the beam and scaled by the 
frequency dependence of the SZ effect, $N_{{\rm pix},\nu p,\nu'p'}$ is the noise
covariance matrix (which is taken to be diagonal in pixel space and $\nu$, i.e.,
$N_{{\rm pix},\nu p,\nu'p'}=\sigma^2_{\nu p}\delta_{\nu\nu'}\delta_{pp'}$), and
$\tilde{C}_{\nu p,\nu' p'}\equiv \sum_{lm}C_lb_{\nu l}b_{\nu' l}Y_{lm,p}Y_{lm,p'}^*$ is the
signal covariance matrix of CMB convolved with the beam
($C_l$ and $b_{\nu l}$ are the CMB power spectrum and the
beam transfer function, respectively). A matrix $F_{\alpha\beta}$ gives
the $1\sigma$ error of $\hat{p}_\alpha$ as
$\sqrt{(F^{-1})_{\alpha\alpha}}$, and is given by
\begin{equation}
 F_{\alpha\beta}
=
(t_\alpha)_{\nu' p'}
[N_{\rm pix}+\tilde{C}]^{-1}_{\nu'p',\nu p}
(t_\beta)_{\nu p}.
\end{equation}

For $d_{\nu p}$, we use the foreground-cleaned V- and W-band
temperature maps at the HEALPix resolution of $N_{\rm side}=1024$,
masked by the {\it KQ75y7} mask. Note that the {\it KQ75y7} mask includes the
7-year source mask, which removes a potential bias in the reconstructed
profile due to any sources which are bright enough to be resolved by
\map, as well as the sources found by other surveys. 
Specifically, the 7-year point source mask includes sources in the
7-year \map\ source catalog \citep{gold/etal:prep}; sources from
\citet{stickel/meisenheimer/kuehr:1994}; sources with 22~GHz fluxes $\ge 0.5$~Jy from
\citet{hirabayashi/etal:2000:la}; flat spectrum objects from
\citet{terasranta/etal:2001}; and sources from the blazar survey of
\citet{perlman/etal:1998} and \citet{landt/etal:2001}.

In Figure~\ref{fig:coma}, we show the measured angular radial profiles of
Coma in 16 angular bins (separated by $\Delta\theta=20$~arcmin), in
units of the Rayleigh-Jeans temperature, for the V- and W-band 
data, as well as for the V+W combined data.
The error bar at a given angular bin is given by
$\sqrt{(F^{-1})_{\alpha\alpha}}$. 

We find that all of these measurements agree well at $\theta\ge 130$~arcmin;
however, at smaller angular scales, $\theta\le 110$~arcmin, the V+W result
shows {\it less} SZ  than both the V- and W-only results.  Does this
make sense? As described in Appendix~\ref{app:sz}, our optimal estimator
uses both the $C^{-1}$-weighted V+W map and the $N^{-1}$-weighted
V$-$W map. While the latter map vanishes for CMB, it does not vanish for
the SZ effect. Therefore, the latter map can be used to separate CMB and
SZ effectively.

This explains why the V+W result and the other results are different
only at small angular scales: at $\theta\ge 130$~arcmin, the measured
signal is $|\Delta T|\lesssim 50~\mu$K. If this was due to SZ, the
difference map, V$-$W, would give $|\Delta T|\lesssim (1-1.56/1.81)\times
50~\mu{\rm K}\simeq 7~\mu$K, which is smaller than the noise level in the
difference map, and thus would not show up. In other words, our
estimator cannot distinguish between CMB and SZ at $\theta\ge
130$~arcmin.

On the other hand, at $\theta\le
110$~arcmin, each of the V- and W-band data shows much bigger signals,
$|\Delta T|\gtrsim 100~\mu$K. If this was due to SZ, the difference map
would give $|\Delta T|\gtrsim 14~\mu$K, which is comparable to or
greater than the noise level in the difference map, and thus would be
visible. We find that the difference map does not detect signals in
$50\le\theta\le 110$~arcmin, which suggests that the measured signal,
$-100~\mu$K, is {\it not} due to SZ, but due to CMB. As a result, the
V$+$W result shows less SZ than the V- and W-only results.

In order to quantify a statistical significance of
detection and interpret the result, we model the SZ profile using a
spherical $\beta$ model \citep{cavaliere/fusco-femiano:1976}:
\begin{equation}
 \Delta T_{\rm SZ}(\theta)
= \Delta T_{\rm SZ}(0)
\left[1+(\theta/\theta_c)^2\right]^{(1-3\beta)/2}.
\end{equation}
To make our analysis consistent with previous measurements described
later, we fix the 
core radius, 
$\theta_c$, and the slope parameter, $\beta$, at $\theta_c=10.5$~arcmin
and $\beta=0.75$ \citep{briel/henry/bohringer:1992}, and vary only the
central decrement, $\Delta T_{\rm SZ}(0)$. 
In this case, the optimal estimator is
\begin{equation}
 \Delta T_{\rm SZ}(0)
= \frac1Ft_{\nu' p'}
[N_{\rm pix}+\tilde{C}]^{-1}_{\nu'p',\nu p}
d_{\nu p},
\end{equation}
where $t_{\nu p}$ is a map of the above $\beta$ model with $\Delta
T_{\rm SZ}(0)=1$, and 
\begin{equation}
 F = t_{\nu' p'}[N_{\rm pix}+\tilde{C}]^{-1}_{\nu'p',\nu p}t_{\nu p},
\label{eq:F}
\end{equation}
gives the $1\sigma$ error as $1/\sqrt{F}$.

For V+W, we find 
$$\Delta T_{\rm SZ,RJ}(0)=-377\pm 105~\mu{\rm K}~\mbox{(68\%~CL)},$$ 
which is a $3.6\sigma$ measurement of the SZ effect toward Coma. 
In terms of the Compton $y$-parameter at the center, we find
\begin{eqnarray*}
y_{WMAP}(0)&=&\frac{-1}{2}\frac{\Delta T_{\rm SZ,RJ}(0)}{T_{\rm cmb}}\\
&=&(6.9\pm 1.9)\times
10^{-5}~\mbox{(68\%~CL)}.
\end{eqnarray*}

Let us compare this measurement with the previous measurements. 
\citet{herbig/etal:1995} used the 5.5-m telescope at the Owens Valley
Radio Observatory (OVRO) to observe Coma at 32~GHz. Using the same
$\theta_c$ and $\beta$ as above, they found the central decrement of $\Delta
T_{\rm SZ,RJ}(0)=-505\pm 92~\mu$K~(68\%~CL), after subtracting
$38~\mu$K due to 
point sources (5C4.81 and 5C4.85). 
These sources have been masked by our point-source mask, and thus we do not
need to correct for point sources.
 
While our estimate of $\Delta T_{\rm SZ,RJ}(0)$ is different from that
of \citet{herbig/etal:1995} only by $1.2\sigma$, and thus is
statistically consistent, we note that \citet{herbig/etal:1995} did not
correct for the CMB fluctuation in the direction of Coma. As the above
results indicate that the CMB fluctuation in the direction of Coma is on
the order of $-100~\mu$K, it is plausible that the OVRO measurement
implies $\Delta T_{\rm SZ,RJ}(0)\sim -400$K, which is an excellent
agreement with the \map\ measurement.

The Coma cluster has been observed also by the Millimetre
and Infrared Testagrigia Observatory (MITO) experiment 
\citep{depetris/etal:2002}.
Using the same $\theta_c$ and $\beta$ as above,
\citet{battistelli/etal:2003} found $\Delta T_{\rm SZ}(0)=-184\pm
39$, $-32\pm 79$, and $+172\pm 36~\mu$K~(68\%~CL) at 143, 214, and
272~GHz, respectively, in units of thermodynamic temperatures.
As MITO has 3 frequencies, they were able to separate SZ, CMB, and the
atmospheric fluctuation. By fitting these 3 data points to the SZ
spectrum, $\Delta T_{\rm SZ}/T_{\rm cmb}=g_\nu y$, we find $y_{\rm MITO}(0)=(6.8\pm 1.0\pm 0.7)\times 10^{-5}$, which is
an excellent agreement with the \map\ measurement.
The first error is statistical and the second error is systematic due to
10\% calibration error of MITO. The calibration error
of the \map\ data \citep[0.2\%;][]{jarosik/etal:prep} is negligible.

Finally, one may try to fit the multi-wavelength data of $T_{\rm SZ}(0)$
to separate the SZ effect and CMB. For this purpose, we fit the \map\
data in V- and W-band to the $\beta$ model {\it without correcting for
the CMB fluctuation}. We find $-381\pm 126~\mu$K and $-523\pm 127~\mu$K in thermodynamic units
(68\%~CL). The OVRO measurement, $T_{\rm SZ,RJ}(0)=-505\pm 92~\mu$K \citep{herbig/etal:1995}, has been scaled to the Rayleigh-Jeans temperature with the SZ spectral dependence correction, and thus we use this measurement at $\nu=0$.
Fitting the \map\ and OVRO data to the SZ effect plus CMB, and the MITO
data only to the SZ effect (because the CMB was already removed from
MITO using their multi-band data), we find $y(0)=(6.8\pm 1.0)\times
10^{-5}$ and $\Delta T_{\rm cmb}(0)=-136\pm 82~\mu$K
(68\%~CL). This result is consistent with our interpretation that the
$y$-parameter 
of the center of Coma is $7\times 10^{-5}$ and the CMB fluctuation is on
the order of $-100~\mu$K. 

The analysis presented here shows that our optimal estimator is an
excellent tool for extracting the SZ effect from multi-frequency data. 

\subsection{Nearby Clusters: Vikhlinin et al.'s low-$z$ sample}
\label{sec:nearby}
\begin{figure*}[t]
\centering \noindent
\includegraphics[width=18cm]{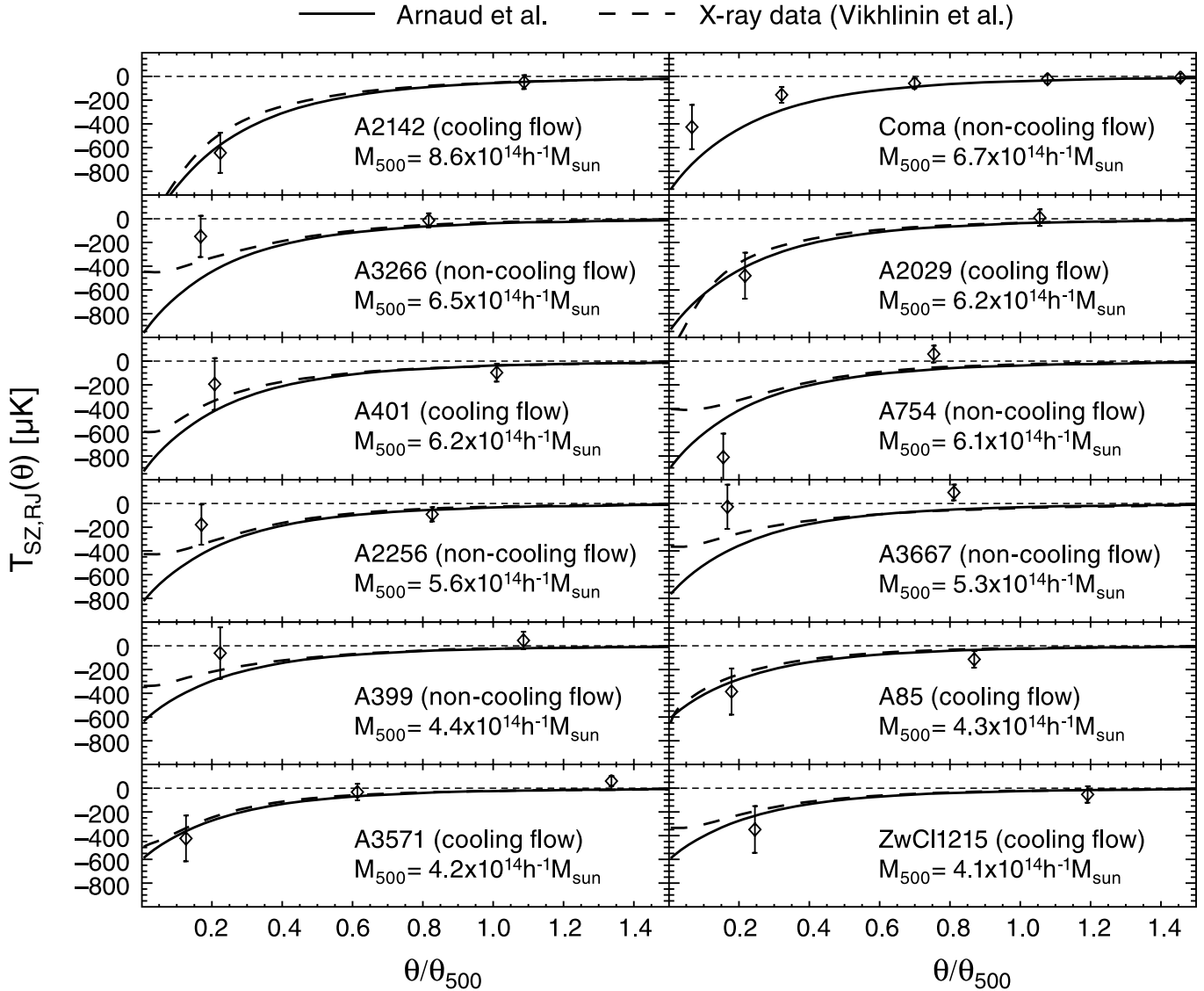}
\caption{%
Angular radial profiles of the SZ effect toward nearby massive clusters
 (with $M_{500}\ge 4\times 10^{14}~h^{-1}~M_\sun$ and $z\le 0.09$), in
 units of the Rayleigh-Jeans (RJ) temperature ($\mu$K). The V- and
 W-band data are combined optimally to separate the CMB and the SZ
 effect. All of these clusters have $\theta_{500}\ge 14'$, i.e., 
 resolved by the \map\ beam. The masses, $M_{500}$, are $M_Y$  given in
 the 6th column of Table~2 in \citet{vikhlinin/etal:2009b}, times
 $h_{\rm vikhlinin}=0.72$ used by them, except for Coma. For Coma, we
 estimate $M_{500}$ using the
 mass-temperature relation given in \citet{vikhlinin/etal:2009b} with
 the temperature of 8.45~keV \citep{wik/etal:2009}.
 The dashed
 lines show the expected SZ 
 effect from the X-ray data on the individual clusters, whereas the
 solid lines show the prediction from the average pressure profile
 found by \citet{arnaud/etal:2010}. Note that Coma is not included in
 the sample of \citet{vikhlinin/etal:2009b}, and thus the X-ray data are
 not shown.
 We find that Arnaud et al.'s profiles overpredict the gas pressure
 (hence the SZ effect) of non-cooling flow clusters. Note that all
 cooling-flow clusters are ``relaxed,'' and all non-cooling-flow
 clusters are ``non-relaxed'' (i.e., morphologically disturbed),
 according to the criterion of \citet{vikhlinin/etal:2009b}.
} 
\label{fig:massive}
\end{figure*}

The Coma cluster is the brightest SZ cluster on the sky. 
There are other clusters that are bright enough to be seen by \map.

\subsubsection{Sample of nearby ($z<0.1$) clusters}

In order to select candidates, we use the sample of 49 nearby
clusters compiled by \citet{vikhlinin/etal:2009b}, which are used by the
cosmological analysis given in  \citet{vikhlinin/etal:2009}. These
clusters are selected from the {\sl ROSAT} All-sky Survey, and 
have detailed follow-up observations by {\sl Chandra}. 
The latter property is especially important, as it allows us to {\it directly}
compare 
the measured SZ effect in the \map\ data and the expected one from the
X-ray data on a 
cluster-by-cluster basis, without relying on any scaling relations.\footnote{
For this reason, the analysis given in this section is ``cleaner'' 
than the one given in Section~\ref{sec:szstacking}, which uses a larger
number of 
clusters but relies on scaling relations. 
Nevertheless,  the results obtained from the
analysis in this section and those in Section~\ref{sec:szstacking} are in good
agreement.}

Not all nearby clusters in \citet{vikhlinin/etal:2009b} are suitable for
our purpose, as some clusters are too small to be resolved by the \map\
beam. We thus select the clusters that have the radius greater than
$14'$ on the sky: specifically, we use the clusters whose $\theta_{500}\equiv
r_{500}/D_A(z)$ is greater than $14'$. Here, $r_{500}$ is the radius
within which 
the mean overdensity is 500 times the critical density of the
universe.
We find that 38 clusters satisfy
this condition. (Note that the Coma cluster is not included in this
sample.)

Of these, 5 clusters have $M_{500}\ge 6\times 10^{14}~h^{-1}~M_\sun$, 7
clusters have $4\times 10^{14}~h^{-1}~M_\sun\le M_{500}< 6\times
10^{14}~h^{-1}~M_\sun$, 13 clusters have 
$2\times 10^{14}~h^{-1}~M_\sun\le M_{500}< 4\times
10^{14}~h^{-1}~M_\sun$, and 13 clusters have
$1\times 10^{14}~h^{-1}~M_\sun\le M_{500}< 2\times
10^{14}~h^{-1}~M_\sun$. 
Here, $M_{500}$ is the mass enclosed within $r_{500}$, i.e.,
$M_{500}\equiv M(r\le r_{500})$.

Finally, we remove the clusters that lie within the {\it KQ75y7} mask
(including the diffuse and the source mask),
leaving 29 clusters for our analysis.
(1 cluster (A478) in $4\times 10^{14}~h^{-1}~M_\sun\le M_{500}< 6\times
10^{14}~h^{-1}~M_\sun$, 4 clusters in $2\times 10^{14}~h^{-1}~M_\sun\le M_{500}< 4\times
10^{14}~h^{-1}~M_\sun$, and 4 clusters in $1\times 10^{14}~h^{-1}~M_\sun\le M_{500}< 2\times
10^{14}~h^{-1}~M_\sun$ are masked, mostly by the point source mask.) 
The highest redshift of this sample is $z=0.0904$ (A2142).

\subsubsection{\map\ versus X-ray: cluster-by-cluster comparison}
\begin{deluxetable}{lccc}
\tablecolumns{4}
\small
\tablewidth{0pt}
\tablecaption{%
Best-fitting Amplitude for the SZ Effect in the \map\ 7-year data
}
\tablehead{
\colhead{Mass Range\tablenotemark{a}}
& \colhead{\# of clusters}
& \colhead{Vikhlinin et al.\tablenotemark{b}}
& \colhead{Arnaud et al.\tablenotemark{c}}
}
\startdata
$6\le M_{500}<9$& 5& $0.90\pm 0.16$ & $0.73\pm 0.13$\nl
$4\le M_{500}<6$& 6& $0.73\pm 0.21$ & $0.60\pm 0.17$ \nl
$2\le M_{500}<4$& 9& $0.71\pm 0.31$ & $0.53\pm 0.25$ \nl
$1\le M_{500}<2$& 9& $-0.15\pm 0.55$ & $-0.12\pm 0.47$ \nl
\hline
$4\le M_{500}<9$& 11& $0.84\pm 0.13$ & $0.68\pm 0.10$\nl
$1\le M_{500}<4$& 18& $0.50\pm 0.27$ & $0.39\pm 0.22$ \nl
\hline
$4\le M_{500}<9$& & & \nl
cooling flow\tablenotemark{d} & 5 & $1.06\pm 0.18$ & $0.89\pm 0.15$\nl
non-cooling flow\tablenotemark{e} & 6 & $0.61\pm 0.18$ & $0.48\pm 0.15$\nl
\hline
$2\le M_{500}<9$& 20& $0.82\pm 0.12$ & $0.660\pm
 0.095$\nl
\hline
$1\le M_{500}<9$& 29& $0.78\pm 0.12$ & $0.629\pm 0.094$
\enddata
\tablenotetext{a}{In units of $10^{14}~h^{-1}~M_\sun$. Coma is not
 included. The masses are derived from the mass-$Y_X$ relation, and are
 given in the 6th column of Table~2 in \citet{vikhlinin/etal:2009b}, times
 $h_{\rm vikhlinin}=0.72$.} 
\tablenotetext{b}{Derived from the X-ray data on the individual clusters
 \citep{vikhlinin/etal:2009b}.}
\tablenotetext{c}{The ``universal pressure profile'' given by
 \citet{arnaud/etal:2010}.}
\tablenotetext{d}{Definition of ``cooling flow'' follows that of
 \citet{vikhlinin/etal:2007}. All of cooling-flow clusters here are also
 ``relaxed,'' according to the criterion of
 \citet{vikhlinin/etal:2009b}.}
\tablenotetext{e}{Definition of ``non-cooling flow'' follows that of
 \citet{vikhlinin/etal:2007}. All of non-cooling-flow clusters here are also
 ``non-relaxed'' (or mergers or morphologically disturbed), according to
 the criterion of \citet{vikhlinin/etal:2009b}.} 
\label{tab:vikh_vs_arnaud}
\end{deluxetable}

In Figure~\ref{fig:massive}, we show the measured SZ effect in the
symbols with error bars, as well as
the expected SZ from the X-ray data in the dashed lines. 

To compute
the expected SZ, we use equation~(\ref{eq:sz}) with $P_e=n_ek_BT_e$,
where $n_e$ and $T_e$ are fits to the X-ray data. 
Specifically, we use
\citep[see equation~(3) and (8) of][]{vikhlinin/etal:2006}\footnote{With
a typo in equation~(8) corrected (A. Vikhlinin 2010, private communication).}
\begin{eqnarray}
\nonumber
 n_e^2(r) &=&
  n_0^2\frac{(r/r_c)^{-\alpha}}{(1+r^2/r_c^2)^{3\beta-\alpha/2}}\frac1{(1+r^\gamma/r_s^\gamma)^{\epsilon/\gamma}}\\
& &+\frac{n_{02}^2}{(1+r^2/r_{c2}^2)^{3\beta_2}},\\
\nonumber
\frac{T_e(r)}{T_{mg}}&=&1.35\frac{(x/0.045)^{1.9}+0.45}{(x/0.045)^{1.9}+1}\\
& &\times\frac1{[1+(x/0.6)^2]^{0.45}},
\end{eqnarray}
where $x\equiv r/r_{500}$.
The parameters in the above equations are found from the {\sl Chandra}
X-ray data, and 
kindly made available to us by A. Vikhlinin.

For a given pressure profile, $P_e(r)$, we compute the SZ temperature profile as
\begin{eqnarray}
\nonumber
\Delta T_{\rm SZ}(\theta)&=&
g_\nu T_{\rm cmb} \frac{\sigma_T}{m_ec^2}P^{\rm 2d}_e(\theta)\\
&\simeq& 273~\mu{\rm K}~g_\nu\left[\frac{P^{\rm 2d}_{e}(\theta)}{25~{\rm
 eV~cm^{-3}~Mpc}}\right],
\label{eq:tszfit}
\end{eqnarray}
where $P^{\rm 2d}_{e}(\theta)$ is the projected electron pressure
profile on the sky: 
\begin{eqnarray}
P^{\rm 2d}_{e}(\theta)
=
\int_{-\sqrt{r_{\rm out}^2-\theta^2D_A^2}}^{\sqrt{r_{\rm out}^2-\theta^2D_A^2}} 
dl~P_e\left(\sqrt{l^2+\theta^2D_A^2}\right).
\end{eqnarray}
Here, we truncate the pressure profile at $r_{\rm out}$. We take this to
be $r_{\rm out}=6r_{500}$.
While the choice of the boundary is somewhat arbitrary, the results are
not sensitive to the exact value because the pressure profile declines
fast enough. 

We find a good agreement between the measured and expected SZ signals
(see Figure~\ref{fig:massive}),
except for A754: A754 is a merging cluster
with a highly disturbed X-ray morphology, and thus the expected SZ
profile, which is 
 derived assuming spherical symmetry (equation~\ref{eq:tszfit}),  may be
 different from the  observed one.

To make the comparison quantitative, we select clusters within a given
mass bin, and fit the expected SZ profiles to
the \map\ data with a single amplitude, $a$, treated as a free
parameter.
The optimal estimator for the normalization of pressure, $a$, is
\begin{equation}
 a = \frac1Ft_{\nu' p'}
[N_{\rm pix}+\tilde{C}]^{-1}_{\nu'p',\nu p}
d_{\nu p},
\label{eq:estimate_a}
\end{equation}
where $t_{\nu p}$ is a map containing the predicted SZ profiles around
clusters, and the $1\sigma$ error is  $1/\sqrt{F}$ where $F$ is given by
equation~(\ref{eq:F}). 

We summarize the results in the second column of
Table~\ref{tab:vikh_vs_arnaud}. 
We find that the amplitudes of {\it all} mass bins are consistent with
unity ($a=1$) to within $2\sigma$ (except for the ``non-cooling flow''
case, for which $a$ is less than unity at $2.2\sigma$; we shall come
back to this important point in the next section). The agreement is
especially good for 
the highest mass bin ($M_{500}\ge 6\times 10^{14}~h^{-1}~M_\sun$),
$a=0.90\pm 0.16$ (68\%~CL). 

Note that this is a $5.6\sigma$ detection of
the SZ effect, just from stacking 5 clusters. 
By stacking 11 clusters with $M_{500}\ge 4\times
10^{14}~h^{-1}~M_\sun$ (i.e., all clusters in Figure~\ref{fig:massive}
but Coma), we find $a=0.84\pm 0.13$ (68\%~CL), a
$6.5\sigma$ detection.
In other words, one does
not need to stack many tens or hundreds of clusters to see the SZ effect in
the \map\ data, contrary to what is commonly done in the literature
\citep{fosalba/gaztanaga/castander:2003,hernandez/martin:2004,hernandez/etal:prep,myers/etal:2004,afshordi/lin/sanderson:2005,lieu/mittaz/zhang:2006,bielby/shanks:2007,afshordi/etal:2007,atrio-barandela/etal:2008,kashlinsky/etal:2008,diego/partridge:2009,melin/etal:prep}. 

From this study, we conclude that the \map\ data and the expectation from the
X-ray data are in good agreement.

\subsubsection{\map\ versus a ``universal pressure profile'' of Arnaud
   et al.: effect of recent mergers}
Recently, \citet{arnaud/etal:2010} derived pressure profiles
of 33 clusters from the X-ray follow-up observations of 
the REXCESS clusters using {\sl XMM-Newton}.
The REXCESS sample contains clusters selected from the {\sl ROSAT}
All-sky Survey \citep{boehringer/etal:2007}. 
By scaling the pressure profiles appropriately by mass and redshift and 
taking the median of the scaled profiles, they produced
a ``universal pressure profile.'' We describe this profile in
Appendix~\ref{sec:arnaud}. 

We show the predicted $\Delta T_{\rm SZ}(\theta)$ from Arnaud et al.'s
pressure profile 
in Figure~\ref{fig:massive} (solid lines). In order to compute their
profile, we need the mass of clusters, $M_{500}$. We take $M_{500}$ from
the the 6th column of Table~2 in \citet{vikhlinin/etal:2009b}, which are
derived from the so-called mass-$Y_X$ relation, the most precise mass
proxy known to date with a scatter of about 5\%.\footnote{
The exception is Coma, which is not included in the nearby sample of
\citet{vikhlinin/etal:2009b}. Therefore, we use the mass-temperature
relation of \citet{vikhlinin/etal:2009b} (the first row of Table 3)
for this cluster: $M_{500}=(3.02\pm 0.11)\times 10^{14}~h^{-1}~M_\sun(T_X/5~{\rm
keV})^{1.53\pm 0.08}/ E(z)$, with $E(z)=1.01$ for Coma's redshift, $z=0.023$.
We use the X-ray temperature of $T_X=8.45\pm 0.06$~keV
\citep{wik/etal:2009}.} Again, we take the outer boundary of the pressure to
be $r_{\rm out}=6r_{500}$.

We fit Arnaud et al's profiles to the \map\
data of 29 clusters. We find that, in all but one of the mass bins, the
best-fitting 
normalization, $a$, is less than unity by more than $2\sigma$. By stacking
11 clusters with $M_{500}\ge 4\times 10^{14}~h^{-1}~M_\sun$, we find
$a=0.68\pm 0.10$ (68\%~CL). This measurement rules out $a=1$ by $3.2\sigma$.
The universal pressure profile overestimates the SZ effect by $\sim 30$\%. 

What causes the discrepancy? The thermal/dynamical state of gas in
clusters may be 
the culprit. From Figure~\ref{fig:massive}, we find that the X-ray data
(hence the SZ effect) and the universal profile agree well for ``cooling
flow'' clusters, but do not agree for non-cooling flow clusters.

The cooling flow clusters have {\it cool cores}, in 
which the cooling time (due to Bremsstrahlung) is shorter than the
Hubble time \citep{fabian:1994}. The clusters shown in
Figure~\ref{fig:massive} are classified as either ``cooling flow'' or
``non-cooling flow'' clusters, following the definition of
\citet{vikhlinin/etal:2007}.

We find that Arnaud et al.'s profiles agree with the X-ray data on the
individual clusters well at $\theta\gtrsim 0.3\theta_{500}$. This agrees
with Figure~8 of \citet{arnaud/etal:2010}. The profiles 
differ significantly in the inner parts of clusters, which is also in 
good agreement with the conclusion of \citet{arnaud/etal:2010}: they
find that cool-core clusters show much steeper inner profiles than
non-cool-core clusters (their Figure~2 and 5).

For cooling-flow clusters, the agreement between the \map\ data and
Arnaud et al.'s profile is good: $a=0.89\pm 0.15$ (68\%~CL). However,
for non-cooling-flow clusters, we find a very low amplitude, $a=0.48\pm
0.15$  (68\%~CL), which rules out Arnaud et al.'s profile by $3.5\sigma$.
A similar trend is also observed for the individual X-ray data of
Vikhlinin 
et al.: $a=1.06\pm 0.18$ and $0.61\pm 0.18$  (68\%~CL) for cooling-flow
and non-cooling-flow clusters, respectively; however, statistical
significance is not large enough to exclude $a=1$.

Based on this study, we conclude that  one must distinguish between
cool-core (cooling flow) and non-cool-core clusters when interpreting
the observed profile of the SZ effect. It is clear (at the $3.2\sigma$
level) that 
Arnaud et al.'s profile is inconsistent with the individual X-ray data
and the SZ data taken by \map, and 
(at the $3.5\sigma$ level) {\it one must distinguish between the
cool-core and non-cool-core clusters}.

Interestingly, {\it all} of cooling-flow clusters are ``relaxed''
clusters, and {\it all}  of non-cooling-flow clusters are
``non-relaxed'' (i.e., morphologically disturbed) clusters, according to
the criterion of \citet{vikhlinin/etal:2009b}. If we interpret this as
non-cooling-flow clusters having undergone recent mergers, then we may
conclude that we are finding the effect of mergers on the SZ effect.

While our conclusion is still based on a limited number of clusters,
it may be valid for a much larger sample of clusters, as we shall show
in Section~\ref{sec:arnaudresults}.

Finally, we note that the current generation of hydrodynamical
simulations predict the pressure profiles that are even steeper
than Arnard et al.'s profile \citep[see Figure~7 of][]{arnaud/etal:2010}.
Therefore, the  simulations also overpredict the amount of
pressure in clusters relative to the \map\ data.
We shall come back to this point in Section~\ref{sec:KSandSim}.

\subsection{Statistical Detection of the SZ Effect}
\label{sec:szstacking}

\begin{figure}[t]
\centering \noindent
\includegraphics[width=7.6cm]{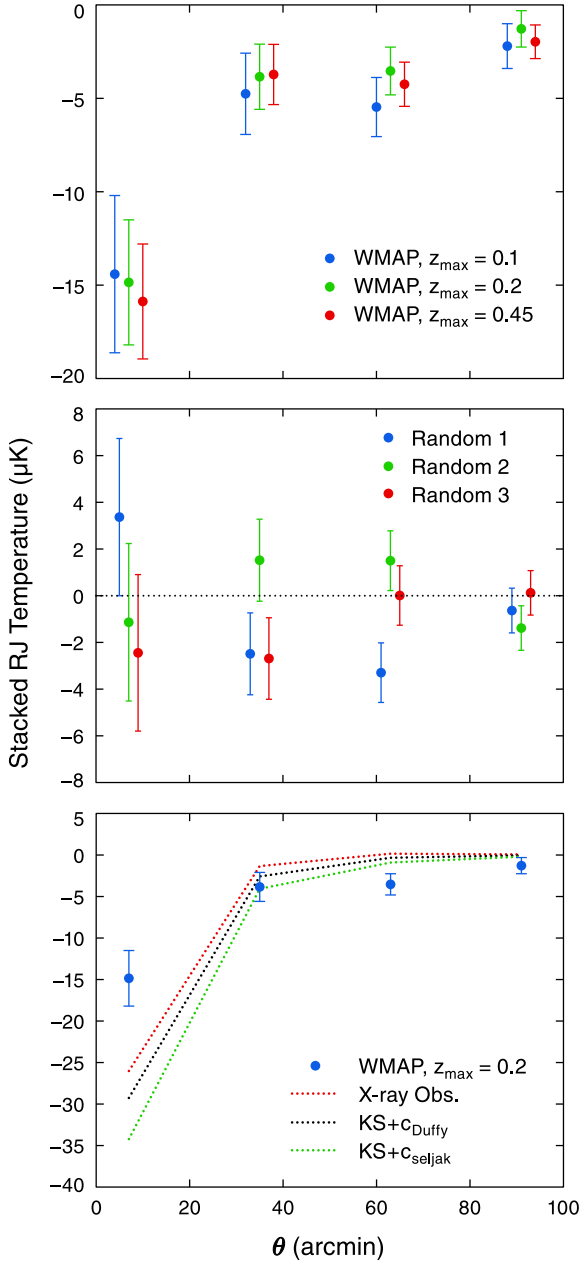}
\caption{%
 Average temperature profile of the SZ effect from the stacking
 analysis, in
 units of the Rayleigh-Jeans (RJ) temperature ($\mu$K), 
 at $\theta=7$, 35, 63, and 91~arcmin.
 The V- and
 W-band data are combined using the optimal estimator.
 (Top) The SZ effect measured from the locations of clusters of galaxies.
 The results with three different maximum redshifts,
 $z_{\rm max}=0.1$ (blue; left), 0.2 (green; middle), and 0.45 (red;
 right), are shown. The error bars include noise due 
 to the CMB fluctuation, and thus are correlated (see
 equation~(\ref{eq:corr}) for the correlation matrix).
 (Middle) A null test showing profiles measured from random locations on
 the sky (for $z_{\rm max}=0.2$; the number of random locations is the
 same as the number of 
 clusters used in the top panel). Three random realizations are
 shown. Our method does not produce biased results.
 (Bottom) The measured profile ($z_{\rm max}=0.2$) is compared with the
 model profiles 
 derived from the median of 33 clusters in the REXCESS sample
 \citep{arnaud/etal:2010}    
  and theoretically calculated from hydrostatic equilibrium
 \citep{komatsu/seljak:2001} with two different concentration
 parameters. Note that the model profiles are calculated also for
 $z_{\rm max}=0.2$ but have not been multiplied by
 the best-fitting normalization factors given in
 Table~\ref{tab:szamplitude}. The theoretical profiles are processed in
 the same manner that the data are processed, using 
 equation~(\ref{eq:phat}). 
} 
\label{fig:szstacking}
\end{figure}

To explore the SZ effect in a large number of clusters, 
we use a galaxy cluster catalog consisting of the {\sl ROSAT}-ESO
flux-limited X-ray (REFLEX) galaxy cluster survey
\citep{boehringer/etal:2004} in the southern hemisphere above the
Galactic plane ($\delta<2.5^\circ$ and $|b|>20^\circ$)
and the extended Brightest Cluster Sample
\citep[eBCS;][]{ebeling/etal:1998,ebeling/etal:2000} in the northern
hemisphere above the Galactic plane ($\delta>0^\circ$ and
$|b|>20^\circ$). Some clusters are contained in both
samples. Eliminating the overlap, this catalog contains 742 clusters of
galaxies.  Of these, 400, 228, and 114 clusters lie in the redshift
ranges of $z\le 0.1$, $0.1<z\le 0.2$, and $0.2<z\le 0.45$,
respectively. 

We use the foreground-reduced V- and W-band maps at the HEALPix
resolution of $N_{\rm side}=1024$, masked by the {\it KQ75y7}
mask, which eliminates the entire Virgo cluster.
Note that this mask also includes the point-source mask, which
masks sources at the locations of some clusters (such as Coma).
After applying the mask, we have
361, 214, and 109 clusters in $z\le 0.1$, $0.1<z\le 0.2$, and $0.2<z\le 0.45$,
respectively. 

We again use equation~(\ref{eq:phat}) to find the angular radial profile in
four angular bins. For this analysis, $(t_{\alpha})_{\nu p}$ is a map
containing many annuli (one annulus around each cluster) corresponding
to a given angular bin $\alpha$, convolved with the beam and scaled by the 
frequency dependence of the SZ effect.

We show the measured profile in the top panel of
  Figure~\ref{fig:szstacking}. We have done this analysis using 3
different choices of the maximum redshift, $z_{\rm max}$, to select
clusters: $z_{\rm max}=0.1$, 0.2, and 0.45. We find that the results are
not sensitive to $z_{\rm max}$. As expected, the results for $z_{\rm
max}=0.1$ have the largest error bars. The error bars for $z_{\rm
max}=0.2$ and 0.45 are similar, indicating that we do not gain much more
information from $z>0.2$.
The error bars have contributions from instrumental noise and CMB
  fluctuations. The latter contribution correlates the errors at
  different angular bins.

The top panel shows a decrement of $-3.6\pm 1.4~\mu$K at a very large
angular distance from the center, $\theta=63$~arcmin, for $z_{\rm
max}=0.2$. As we do not expect to have such an extended gas distribution
around clusters, one may wonder if this result 
implies that we have a bias in the zero level. 
In order to check for a potential systematic bias, we perform the
following null test: instead of measuring the SZ signals from the locations of
clusters, we measure them from random locations in the \map\ data. In
the middle panel of Figure~\ref{fig:szstacking}, we show that our 
method passes a null test. We find that the measured profiles are
consistent with zero; thus, our method does not introduce a bias.

Is this signal at a degree scale real? For example, are there nearby
massive clusters 
(such as Coma) which give a significant SZ effect at a degree scale? 
While the Virgo cluster has the largest angular size on the sky, the
{\it KQ75y7} mask eliminates Virgo.
In order to see if other nearby clusters give significant
contributions, we remove all clusters at $z\le 0.03$ (where there are 57
clusters) and remeasure the SZ
profile. We find that the changes are small, less than $1~\mu$K at all
angular bins. At $\theta=63~$arcmin, the change is especially small,
$\sim 0.1~\mu$K, and thus nearby clusters do not make much contribution to
this bin.

 The apparent decrement at $\theta=63$~arcmin is probably due to a statistical
 fluctuation. The angular bins are correlated with the following
 correlation matrix:
\begin{equation}
 \left(
\begin{array}{cccc}
     1     &   0.5242 &   0.0552 &   0.0446 \\
  0.5242 &      1     &   0.4170 &   0.0638 \\
  0.0552 &   0.4170 &      1    &   0.4402 \\
  0.0446 &   0.0638 &   0.4402 &    1    
\end{array}
\right),
\label{eq:corr}
\end{equation}
where the columns correspond to $\theta=7$, 35, 63, and 91~arcmin,
respectively. The decrements at the first two bins (at $\theta=7$ and
35~arcmin) can drive the third bin at $\theta=63$~arcmin to be more negative.
Note also that one of the realizations shown in the bottom
 panel (``Random 1'' in 
the middle panel of 
Figure~\ref{fig:szstacking}) shows $\sim -3.5~\mu$K at
$\theta=63$~arcmin. The second bin is also negative with a similar amplitude.
On the other hand, ``Random 2'' shows both positive temperatures at
the second 
and third bins, which is also consistent with a positive correlation
between these bins. 

Finally, in the bottom
panel of Figure~\ref{fig:szstacking}, we compare the measured SZ profile
with the expected profiles from various cluster gas models (described in
Section~\ref{sec:szimplications}). None of them show a significant
signal at $\theta=63$~arcmin, which is also consistent with our
interpretation that it is a statistical fluctuation.

\subsection{Interpretations}
\label{sec:szimplications}
\subsubsection{General idea}

In order to interpret the measured SZ profile, we need a model for the
electron pressure profile, $P_e(r)$ (see equation~(\ref{eq:sz})). 
For fully ionized gas, the electron pressure is related to the {\it gas}
(baryonic) pressure, 
$P_{\rm gas}(r)$, by
\begin{equation}
 P_e(r) = \left(\frac{2+2X}{3+5X}\right)P_{\rm gas}(r),
\label{eq:PgasPe}
\end{equation}
where $X$ is the abundance of hydrogen in clusters.
For $X=0.76$, one finds $P_e(r)=0.518P_{\rm gas}(r)$.

We
explore three 
possibilities: (i) Arnaud et al.'s profile that we have used in
Section~\ref{sec:nearby},  (ii)
theoretical profiles derived by assuming that the gas pressure is in hydrostatic
equilibrium with gravitational potential  given by an Navarro-Frenk-White
\citep[NFW;][]{{navarro/frenk/white:1997}} mass density profile
\citep{komatsu/seljak:2001}, and (iii) theoretical profiles from
hydrodynamical simulations of clusters of galaxies with and without gas
cooling and star formation \citep{nagai/kravtsov/vikhlinin:2007}.

The case (ii) is relevant because this profile
is used in the calculation of the SZ power spectrum
\citep{komatsu/seljak:2002} that has been used as a template to
marginalize over in the cosmological parameter estimation since the
3-year analysis
\citep{spergel/etal:2007,dunkley/etal:2009,larson/etal:prep}. 
Analytical models and hydrodynamical simulations for the SZ signal are
also the basis for planned efforts 
to use the SZ signal to constrain cosmological models.
 
As we have shown in the previous section using 29 nearby clusters, 
Arnaud et al.'s pressure profile overpredicts the SZ effect in
the \map\ data by $\sim 30$\%. An interesting question is whether this
trend extends to 
a larger number of clusters. 

\subsubsection{Komatsu-Seljak profile}
\begin{figure*}[t]
\centering \noindent
\includegraphics[width=15cm]{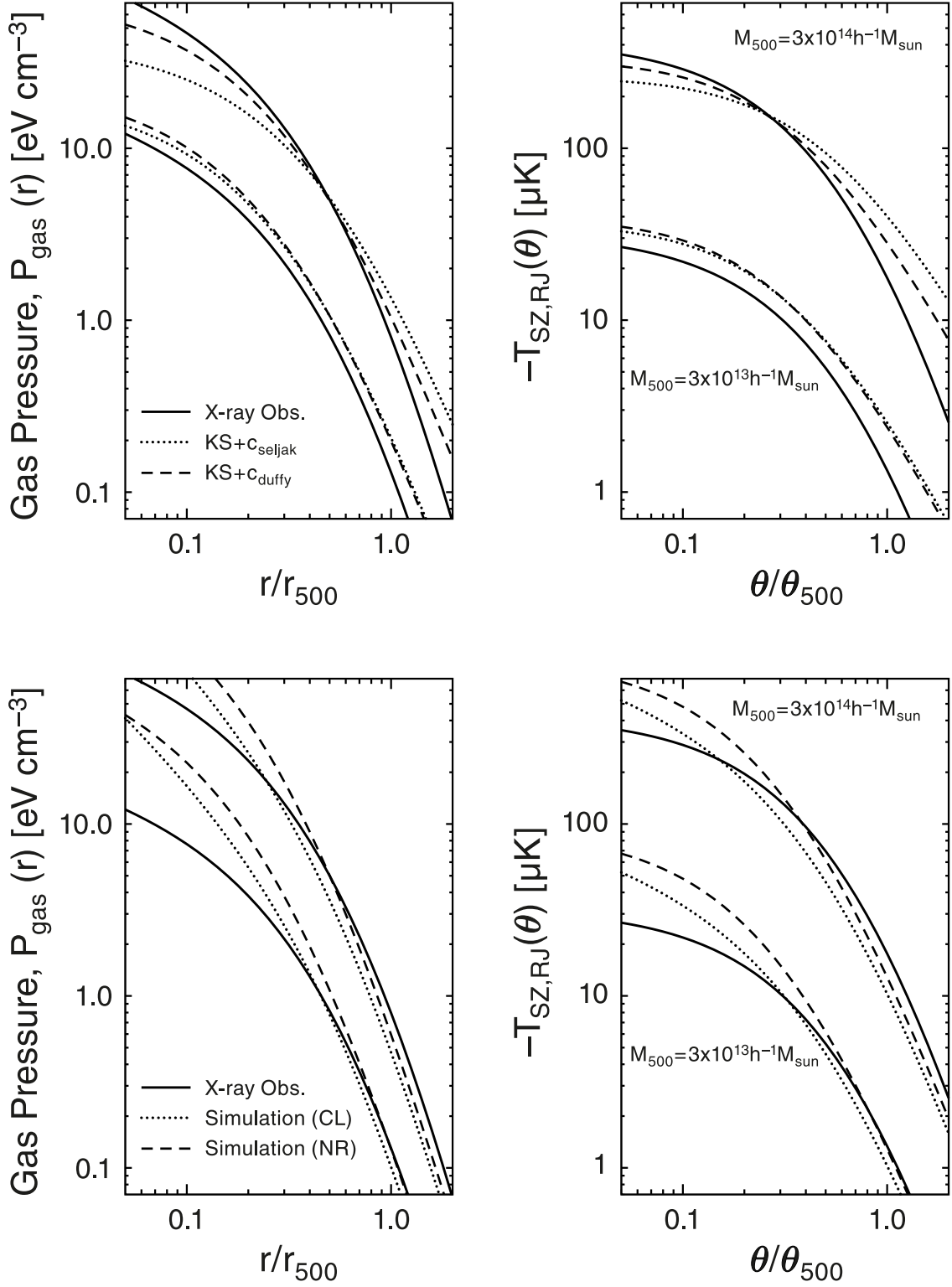}
\caption{%
 Gas pressure profiles of clusters of galaxies, $P_{\rm
 gas}(r)$, at $z=0.1$, and the projected profiles of the SZ effect,
 $\Delta T_{\rm SZ}(\theta)$  (Rayleigh-Jeans temperature in $\mu$K).
 (Top Left) The gas pressure profiles. 
 The upper and bottom set of curves show $M_{500}=3\times 10^{14}$ and
 $3\times 10^{13}~h^{-1}~M_\sun$, respectively. The horizontal axis
 shows radii scaled by the corresponding  
 $r_{500}=0.78$ and $0.36~h^{-1}~{\rm Mpc}$, respectively.
 The solid lines show $P_{\rm gas}(r)=P_e(r)/0.518$ derived from X-ray
 observations \citep{arnaud/etal:2010}, while the dotted and dashed
 lines show $P_{\rm gas}(r)$ predicted from hydrostatic equilibrium
 \citep{komatsu/seljak:2001} with NFW concentration parameters of
 \citet{seljak:2000} and \citet{duffy/etal:2008}, respectively. 
 (Top Right) The projected SZ profiles computed from the corresponding
 curves in the top left  panel and  equation~(\ref{eq:tszfit}). The
 horizontal axis shows angular radii  
 scaled by $\theta_{500}=r_{500}/D_A$,
 which is 10 and 4.7 arcmin for $M_{500}=3\times 10^{14}$ and $3\times
 10^{13}~h^{-1}~M_\sun$, respectively. 
 (Bottom Left) Same as the top left panel, but 
 the dotted and dashed
 lines show $P_{\rm gas}(r)$ predicted from ``Cooling+Star Formation''
 and ``Non-radiative'' simulation runs by
 \citet{nagai/kravtsov/vikhlinin:2007}.
 (Bottom Right) Same as the top right panel, but the dotted and dashed
 lines are computed from the corresponding
 curves in the bottom left  panel and  equation~(\ref{eq:tszfit}).
} 
\label{fig:profile}
\end{figure*}
\begin{figure}[t]
\centering \noindent
\includegraphics[width=8cm]{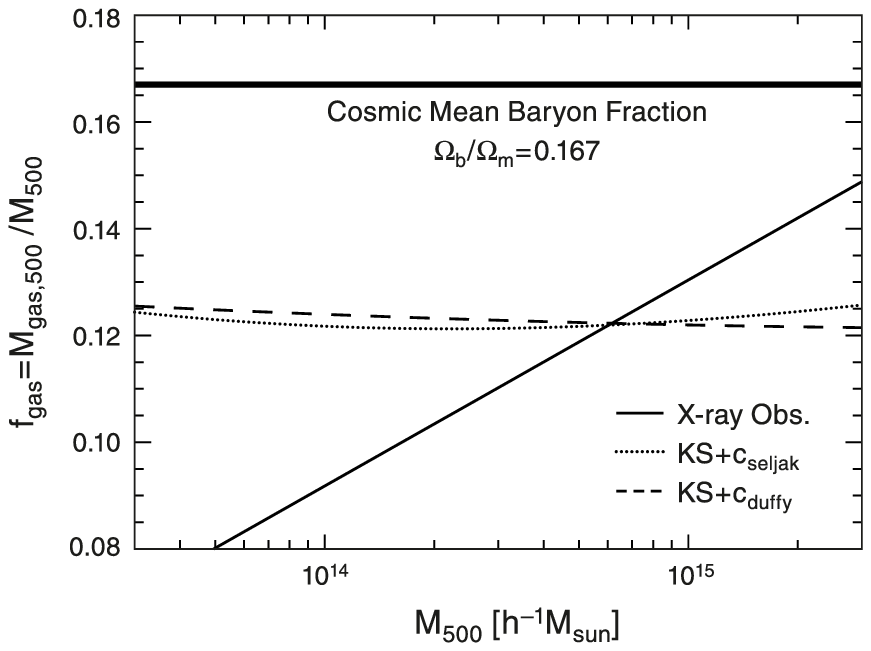}
\caption{%
 Gas mass fraction as a function of $M_{500}$. The thick horizontal line
 shows the 
 cosmic mean baryon fraction, $\Omega_b/\Omega_m=0.167$.
 The solid line shows the gas mass fraction, $f_{\rm gas}=M_{\rm
 gas,500}/M_{500}$, derived from X-ray observations
 \citep{vikhlinin/etal:2009b}, while the dotted and dashed  lines show
 $f_{\rm gas}$ predicted from hydrostatic equilibrium
 \citep{komatsu/seljak:2001} with NFW concentration parameters of
 \citet{seljak:2000} and \citet{duffy/etal:2008}, respectively. 
} 
\label{fig:fgas}
\end{figure}

The
normalization of the KS profile has been fixed by assuming that the gas
density at the virial radius is equal to the cosmic mean baryon
fraction, $\Omega_b/\Omega_m$, times the total mass density at the
virial radius. This is an upper limit: for example, star formation turns
gas into stars, reducing the amount of gas. 
KS also assumes that the gas is virialized and in thermal equilibrium
(i.e., electrons and protons share the same temperature) everywhere in a
cluster, that virialization converts potential energy of the cluster into
thermal energy only, and that the pressure contributed by bulk flows,
cosmic rays, and magnetic fields are unimportant. 

We give details of the gas pressure profiles in
Appendix~\ref{sec:pressure}. In the top left panel of
Figure~\ref{fig:profile}, we show Arnaud et al.'s pressure 
profiles (see Appendix~\ref{sec:arnaud}) in the
solid lines, and the KS profiles (see Appendix~\ref{sec:ks})
in the dotted and dashed lines. One of the inputs for the KS profile is
the so-called concentration parameter of the NFW profile. The dotted
line is for the concentration parameter of $c=10(M_{\rm
vir}/3.42\times 10^{12}~h^{-1}~M_\sun)^{-0.2}/(1+z)$
\citep{seljak:2000}, which was used by \citet{komatsu/seljak:2002} for
their calculation of the SZ power spectrum. Here, $M_{\rm vir}$ is the
virial mass, i.e., mass enclosed within the virial radius. The
dashed line is for 
$c=7.85(M_{\rm 
vir}/2\times 10^{12}~h^{-1}~M_\sun)^{-0.081}/(1+z)^{0.71}$, which was
found from recent N-body simulations with the \map\ 5-year cosmological
parameters \citep{duffy/etal:2008}.

We find that the KS profiles and Arnaud et al.'s profiles generally
agree. The agreement is quite good especially for the KS profile with
the concentration parameter of \citet{duffy/etal:2008}.
The KS profiles tend to overestimate the gas pressure relative to 
Arnaud et al.'s one for low-mass clusters ($M_\sun\lesssim
10^{14}~h^{-1}~M_\sun$). 
Can we explain this trend by a smaller gas mass fraction in clusters than
the cosmic mean? To answer this, we compute the gas mass fraction by
integrating the gas density profile:
\begin{equation}
 f_{\rm gas}\equiv \frac{M_{\rm gas,500}}{M_{500}}
= \frac{4\pi\int_0^{r_{500}} r^2dr~\rho_{\rm gas}(r)}{M_{500}},
\end{equation}
where $M_{500}$ and $M_{\rm gas,500}$ are the total mass and gas mass
contained within $r_{500}$, respectively. 

In Figure~\ref{fig:fgas}, we show $f_{\rm gas}$ from X-ray observations
\citep{vikhlinin/etal:2009b}:
\begin{equation}
 f_{\rm gas}(h/0.72)^{3/2}=0.125+0.037\log_{10}(M_{500}/10^{15}~h^{-1}~M_\sun),
\end{equation}
for $h=0.7$, and $f_{\rm gas}$ from the KS profiles with the
concentration parameters of \citet{seljak:2000} and
\citet{duffy/etal:2008}. We find that the KS predictions, $f_{\rm
gas}\simeq 0.12$, are always much smaller than the cosmic mean baryon fraction,
$\Omega_b/\Omega_m=0.167$, and are nearly independent of mass. A slight
dependence on mass is due to the dependence of the concentration
parameters on mass. While the
KS profile is normalized such that the gas density {\it at} the virial
radius is $\Omega_b/\Omega_m$ times the total mass density, the gas mass
{\it within} $r_{500}$ is much smaller than $\Omega_b/\Omega_m$ times
$M_{500}$, as the gas density and total matter density profiles are very
different near the center: while the gas density profile has a
constant-density core, the total matter density, which is dominated by
dark matter, increases as $\rho_m\propto 1/r$ near the center. 

However, the behavior of $f_{\rm gas}$  measured from X-ray observations
is very different. It has a much steeper dependence on mass than
predicted by KS. The reason for such a steep dependence on mass is not
yet understood. It could be due to star formation occurring more
effectively in lower mass clusters. 
In any case, for $M_{500} = 3\times 10^{14}~h^{-1}~M_\sun$, the observed
gas mass fraction is $f_{\rm gas}\simeq 0.11$, which is only 10\% smaller
than the KS value, $0.12$. For $M_{500} = 3\times
10^{13}~h^{-1}~M_\sun$, the observed
gas mass fraction, $0.08$, is about 30\% smaller than the KS value. This
is consistent with the difference between the KS and Arnaud et al.'s pressure
profiles that we see in Figure~\ref{fig:profile}; thus, 
once the observed mass dependence of  $f_{\rm gas}$ is taken into
account, these profiles agree well. 

To calibrate the amplitude of gas pressure, we shall
use the KS pressure profile (without any modification to $f_{\rm gas}$)
as a template, 
and find its normalization, 
$a$, from 
the \map\ data using the estimator given in equation~(\ref{eq:estimate_a}).
We shall present the results for hydrodynamical
simulations later.

For a given gas pressure profile, $P_{\rm gas}(r)$, we compute the
electron pressure as
$P_e=0.518P_{\rm gas}$ (see equation~(\ref{eq:PgasPe})).
We then use equation~(\ref{eq:tszfit}) to calculate the expected SZ
profile, $\Delta T_{\rm SZ}(\theta)$. We take the outer boundary of the
pressure to
be 3 times the virial radius, $r_{\rm out}=3r_{\rm vir}$, which is the
same as the parameter used by \citet{komatsu/seljak:2002}.
In the right panels of Figure~\ref{fig:profile}, we show the predicted
$\Delta T_{\rm SZ}(\theta)$, which will be used as templates, i.e.,
$t_{\nu p}$. 

\subsubsection{Luminosity-size relation}

\begin{figure}[t]
\centering \noindent
\includegraphics[width=8.5cm]{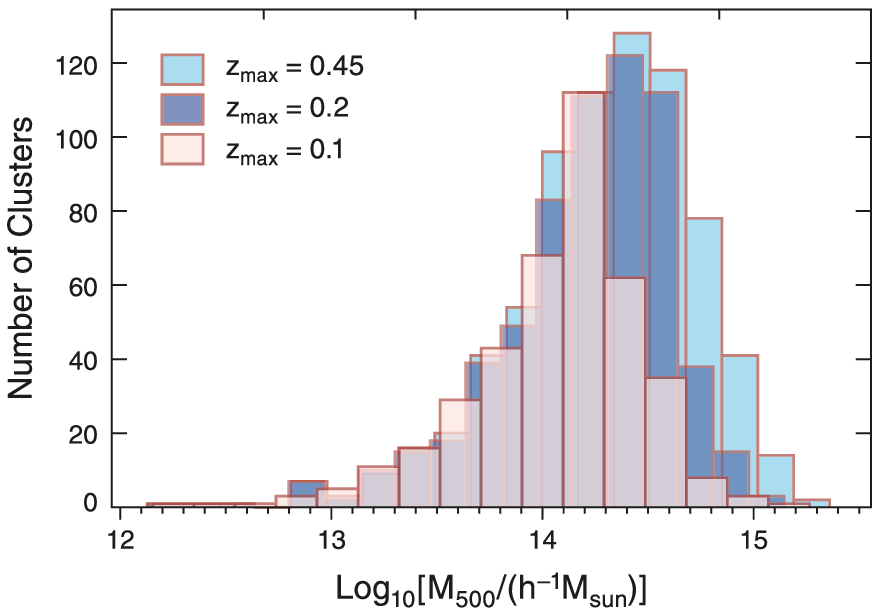}
\caption{%
 Distribution of $M_{500}$ estimated from clusters in the catalog using
 the measured X-ray luminosities in 0.1-2.4~keV band, $L_X$, and
 equations~(\ref{eq:r500}) and (\ref{eq:m500}). The light blue, dark
 blue, and pink histograms show $z_{\rm max}=0.45$, 0.2, and 0.1,
 respectively. 
} 
\label{fig:counts}
\end{figure}

Now, in order to compute the expected pressure profiles from each
cluster in the catalog, we need to know $r_{500}$. We calculate
$r_{500}$ from the observed X-ray luminosity in {\sl ROSAT}'s
0.1-2.4~keV band, $L_X$, as 
\begin{eqnarray}
\nonumber
r_{500} &=& \frac{(0.753\pm 0.063)~h^{-1}~{\rm Mpc}}{E(z)}\\
& &\times \left(\frac{L_{\rm
				       X}}{10^{44}~h^{-2}~{\rm
				       erg~s^{-1}}}\right)^{0.228\pm 0.015},
\label{eq:r500}
\end{eqnarray}
where $E(z)\equiv H(z)/H_0 =
\left[\Omega_m(1+z)^3+\Omega_\Lambda\right]^{1/2}$ for a $\Lambda$CDM
model. This is an empirical relation found from X-ray observations \citep[see
equation~(2) of][]{boehringer/etal:2007} based upon the
temperature-$L_X$ relation from \citet{ikebe/etal:2002} and the
$r_{500}$-temperature relation from
\citet{arnaud/pointecouteau/pratt:2005}. The error bars have been
calculated by propagating the errors in the
temperature-$L_X$ and $r_{500}$-temperature relations.
Admittedly,
there is a significant scatter around this relation, which is the most
dominant source of systematic error in this  type of analysis. 
(The results presented in Section~\ref{sec:nearby} do not suffer from
this systematic error, as they do not rely on $L_X$-$r_{500}$ relations.)
As $M_{500}\propto
r_{500}^3$, a $\approx
10$\% error in the predicted values of $r_{500}$ gives 
the mass calibration error of $\approx 30$\%. Moreover, the
SZ effect is given by $M_{500}$ times the gas temperature, the latter
being proportional to $M_{500}^{2/3}$ according to the virial
theorem. Therefore, the total calibration error can be as big as
$\approx 50$\%. 

In order to quantify this systematic error, we repeat our analysis for 3
different size-luminosity relations: (i) the central 
values, (ii) the normalization and slope shifted up by $1\sigma$ to
$0.816$ and $0.243$, and (iii) the normalization and slope shifted down
by $1\sigma$ to $0.690$ and $0.213$. 
We adopt this as an estimate for the systematic error in
our results due to the size-luminosity calibration error. For how this error
would affect our conclusions, see Section~\ref{sec:sz_syserror}. 

Note that this estimate of the systematic error is conservative, as we
allowed {\it all} clusters to deviate from the best-fit scaling
relation at once by $\pm 1\sigma$. In reality, the nature of this error is
random, and thus the actual error caused by the scatter in the scaling
relation would probably be smaller. \citet{melin/etal:prep} performed such an
analysis, and found that the systematic error is sub-dominant compared
to the statistical error.

Nevertheless, we shall adopt our conservative estimate of the
systematic error, as 
the mean scaling relation also varies from authors to authors. 
The mean scaling relations used by \citet{melin/etal:prep} are within
the error bar of the scaling relation that we use (equation~(\ref{eq:r500})).

In Figure~\ref{fig:counts}, we show the distribution of $M_{500}$
estimated from clusters in the catalog using the measured values of
$L_X$ and equations~(\ref{eq:r500}) and (\ref{eq:m500}). The
distribution peaks at $M_{500}\sim 3\times 10^{14}~h^{-1}~M_\sun$ for
$z_{\rm max}=0.2$ and 0.45, while it peaks at $M_{500}\sim 1.5\times
10^{14}~h^{-1}~M_\sun$ for $z_{\rm max}=0.1$. 

\subsubsection{Results: Arnaud et al.'s profile}
\label{sec:arnaudresults}
\begin{deluxetable*}{lccccc}
\tablecolumns{4}
\small
\tablewidth{0pt}
\tablecaption{%
Best-fitting Amplitude of Gas Pressure Profile\tablenotemark{a}
}
\tablehead{
\colhead{Gas Pressure Profile}
& \colhead{Type}
& \colhead{$z_{\rm max}=0.1$}
& \colhead{$z_{\rm max}=0.2$}
& \colhead{High $L_X$\tablenotemark{b}}
& \colhead{Low $L_X$\tablenotemark{c}}
}
\startdata
\citet{arnaud/etal:2010}
& X-ray Obs. (Fid.)\tablenotemark{d}
& $0.64\pm 0.09$
& $0.59\pm 0.07^{+0.38}_{-0.23}$
& $0.67\pm 0.09$
& $0.43\pm 0.12$
\nl
\citet{arnaud/etal:2010}
& REXCESS scaling\tablenotemark{e}
& N/A
& $0.78\pm 0.09$
& $0.90\pm 0.12$
& $0.55\pm 0.16$
\nl
\citet{arnaud/etal:2010}
& intrinsic scaling\tablenotemark{f}
& N/A
& $0.69\pm 0.08$
& $0.84\pm 0.11$
& $0.46\pm 0.13$
\nl
\citet{arnaud/etal:2010}
& $r_{\rm out}=2r_{500}$\tablenotemark{g}
& N/A
& $0.59\pm 0.07$
& $0.67\pm 0.09$
& $0.43\pm 0.12$
\nl
\citet{arnaud/etal:2010}
& $r_{\rm out}=r_{500}$\tablenotemark{h}
& N/A
& $0.65\pm 0.08$
& $0.74\pm 0.09$
& $0.44\pm 0.14$
\nl
\citet{komatsu/seljak:2001}
& equation~(\ref{eq:c_seljak})
& $0.59\pm 0.09$
& $0.46\pm 0.06^{+0.31}_{-0.18}$
& $0.49\pm 0.08$
& $0.40\pm 0.11$
\nl
\citet{komatsu/seljak:2001}
& equation~(\ref{eq:c_duffy})
& $0.67\pm 0.09$
& $0.58\pm 0.07^{+0.33}_{-0.20}$
& $0.66\pm 0.09$
& $0.43\pm 0.12$
\nl
\citet{nagai/kravtsov/vikhlinin:2007}
& Non-radiative
& N/A
& $0.50\pm 0.06^{+0.28}_{-0.18}$
& $0.60\pm 0.08$
& $0.33\pm 0.10$
\nl
\citet{nagai/kravtsov/vikhlinin:2007}
& Cooling+SF
& N/A
& $0.67\pm 0.08^{+0.37}_{-0.23}$
& $0.79\pm 0.10$
& $0.45\pm 0.14$
\enddata
\tablenotetext{a}{
 The quoted error bars show 68\%~CL. The first error is statistical,
 while the second error is systematic. The systematic error is caused by 
 the calibration error in the
 size-luminosity relation ($r_{500}$-$L_X$ relation; see
 equation~(\ref{eq:r500}) and discussion below it).
  While we quote the systematic error
 in the amplitudes only for
 $z_{\rm max}=0.2$, the amplitudes for $z_{\rm max}=0.1$ also have
 similar levels of the systematic error.
 Due to a potential contamination from unresolved radio
 sources, 
 the best-fitting amplitudes could also be underestimated by $\approx 5$ to
 10\%. This is not included in the systematic error budget because it is
 sub-dominant. See Section~\ref{sec:sz_syserror} for discussion on the
 point source contamination.}
\tablenotetext{b}{
 ``High $L_X$'' uses clusters with $4.5<L_X/(10^{44}~{\rm
 erg~s^{-1}})<45$ and $z\le 0.2$. Before masking, there are 82 clusters.
 The quoted errors are statistical.
}
\tablenotetext{c}{
 ``Low $L_X$'' uses clusters with $0.45<L_X/(10^{44}~{\rm
 erg~s^{-1}})<4.5$  and $z\le 0.2$. Before masking, there are 417
 clusters. Clusters less luminous than these (129 clusters are fainter
 than $0.45\times 10^{44}~{\rm
 erg~s^{-1}}$) do not yield 
 a statistically significant detection. 
 The quoted errors are statistical.
}
\tablenotetext{d}{
With the fiducial scaling relation between $r_{500}$ and $L_X$,
$r_{500}=\frac{0.753~h^{-1}~{\rm
Mpc}}{E(z)}[L_X/(10^{44}~h^{-2}~{\rm erg~s^{-1}})]^{0.228}$
\citep{boehringer/etal:2007}.
For this scaling relation,
 $L_X=4.5\times 10^{44}~{\rm erg~s^{-1}}$ corresponds to 
$M_{500}=4.1$ and $3.9\times 10^{14}~h^{-1}~M_\sun$ for $z=0.1$ and 0.2,
and 
 $L_X=0.45\times 10^{44}~{\rm erg~s^{-1}}$ corresponds to 
$M_{500}=0.84$ and $0.80\times 10^{14}~h^{-1}~M_\sun$ for $z=0.1$ and 0.2,
 respectively. 
}
\tablenotetext{e}{
With the ``REXCESS'' scaling relation,
$r_{500}=\frac{0.717~h^{-1}~{\rm
Mpc}}{E^{1.19}(z)}[L_X/(10^{44}~h^{-2}~{\rm erg~s^{-1}})]^{0.222}$,
used by \citet{melin/etal:prep}.
For this scaling relation,
 $L_X=4.5\times 10^{44}~{\rm erg~s^{-1}}$ corresponds to 
$M_{500}=3.4$ and $3.1\times 10^{14}~h^{-1}~M_\sun$ for $z=0.1$ and 0.2,
and 
$L_X=0.45\times 10^{44}~{\rm erg~s^{-1}}$ corresponds to 
$M_{500}=0.73$ and $0.68\times 10^{14}~h^{-1}~M_\sun$ for $z=0.1$ and 0.2,
 respectively.
The quoted errors are statistical.
}
\tablenotetext{f}{
With the ``intrinsic'' scaling relation, 
 $r_{500}=\frac{0.745~h^{-1}~{\rm
Mpc}}{E^{1.15}(z)}[L_X/(10^{44}~h^{-2}~{\rm erg~s^{-1}})]^{0.207}$,
used by \citet{melin/etal:prep}.
For this scaling relation,
 $L_X=4.5\times 10^{44}~{\rm erg~s^{-1}}$ corresponds to 
$M_{500}=3.7$ and $3.4\times 10^{14}~h^{-1}~M_\sun$ for $z=0.1$ and 0.2,
and 
$L_X=0.45\times 10^{44}~{\rm erg~s^{-1}}$ corresponds to 
$M_{500}=0.88$ and $0.82\times 10^{14}~h^{-1}~M_\sun$ for $z=0.1$ and 0.2,
 respectively.
The quoted errors are statistical.
}
\tablenotetext{g}{
The gas extension is truncated at $r_{\rm out}=2r_{500}$, instead of
 $6r_{500}$. The fiducial $r_{500}$-$L_X$ relation is used. The quoted
 errors are statistical. 
}
\tablenotetext{h}{
The gas extension is truncated at $r_{\rm out}=r_{500}$, instead of
 $6r_{500}$. The fiducial $r_{500}$-$L_X$ relation is used.
The quoted errors are statistical.
}
\label{tab:szamplitude}
\end{deluxetable*}

For Arnaud et al.'s pressure profile, we find the best-fitting
amplitudes of $a=0.64\pm 0.09$ and $0.59\pm 0.07$ (68\%~CL) for $z_{\rm
max}=0.1$ and 0.2, respectively. The former result is fully consistent
with what we find from the nearby clusters in
Section~\ref{sec:nearby}: $a=0.63\pm 0.09$ (68\%~CL; for $1\times
10^{14}~h^{-1}~M_\sun\le M_{500}<9\times 10^{14}~h^{-1}~M_\sun$ and
$z\le 0.09$).

The significance level of statistical detection of the SZ effect is about
8$\sigma$ for $z_{\rm max}=0.2$.
With the systematic error included, we
find $a=0.59\pm 0.07^{+0.38}_{-0.23}$ for $z_{\rm max}=0.2$; however,
the above agreement may suggest that the fiducial scaling relation
(equation~(\ref{eq:r500})) is, in fact, a good one. 

As we have shown in Section~\ref{sec:nearby}, the measured SZ effects and
the predictions from the X-ray data agree on a cluster-by-cluster
basis. A plausible explanation for the discrepancy between the \map\ data and
Arnaud et al.'s profile is that Arnaud et al.'s profile does not distinguish
between cooling-flow and non-cooling-flow clusters.

Nevertheless, this result, which shows that the SZ effect seen in the
\map\ data is  
{\it less} than the average ``expectation'' from X-ray observations, agrees
qualitatively with 
some of the previous work 
\citep{lieu/mittaz/zhang:2006,bielby/shanks:2007,diego/partridge:2009}.
The other work showed that 
the SZ effect seen in the \map\ data is 
consistent with expectations from X-ray observations
 \citep{afshordi/etal:2007,melin/etal:prep}.
	
These authors used widely different methods and cluster catalogues. 
\citet{lieu/mittaz/zhang:2006} were the first to claim that 
the SZ effect seen in the \map\ data is significantly less than
expected from X-ray data,  by
using 31 clusters compiled by
\citet{bonamente/etal:2002}.
\citet{bielby/shanks:2007} extended the analysis of
\citet{lieu/mittaz/zhang:2006} by using 38 clusters compiled by
\citet{bonamente/etal:2006}, for which the observational data of the SZ
effect from OVRO and Berkeley Illinois Maryland Association (BIMA) are
available. They did not use scaling relations, but used a spherical
isothermal $\beta$ model to fit the X-ray 
surface brightness profile of each cluster in the catalog, and calculated the
expected SZ signals, assuming that the intracluster gas is isothermal. 
\citet{lieu/mittaz/zhang:2006} found that the measured signal is smaller
than expected from X-ray 
data by a factor of 3--4, and \citet{bielby/shanks:2007} found a
similar result for the cluster catalog of \citet{bonamente/etal:2006}.

\citet{diego/partridge:2009} used the same cluster catalog that we use
(REFLEX+eBCS), but used a different scaling relation: they related the
cluster core radius to the X-ray luminosity (we relate $r_{500}$ to the
X-ray luminosity). They found a large discrepancy \citep[similar
to][]{lieu/mittaz/zhang:2006,bielby/shanks:2007} when a spherical
isothermal $\beta$ model was used to predict the SZ signal, while they
found a smaller discrepancy (similar to our results) when more
realistic gas models were used. 
\citet{afshordi/etal:2007} used 193 clusters selected from the XBAC
catalog. Their catalog consisted of the clusters that have measured X-ray
temperatures ($>3$~keV). They then used a scaling relation
between $r_{200}$ and the X-ray 
temperature. They found that the measured SZ signal and X-ray data are
consistent. 

\citet{melin/etal:prep} used 
the 5-year \map\ data and a bigger sample of 893 clusters and a scaling
relation between $r_{500}$ and the X-ray luminosity taken from 
\citet{pratt/etal:2009,arnaud/etal:2010}.
They compared the measured integrated pressure from the
\map\ data to the expectation from Arnaud et al.'s profile, and
concluded that they agree very well. 
(The normalization is consistent with unity within the statistical
uncertainty.) We find, on the other hand, that the normalization is
significantly less than unity compared to the statistical uncertainty.
How can we reconcile these results?

One possibility would be the difference in the scaling
relations. The scaling relation shifted down by $1\sigma$ would make the
predicted SZ signals smaller, which would then increase the best-fitting
amplitude. Given the size of the systematic error, $a=0.59\pm
0.07^{+0.38}_{-0.23}$, $a\approx 1$ may not be inconsistent with the
data.
Specifically, they used two scaling relations:
\begin{itemize}
 \item[1.] $r_{500}=\frac{0.717~h^{-1}~{\rm
Mpc}}{E^{1.19}(z)}[L_{500}/(10^{44}~h^{-2}~{\rm erg~s^{-1}})]^{0.222}$,
 \item[2.] $r_{500}=\frac{0.745~h^{-1}~{\rm
Mpc}}{E^{1.15}(z)}[L_{500}/(10^{44}~h^{-2}~{\rm erg~s^{-1}})]^{0.207}$,
\end{itemize}
where the relations 1 and 2
correspond to the ``REXCESS'' and ``intrinsic'' relations in
\citet{melin/etal:prep}, respectively.
Here, $L_{500}$ is the X-ray luminosity measured within $r_{500}$, which
is calculated from $L_X$. 
While we do not have the conversion factors
they used, a typical magnitude of the conversion factors is about 10\%,
according to \citet{melin/etal:prep}. A 10\% change in $L_X$ gives a 2\%
change in $r_{500}$, which is negligible compared to the other
uncertainties; thus, we shall assume that $L_X$ and $L_{500}$ are the
same, and repeat our 
analysis using these scaling relations. 
We find the 
amplitudes of $a=0.78\pm 0.09$ and $0.69\pm 0.08$ ($z_{\rm max}=0.2$; 68\%~CL) 
for the relations 1 and 2, respectively; thus, 
while these scaling relations give larger amplitudes, 
they cannot completely explain the difference between 
the results of \citet{melin/etal:prep}
($a\simeq 1$) and our results.
However, we find that the discrepancy is much less for high X-ray
luminosity clusters. 
See Section~\ref{sec:lum_bin}.

While the method of \citet{melin/etal:prep} and our method are similar,
they are different in details.
We compare the predicted angular radial profiles of the SZ effect to the
\map\ data to find the best-fitting amplitude. 
\citet{melin/etal:prep} measured the {\it integrated} pressure within 5 times
$r_{500}$, and  converted it to the integrated
pressure within $r_{500}$, $Y_{r500}$, assuming the
distribution of pressure beyond $r_{500}$ is described by the profile of
\citet{arnaud/etal:2010}. Whether the difference in methodology can
account for the difference between our results and their results is
unclear, and requires further investigation.\footnote{There are also differences in the estimators used.
In \citet{melin/etal:prep}, a ``matched-filter estimator'' proposed by
\citet{herranz/etal:2002} was used for estimating the normalization of
the Arnaud et al. profile. 
 Their estimator is essentially the same as the optimal
estimator we derive in Appendix~\ref{app:sz}, with some differences in details
of the implementation.
Their estimator is given by, in our notation,
\begin{equation}
 \hat{p} = \frac1F\int d^2{\mathbf l}~ \tilde{t}_{\nu'{\mathbf
  l}}(P^{-1})_{\nu'{\mathbf l},\nu {\mathbf l}}\tilde{d}_{\nu
  {\mathbf l}}, 
\end{equation}
where 
\begin{equation}
 F\equiv \int d^2{\mathbf l}~ \tilde{t}_{\nu'{\mathbf l}}
(P^{-1})_{\nu'{\mathbf l},\nu{\mathbf l}}\tilde{t}_{\nu{\mathbf l}}.
\end{equation}
Here, $\tilde{t}$ and $\tilde{d}$ are the 2d Fourier transforms of a
template map, $t$, and the data map, $d$, respectively, and 
$P_{\nu'{\mathbf l},\nu\mathbf{l}}$ is the power spectrum of the CMB
signal plus instrumental noise, both of which are assumed to be diagonal
in Fourier space. 
The summation over the repeated indices is understood. 
For comparison, our estimator for the same quantity is given by
\begin{equation}
 \hat{p} = 
\frac1Ft_{\nu' p'}(C^{-1}_{\rm tot})_{\nu'p',\nu p}
d_{\nu p},
\end{equation}
where  $C_{\rm tot}=N_{\rm pix}+AS_{\rm harm}A^{T}$ is the pixel-space
covariance matrix of the CMB 
signal plus instrumental noise (see equation~\ref{eq:hp_def}), and
\begin{equation}
 F = t_{\nu' p'}(C^{-1}_{\rm tot})_{\nu'p',\nu p}t_{\nu p}.
\end{equation}
There are two differences in the implementation: 
\begin{itemize}
 \item [1.] \citet{melin/etal:prep} re-project the \map\ data onto 504
       square ($10^\circ\times 10^\circ$) tangential overlapping flat
       patches, and calculate the above 2d Fourier transform on each
       flat patch. We perform the analysis on the full sky by
       calculating the covariance matrix with the spherical harmonics.
 \item [2.] \citet{melin/etal:prep} calculate $P$ from the data. We
       calculate $C$ from the best-fitting $\Lambda$CDM model for the
       CMB signal and the noise model.
\end{itemize}}

In any case, we emphasize once again that the SZ effect measured by the
\map\ and the predictions from X-ray data agree well, {\it when the actual
X-ray profile of individual clusters, rather than the average (or median)
profile, is used}, and there is a reason why Arnaud et al.'s profile
would overpredict the pressure (i.e., cooling flows; see
Section~\ref{sec:nearby}). Therefore, it 
is likely that the
difference between our results and \citet{melin/etal:prep} simply
points to the fundamental limitation of the analysis using many clusters
(with little or no X-ray data) and scaling relations. 

\subsubsection{Results: KS profile and hydrodynamical simulation}
\label{sec:KSandSim}
Let us turn our attention to the analytical KS profile.
For the KS profile with the concentration parameter of
\cite{seljak:2000}, we find the best-fitting amplitudes of 
$a=0.59\pm 0.09$ and $0.46\pm 0.06^{+0.31}_{-0.18}$~(68\%~CL) for $z_{\rm max}=0.1$ and 0.2,
respectively.
For the KS profile with the concentration parameter of
\cite{duffy/etal:2008}, we find 
$a=0.67\pm 0.09$ and $0.58\pm 0.07^{+0.33}_{-0.20}$~(68\%~CL) for $z_{\rm max}=0.1$ and 0.2,
respectively.
These results are consistent with those for Arnaud et al.'s pressure
profiles. 

Recently, the SPT collaboration detected the SZ
power spectrum at $l\gtrsim 3000$. By fitting their SZ power spectrum
data to the theoretical model of \citet{komatsu/seljak:2002}, they found
the best-fitting 
amplitude of $A_{\rm SZ}=0.37\pm 0.17$
\citep[68\%~CL;][]{lueker/etal:2010}. 
The calculation of
\citet{komatsu/seljak:2002} is based on the KS gas pressure profile.
As the amplitude of SZ power spectrum is proportional to the gas
pressure squared, i.e., $A_{\rm SZ}\propto a^2$, our result for the KS
profiles, $a\approx 0.5-0.7$, is
consistent with $A_{\rm SZ}= 0.37\pm 0.17$ found from SPT.
The ACT collaboration placed an upper limit of $A_{\rm SZ}<1.63$ 
\citep[95\%~CL;][]{fowler/etal:2010}, which is consistent with the SPT result.

What do hydrodynamical simulations tell us?
As the analytical calculations such as \citet{komatsu/seljak:2001} are
limited, we also fit the pressure profiles derived from hydrodynamical
simulations of \citet{nagai/kravtsov/vikhlinin:2007} to the \map\ data.
In the bottom panels of Figure~\ref{fig:profile}, we show the gas
pressure profiles from ``Non-radiative'' and ``Cooling+Star Formation
(SF)'' runs. 

By fitting the  SZ templates constructed from these simulated profiles
to the \map\ data, we find the best-fitting amplitudes of 
$0.50\pm 0.06^{+0.28}_{-0.18}$ and $0.67\pm
0.08^{+0.37}_{-0.23}$~(68\%~CL) for Non-radiative and 
Cooling+SF runs, respectively, which are consistent with the amplitudes
found for the KS profiles and Arnaud et al.'s profiles. See
Table~\ref{tab:szamplitude} for a summary 
of the best-fitting amplitudes.

That the KS, simulation, and Arnaud et al.'s profiles yield similar
results indicates that {\it all} of these profiles overpredict the
amount of SZ effect seen in the \map\ data by $\sim 30-50$\%. 
This conclusion is made robust by the results we presented in
Section~\ref{sec:nearby}: the analysis that does {\it not} use scaling
relations between $L_X$ and $r_{500}$, but uses only a subset of clusters
that have the detailed follow-up observations by {\it Chandra}, yields
the same result. This is one of the main results of our SZ analysis.

\subsection{Luminosity bin analysis}
\label{sec:lum_bin}
To see the dependence of the best-fitting normalization on X-ray
luminosities (hence $M_{500}$), we divide the cluster samples
into 3 luminosity bins: (i) ``High $L_X$'' with $4.5<L_X/(10^{44}~{\rm
erg~s^{-1}})\le 45$, (ii) ``Low $L_X$'' with $0.45<L_X/(10^{44}~{\rm
erg~s^{-1}})\le 4.5$, and (iii) clusters fainter than (ii). 
There are 82, 417, and 129 clusters in (i), (ii), and (iii),
respectively. In Table~\ref{tab:szamplitude}, we show that we detect 
significant SZ signals in (i) and (ii), despite the smaller number of
clusters used in each luminosity bin.  We do not have a statistically
significant detection in (iii). 

The high $L_X$ clusters have $M\gtrsim 4\times 10^{14}~h^{-1}~M_\sun$.
For these clusters, the agreement between the \map\ data and 
the expected SZ signals is much better. In particular, 
for the REXCESS scaling relation, we find $a=0.90\pm 0.12$, which is
consistent with unity within the $1\sigma$ statistical error. 
This implies that, at least for high X-ray luminosity clusters, our
results and the results of \citet{melin/etal:prep} agree within the
statistical uncertainty.

On the other hand, we find that less luminous clusters tend to have
significantly lower best-fitting 
amplitudes for all models of gas-pressure profiles and scaling relations
that we have explored.
This trend is consistent with, for example, 
the gas mass fraction being lower for lower mass clusters. It is also
consistent with radio point sources filling some of the SZ effect seen
in the \map\ data. For the point source contamination, see
Section~\ref{sec:sz_syserror}. 

\subsection{Systematic Errors}
\label{sec:sz_syserror}

The best-fitting amplitudes may be shifted up and down by $\approx 50$\%
due to the calibration
error in the size-luminosity relation (equation~(\ref{eq:r500})). 
As we have shown already, 
the best-fitting amplitudes for the KS
profiles can be shifted up to $0.77$ and $0.91$ for the
concentration parameters of \citet{seljak:2000} and
\citet{duffy/etal:2008}, respectively. Similarly, the amplitude for 
Arnaud et al.'s profile can be shifted up to $0.97$.
As this calibration error shifts all amplitudes given in
Table~\ref{tab:szamplitude} by the same amount, it
does not affect our conclusion that all of the gas
pressure profiles considered above yield similar results.

This type of systematic error can be reduced by using a
subset of 
clusters of galaxies for which the scaling relations are more tightly
constrained \citep[see,
e.g.,][]{pratt/etal:2009,vikhlinin/etal:2009b,mantz/etal:2010c};
however, reducing the number of samples increases the statistical error.  
Indeed, the analysis presented in Section~\ref{sec:nearby} does not
suffer from the ambiguity in the scaling relations.

How important are radio point sources? 
While we have not attempted to correct for potential contamination
from unresolved radio point sources, we estimate the magnitude of
effects here. If, on 
average, each cluster has a $F_{\rm src}=10$~mJy source, then the 
corresponding temperatures,
\begin{equation}
 \Delta T_{\rm src}
= 40.34~\mu{\rm K}
\left[\frac{\sinh^2(x/2)}{x^4}\frac{F_{\rm src}}{10~{\rm
mJy}}\frac{10^{-5}~{\rm sr}}{\Omega_{\rm beam}}\right],
\end{equation}
 are 2.24, 2.29, and 2.19~${\mu}$K in Q, V, and W bands, respectively. 
Here, $x=\nu/(56.78~{\rm GHz})$, and $\Omega_{\rm beam}=9.0\times
10^{-5}$, $4.2\times 10^{-5}$, and $2.1\times 10^{-5}$~sr are the solid
angles of beams in Q, V, and W bands, respectively
\citep{jarosik/etal:prep}.
Using the radio sources observed in clusters of galaxies 
by \citet{lin/etal:2009}, \citet{diego/partridge:2009} estimated that the
mean flux of sources in Q band is 10.4~mJy, and that at 90~GHz (which is
close to 94~GHz of W band) is $\approx 4$ to 6~mJy. 
Using these estimates, we expect the source contamination at the level
of $\approx 1$ to 2~$\mu$K in V and W bands, which is $\approx 5$ to
10\% of the measured SZ temperature. Therefore, the best-fitting amplitudes
reported in Table~\ref{tab:szamplitude} could be underestimated by
$\approx 5$ to 10\%.

\subsection{Discussion}
\label{sec:sz_discussion}
The gas pressure profile is not the only factor that determines the SZ
power spectrum. The other important factor is the mass function, $dn/dM$:
\begin{equation}
 C_l\propto \int dz\frac{dV}{dz}\int dM\frac{dn}{dM}
|\tilde{P}^{\rm 2d}_l|^2,
\end{equation}
where $V(z)$ is the comoving volume of the universe and $\tilde{P}^{\rm
2d}_l$ is the 2d Fourier transform of $P^{\rm 2d}(\theta)$. 
Therefore, a lower-than-expected $A_{\rm SZ}$ may imply either 
a lower-than-expected amplitude of matter density fluctuations, i.e.,
$\sigma_8$, or a lower-than-expected gas pressure, or both. 

As the predictions for the SZ power spectrum available today 
\citep[see, e.g.,][and references therein]{shaw/etal:2009,sehgal/etal:2010}
are similar to the prediction of \citet{komatsu/seljak:2002}
(for example, \citet{lueker/etal:2010} found $A_{\rm SZ}=0.55\pm 0.21$
for the prediction of \citet{sehgal/etal:2010}, which is based on the gas
model of \citet{bode/ostriker/vikhlinin:2009}), a plausible explanation
for a lower-than-expected $A_{\rm SZ}$ is a lower-than-expected gas
pressure. 

\citet{arnaud/pointecouteau/pratt:2007} find that the X-ray observed
integrated pressure enclosed within 
$r_{500}$, $Y_{X}\equiv M_{{\rm gas},500}T_X$, for a given $M_{500}$ is
about a factor of 0.75 times the prediction from the 
Cooling+SF simulation of \citet{nagai/kravtsov/vikhlinin:2007}. 
This is in good agreement with our corresponding result for the ``High $L_X$''
samples, $0.79\pm 0.10$ (68\%~CL; statistical error only). 

While the KS profile is generally in good agreement with Arnaud et
al.'s profile, the former is more extended than the latter (see
Figure~\ref{fig:profile}), which makes the KS prediction for the 
projected SZ profiles bigger. Note, however, that the outer slope
of the fitting formula given by 
\citet{arnaud/etal:2010} (equation~(\ref{eq:gnfw})) has been forced to
match that from 
hydrodynamical simulations of \citet{nagai/kravtsov/vikhlinin:2007} in
$r\ge r_{500}$.
See the bottom panels of Figure~\ref{fig:profile}. The
steepness of the profile at $r\gtrsim r_{500}$ from the simulation may
be attributed to a significant non-thermal pressure support from $\rho
v^2$, which makes it possible to balance gravity by less thermal
pressure at larger radii. In other words, the {\it total} pressure
(i.e., thermal plus $\rho v^2$) profile would probably be closer to the
KS prediction, but the thermal pressure would decline more rapidly than
the total pressure would. 

If the SZ effect seen in the \map\ data is less than theoretically
expected, what 
would be the implications?  
One possibility is that protons and electrons do not share the same
temperature. The electron-proton equilibration time is longer than the
Hubble time at the virial radius, so that the electron temperature may
be lower than the proton temperature in the outer regions of clusters
which contribute a significant fraction of the predicted SZ flux
\citep{rudd/nagai:2009,wong/sarazin:2009}. 
The other sources of non-thermal pressure support in outskirts of
the cluster (turbulence, magnetic
field, and cosmic rays) would reduce the thermal SZ effect relative
to the expectation, if these effects are not taken into account in
modeling the intracluster medium. 
Heat conduction may also play some role in suppressing the gas
pressure \citep{loeb:2002,loeb:2007}.

In order to explore the impact of gas pressure at $r>r_{500}$, we cut
the pressure 
profile at 
$r_{\rm out}=r_{500}$ (instead of $6r_{500}$) and
repeat the analysis. We find $a=0.74\pm 0.09$ and $0.44\pm 0.14$ for
high and low $L_X$ 
clusters, respectively. (We found 
$a=0.67\pm 0.09$ and $0.43\pm 0.12$ for $r_{\rm out}=6r_{500}$. See
Table~\ref{tab:szamplitude}.) 
These results are somewhat puzzling - the X-ray observations directly
measure gas out to $r_{500}$, and thus we would expect to find $a\approx
1$ at least out to $r_{500}$. This result may suggest that, as we have
shown in Section~\ref{sec:nearby}, the problem
is not with the outskirts of the cluster, but with the inner parts
where the cooling flow has the largest effect.

The {\it relative}
amplitudes between high and low $L_X$ clusters suggest that a
significant amount of pressure is missing in low mass ($M_{500}\lesssim
4\times 10^{14}~h^{-1}~M_\sun$) clusters, even if we scale all the
results such that high-mass clusters are forced to have $a=1$. 
A similar trend is also seen in Figure~3 of \citet{melin/etal:prep}.
This interpretation is consistent with the SZ power spectrum being lower
than theoretically expected. The SPT measures the SZ power spectrum at $l\gtrsim
3000$. At such high multipoles, the contributions to the SZ power
spectrum are dominated by relatively low-mass clusters, 
$M_{500}\lesssim
4\times 10^{14}~h^{-1}~M_\sun$ \citep[see Figure~6
of][]{komatsu/seljak:2002}. Therefore, a plausible explanation for the
lower-than-expected SZ power spectrum is a missing pressure (relative to
theory) in lower mass clusters.  

Scaling relations, gas pressure, and 
entropy of low-mass clusters and 
groups have been studied in the literature.\footnote{A systematic study of the thermodynamic properties of low-mass clusters
and groups is given in \citet{finoguenov/etal:2007} \citep[also
see][]{finoguenov/etal:2005,finoguenov/boehringer/zhang:2005}.}
\citet{leauthaud/etal:2010} obtained a relation between
$L_X$ of 206 X-ray-selected galaxy groups and the mass ($M_{200}$)
derived from the 
stacking analysis of weak lensing measurements. 
Converting their best-fitting relation to $r_{200}$--$L_X$ relation, we
find $r_{200}=\frac{1.26~h^{-1}~{\rm Mpc}}{E^{0.89}(z)}
[L_X/(10^{44}~h^{-2}~{\rm erg~s^{-1}})]^{0.22}$.
(Note that the pivot luminosity of the original scaling relation is
$2.6\times 10^{42}~h^{-2}~{\rm erg~s^{-1}}$.)
As $r_{500}\approx
0.65r_{200}$, their relation is $\approx 1\sigma$ higher than the
fiducial scaling relation that we adopted (equation~(\ref{eq:r500})).
Had we used their scaling relation, we would find even lower
normalizations. 

The next generation of simulations or analytical calculations of the SZ
effect should be focused more on understanding the gas pressure
profiles, both the amplitude and the shape, especially in
low-mass clusters.  New measurements of the SZ
effect toward many individual clusters with unprecedented sensitivity
are now becoming available
\citep{staniszewski/etal:2009,hincks/etal:prep,plagge/etal:2010}. These
new measurements would be important for understanding the gas pressure
in low-mass clusters. 

\section{Conclusion}
\label{sec:conclusion}
\begin{figure}[t]
\centering \noindent
\includegraphics[width=8cm]{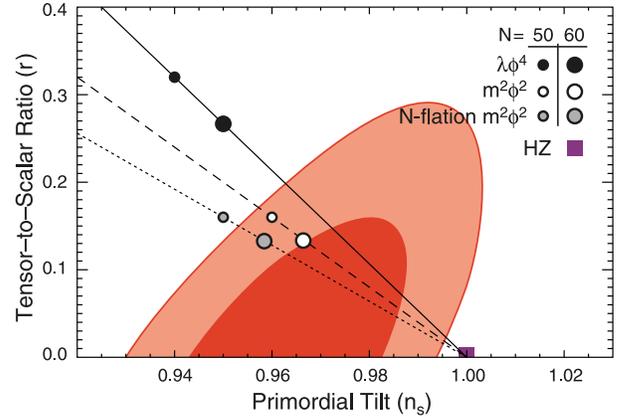}
\caption{%
 Two-dimensional joint marginalized constraint (68\% and
 95\% CL) on the primordial tilt, $n_s$, and the tensor-to-scalar ratio,
 $r$, derived from the data combination of \map+BAO+$H_0$.
 The symbols show the  predictions from ``chaotic'' inflation
 models whose potential is given by 
 $V(\phi)\propto \phi^\alpha$ \citep{linde:1983}, with 
 $\alpha=4$ (solid) and 
 $\alpha=2$ (dashed) for single-field models, and $\alpha=2$ for
 multi-axion field models with $\beta=1/2$
 \citep[dotted;][]{easther/mcallister:2006}.
} 
\label{fig:nsr}
\end{figure}

With the \map\ 7-year temperature and
polarization data, new measurements of $H_0$ \citep{riess/etal:2009}, and
improved large-scale structure data \citep{percival/etal:2009}, we have been able to rigorously test the standard cosmological model.
The model continues to be an exquisite fit to the existing data.  Depending on the parameters, we also use the
other data sets such as the
small-scale CMB temperature power spectra
\citep[for the primordial helium
abundance]{brown/etal:2009,reichardt/etal:2009}, the 
power spectrum of LRGs derived from {\sl SDSS} \citep[for neutrino
properties]{reid/etal:2010},
the Type Ia supernova data \citep[for dark energy]{hicken/etal:2009},
and the time-delay
distance to  the lens system B1608+656 \citep[for dark energy and
spatial curvature]{suyu/etal:2010}. The combined
data sets enable improved constraints  over the \map-only constraints on the cosmological parameters
presented in \citet{larson/etal:prep}  on
physically-motivated extensions of the standard model.

We summarize the most
significant findings from our analysis (also see
Table~\ref{tab:deviation}, \ref{tab:yhe}, and \ref{tab:darkenergy}):

\begin{itemize}
\item[1.] {\bf Gravitational waves and primordial power spectrum}.
Our best estimate of the spectral index of a power-law primordial power
	  spectrum of curvature perturbations is $n_s=0.968\pm
	  0.012$~(68\%~CL). 
	  We find
	  no evidence for tensor modes: the 95\%~CL
	  limit is $r<0.24$.\footnote{This is the 7-year
	  \map+BAO+$H_0$ limit. The 5-year \map+BAO+SN limit
	  was $r<0.22$~(95\%~CL). For comparison, the 7-year
	  \map+BAO+SN limit is $r<0.20$~(95\%~CL). These limits do not
	  include systematic errors in the supernova data.} There is no
	  evidence for the running spectral index,
\ensuremath{dn_s/d\ln{k} = -0.022\pm 0.020}~(68\%~CL).
Given that the improvements on $n_s$, $r$, and $dn_s/d\ln k$ from the
	  5-year results are modest, their implications for models of
	  inflation
	  are similar to those discussed in Section~3.3 of
	  \citet{komatsu/etal:2009}. Also see \citet{kinney/etal:2008},
	  \citet{peiris/easther:2008} 
	  and \citet{finelli/etal:2010} for more recent surveys of
	  implications for inflation. In Figure~\ref{fig:nsr}, we compare
	  the 7-year \map+BAO+$H_0$ limits on $n_s$ and $r$ to the
	  predictions from inflation models with monomial potential,
	  $V(\phi)\propto \phi^\alpha$.

\item[2.] {\bf Neutrino properties}. Better determinations of the
	  amplitude of the third
	  acoustic peak of the temperature power spectrum and $H_0$
	  have led to improved limits on the
	  total mass of neutrinos,
\ensuremath{\sum m_\nu < 0.58\ \mbox{eV}\ \mbox{(95\% CL)}}, and
	  the effective number of neutrino species,
\ensuremath{N_{\rm eff} = 4.34^{+ 0.86}_{- 0.88}}~(68\%~CL), both of which are derived from
	  \map+BAO+$H_0$ without any information on the growth of
	  structure.
When BAO is replaced by the LRG power
	  spectrum, we find 
	  \ensuremath{\sum m_\nu < 0.44\ \mbox{eV}\ \mbox{(95\% CL)}},
and the effective number of neutrino species,
\ensuremath{N_{\rm eff} = 4.25^{+ 0.76}_{- 0.80}}~(68\%~CL).

\item[3.] {\bf Primordial helium abundance}. By combining the \map\ data
	  with the small-scale CMB data, we have detected, by more
	  than $3\sigma$, a change
	  in the Silk damping on small angular scales ($l\gtrsim 500$)
	  due to the effect of primordial helium on the temperature
	  power spectrum. We find $Y_p=0.326\pm 0.075$~(68\%~CL).
	  The
	  astrophysical measurements of helium abundance in stars
	  or HII regions provide tight upper limits on $Y_p$, whereas the CMB
	  data can be used to provide a lower limit. With a
	  conservative hard prior on $Y_p<0.3$, we find
	  $0.23<Y_p<0.3$~(68\%~CL). Our detection of helium at $z \sim
	  1000$ contradicts 
	versions of
	the ``cold big bang model,'' where most of the cosmological
	  helium is produced by the first generation 
	of stars \citep{aguirre:2000}.

\item[4.] {\bf Parity violation}. The 7-year polarization data have
	  significantly improved over the 5-year data. This has led to a
	  significantly improved limit on the rotation angle of
	  the polarization plane due to potential parity-violating
	  effects. Our best limit is $\Delta\alpha=-1.1^\circ\pm
1.4^\circ~(\rm statistical)\pm 1.5^\circ~(\rm systematic)$
(68\%~CL).

\item[5.] {\bf Axion dark
	  matter}. The 7-year \map+BAO+$H_0$ limit on the non-adiabatic
	  perturbations
		  that are uncorrelated with curvature perturbations,
	  \ensuremath{\alpha_{0} < 0.077\ \mbox{(95\% CL)}},
	  constrains the parameter space of axion dark matter in the
	  context of the misalignment scenario. It continues to suggest
	  that a future detection of tensor-to-scalar ratio, $r$, at the
	  level of $r=10^{-2}$ would
	  require a fine-tuning of parameters such as the misalignment
	  angle, $\theta<3\times 10^{-9}$, a significant amount of
	  entropy production between the QCD phase transition and the
	  big bang nucleosynthesis, $\gamma<0.9\times 10^{-9}$, a
	  super-Planckian axion decay constant, $f_a>3\times
	  10^{32}$~GeV, an axion contribution to the matter density of
	  the universe being totally
	  sub-dominant, or a combination of all of the above with less
	  tuning in each \citep[also
	  see Section~3.6.3 of][]{komatsu/etal:2009}. The 7-year
	  \map+BAO+$H_0$ limit on correlated isocurvature perturbations,
	  which is relevant to the curvaton dark matter, is
	  \ensuremath{\alpha_{-1} < 0.0047\ \mbox{(95\% CL)}}.

\item[6.] {\bf Dark energy}. With \map+BAO+$H_0$ but without
	  high-redshift Type Ia supernovae, we find
	  \ensuremath{w = -1.10\pm 0.14}~(68\%~CL)
	  for a flat universe. Adding the supernova data reduces the
	  error bar by about a half. For a curved universe, addition of
	  supernova data reduces the error in $w$ dramatically (by a
	  factor of more than four), while the error in curvature is
	  well constrained by \map+BAO+$H_0$. In Figure~\ref{fig:w0wa},
	  we show the 7-year limits on a time-dependent
	  equation of state in the form of $w=w_0+w_a(1-a)$.
          We find $w_0=-0.93\pm 0.13$ and $w_a=-0.41^{+0.72}_{-0.71}$
(68\%~CL) from \map+BAO+$H_0$+SN. 
	  The data are consistent with a flat universe
	  dominated by a cosmological constant.

\item[7.] {\bf Primordial non-Gaussianity}. The 95\%~CL limits on
	  physically-motivated primordial non-Gaussianity parameters are
$-10<\fnlKS<74$, $-214<\fnleq<266$, and $-410<\fnlor<6$. 
When combined with the limit on $\fnlKS$ from {\sl SDSS},
 $-29<\fnlKS<70$ \citep{slosar/etal:2008}, we find $-5<\fnlKS<59$.
The data are
	  consistent with Gaussian primordial curvature perturbations.
\item[8.] {\bf Sunyaev--Zel'dovich effect}. Using the optimal
	  estimator, we have measured the SZ effect towards clusters of
	  galaxies. 
	  We have detected the SZ effect toward the Coma
	  cluster at $3.6\sigma$, and made the statistical detection of 
	  the SZ effect  by optimally stacking the \map\ data at the
	  locations of known clusters of galaxies. By stacking 11 nearby
	  massive clusters, we detect the SZ effect at $6.5\sigma$, and
	  find that the measured SZ signal and the predictions from the
	  X-ray data agree well. On the other hand, we find that the
	  SZ signal from the stacking analysis is about $0.5-0.7$ times
	  the predictions from the current
	  generation of analytical calculations, hydrodynamical
	  simulations, and the ``universal pressure profile'' of
	  \citet{arnaud/etal:2010}. 
	  We detect the expected SZ signal in relaxed clusters that have
	  cool cores. We find that the SZ signal from non-relaxed
	  clusters have SZ signals that are 50\% of the signal predicted  
          by Arnaud et al.'s profile. The discrepancy with theoretical
	  predictions presents a puzzle. 
	  This lower-than-theoretically-expected SZ signal is consistent
	  with the 
	  lower-than-theoretically-expected SZ power spectrum recently
	  measured by the 
	  SPT collaboration 
	  \citep{lueker/etal:2010}. 
	  While we find a better agreement between the \map\ data
	  and the expectations for massive clusters with $M_{500}\gtrsim
	  4\times 10^{14}~h^{-1}~M_\sun$, a significant amount of
	  pressure (relative to theory) is missing
	  in lower mass clusters. Our results
	  imply that we may not fully understand the gas pressure in
	  low-mass clusters. 
	  This issue would become particularly important when
	  the SZ effect is used as a cosmological probe. 
\end{itemize}

We also reported a novel analysis of the \map\ temperature and
polarization data that enable us to  directly ``see'' the imprint of adiabatic scalar fluctuations in
the maps of polarization directions around temperature hot and cold
spots.  These give a striking confirmation of our understanding of the physics
at the decoupling epoch in the form of
radial and tangential polarization patterns at two characteristic
angular scales that are important for the physics of acoustic
oscillation: the compression phase at $\theta=2\theta_A$ and the
reversal phase at $\theta=\theta_A$.

The CMB data have provided us with many stringent constraints on various
properties of our universe. One of many lessons that we have learned
from the CMB data is that, given the data that we have, inventions of
new, physically-motivated, observables beyond the spherically-averaged
power spectrum often lead to new insights into the
physics of the
universe. Well-studied examples include primordial non-Gaussianity
parameters ($f_{\rm NL}$ from the bispectrum),
parity-violation angle ($\Delta\alpha$ from the TB and EB correlations),
modulated primordial power
spectrum \citep[$g(k)$ from direction-dependent power spectra;][see
Bennett et al.~2010 for the 7-year
limits]{ackerman/carroll/wise:2007,hanson/lewis:2009,groeneboom/etal:2010}.

The data continue to improve, including more integration of the \map\
observations. At the same rate, it is important to find more ways to
subject the data to various properties of the universe that have not
been explored yet.

\acknowledgements
The \map\ mission is made possible by the support of the Science Mission
Directorate Office at NASA Headquarters.  This research was additionally
supported by NASA grants NNG05GE76G, NNX07AL75G S01, LTSA03-000-0090,
ATPNNG04GK55G, ADP03-0000-092, and NNX08AL43G, and NSF grants
AST-0807649 and PHY-0758153. 
EK acknowledges support from an Alfred P. Sloan Research Fellowship.
JD is partly supported by an Research Councils UK (RCUK) fellowship.
We thank Mike Greason  for his help on the analysis of the \map\ data, and 
T.~B. Griswold for the artwork. 
We thank D. Jeong for his help on calculating the peak bias presented in
Section~\ref{sec:pol}, B.~A. Reid for discussion on
the treatment of massive neutrinos in the expansion rate which has led
to our exact treatment in Section~\ref{sec:exactnu}, A.~G. Riess for
discussion on the Type Ia supernova data set and the $H_0$ measurement, 
M. Sullivan for discussion on the Type Ia supernova data set, 
S.~H. Suyu and P.~J. Marshall for providing
us with the likelihood function for the time-delay distance and 
discussion on strong lensing measurements, 
A. Vikhlinin for the X-ray data on his nearby cluster samples (used in
Section~\ref{sec:nearby}), 
N. Afshordi, P. Bode, R. Lieu, Y.-T. Lin, D. Nagai, N. Sehgal, L. Shaw
and H. Trac for discussion and 
feedback on Section~\ref{sec:SZ} (SZ effect), and 
F. Takahashi for his help on refining the results on axion dark
matter presented in Section~\ref{sec:AD}.
Computations for the analysis of non-Gaussianity in Section~\ref{sec:NG} were
carried out by the Terascale Infrastructure for Groundbreaking
Research in Engineering and Science (TIGRESS) at 
the Princeton Institute for Computational Science and Engineering
(PICSciE). 
This research has made use of NASA's Astrophysics Data System
Bibliographic Services.  We acknowledge use of the HEALPix
\citep{gorski/etal:2005}, CAMB \citep{lewis/challinor/lasenby:2000}, and
CMBFAST \citep{seljak/zaldarriaga:1996} packages. 

\appendix
\section{Effects of the improved recombination history on the
 $\Lambda$CDM parameters}
\label{app:comparison}
The constraints on the cosmological parameters reported in the original version
of this paper were based on a version of {\sf CAMB} which used a
recombination history calculated by the {\sf RECFAST} version 1.4.2
\citep{seager/sasselov/scott:1999,seager/sasselov/scott:2000,wong/moss/scott:2008,scott/moss:2009}. Shortly
after the submission of the original version, a new version {\sf CAMB}
was released with the {\sf RECFAST} version 1.5. This revision
incorporates the improved treatment of the hydrogen and helium
recombination, following numerous work done over the last several
years \citep[see][and references therein]{rubino-martin/etal:2010}.
Specifically, the code multiplies the ionization fraction, $x_e(z)$,
by a cosmology-independent ``fudge function,'' $f(z)$, found by
\citet{rubino-martin/etal:2010}. 
A change in the recombination history mostly affects the time and duration of
the photon decoupling which, in turn, affects the amount of Silk
damping. Therefore, it is expected to affect the cosmological parameters
such as $n_s$ and $\Omega_bh^2$ \citep{rubino-martin/etal:2010}. 

In order to see the effects of the improved recombination code on the
cosmological parameters, we have re-run the $\Lambda$CDM chain with the
latest {\sf CAMB} code that includes {\sf RECFAST} version 1.5. We find
that the effects are small, and in most cases negligible compared to the
error bars; however, we
find that the significance at which $n_s=1$ is excluded is no longer
more than $3\sigma$: with the improved recombination code, we find
$n_s=0.968\pm 0.012$~(68\%~CL), and $n_s=1$ is excluded at 99.5\%~CL.

Finally, the \map\ likelihood code has also changed from the
initial version (4.0), which used a temperature power spectrum with a
slightly incorrect estimate for the residual point-source amplitude, and
a TE power spectrum with a slightly incorrect $f_{\rm sky}$ factor. The
new version (4.1) corrects both errors; however, the change in the
parameters is largely driven by the above modification of the
recombination history.

Throughout the main body of this paper, we have adopted the new
parameters for the simplest 6-parameter $\Lambda$CDM model, but we
have kept the previous parameters for all the other models because the
changes are too small to report. We compare the $\Lambda$CDM parameters
derived from \map+BAO+$H_0$ in Table~\ref{tab:comparison}. See
\citet{larson/etal:prep} for the comparison of \map-only parameters.

\begin{deluxetable*}{lccccc}
\tablecolumns{6}
\small
\tablewidth{0pt}
\tablecaption{%
Comparison of the $\Lambda$CDM parameters (\map+BAO+$H_0$): RECFAST version
 1.4.2 versus 1.5 
}
\tablehead{\colhead{Class} &
\colhead{Parameter}
&\colhead{ML (1.5)}
&\colhead{ML (1.4.2)}
&\colhead{Mean (1.5)}
&\colhead{Mean (1.4.2)}
}
\startdata
Primary &
$100\Omega_bh^2$
&2.253  & 2.246
&$2.255\pm 0.054$ & \ensuremath{2.260\pm 0.053}
\nl
&
$\Omega_ch^2$
&0.1122  & 0.1120
&$0.1126\pm 0.0036$  & \ensuremath{0.1123\pm 0.0035} 
\nl
&
$\Omega_\Lambda$
&0.728  & 0.728
&$0.725\pm 0.016$  & \ensuremath{0.728^{+ 0.015}_{- 0.016}} 
\nl
&
$n_s$
&0.967  & 0.961
&$0.968\pm 0.012$  & \ensuremath{0.963\pm 0.012} 
\nl
&
$\tau$
&0.085 & 0.087
&$0.088\pm 0.014$ & \ensuremath{0.087\pm 0.014}
\nl
&
$\Delta^2_{\cal R}(k_0)$
&$2.42\times 10^{-9}$  & $2.45\times 10^{-9}$
&$(2.430\pm 0.091)\times 10^{-9}$ & \ensuremath{(2.441^{+ 0.088}_{- 0.092})\times 10^{-9}} 
\nl
\hline
Derived &
$\sigma_8$
&0.810  & 0.807 
&$0.816\pm 0.024$  & \ensuremath{0.809\pm 0.024} 
\nl
&
$H_0$
&$70.4~{\rm km/s/Mpc}$  & $70.2~{\rm km/s/Mpc}$
&$70.2\pm 1.4$~{km/s/Mpc}  & \ensuremath{70.4^{+ 1.3}_{- 1.4}\ \mbox{km/s/Mpc}} 
\nl
&
$\Omega_b$
&0.0455 & 0.0455
&$0.0458\pm 0.0016$  & \ensuremath{0.0456\pm 0.0016} 
\nl
&
$\Omega_c$
&0.226  & 0.227
&$0.229\pm 0.015$  & \ensuremath{0.227\pm 0.014} 
\nl
&
$\Omega_mh^2$
&0.1347  & 0.1344
&$0.1352\pm 0.0036$  & \ensuremath{0.1349\pm 0.0036} 
\nl
&
$z_{\rm reion}$
&10.3 & 10.5
&$10.6\pm 1.2$ & \ensuremath{10.4\pm 1.2}
\nl 
&
$t_0$
&13.76~Gyr & 13.78~Gyr
&$13.76\pm 0.11$~Gyr & \ensuremath{13.75\pm 0.11\ \mbox{Gyr}} 
\enddata
\label{tab:comparison}
\end{deluxetable*}

\section{Stacked Profiles of Temperature and Polarization of the CMB}
\label{app:stacking}
\subsection{Formulae of Stacked Profiles from Peak Theory}

In order to calculate the stacked profiles of temperature and
polarization of the CMB at the locations of temperature peaks, we need to 
relate the peak number density contrast,
$\delta_{\rm pk}$, to the underlying temperature fluctuation, $\Delta
T$. 

One
often encounters a similar problem in the large-scale structure of the
universe: how can we relate the number density contrast of galaxies to the
underlying matter density fluctuation? It is often assumed that the 
number density contrast of peaks with a given peak height $\nu$ is
simply proportional to the 
underlying density field. If one adopted such a linear and {\it
scale-independent} bias 
prescription, one would find\footnote{For convenience, we write the bias
parameters in units of [temperature]$^{-1}$.}
\begin{equation}
\delta_{\rm pk}(\hat{\mathbf n})  = b_\nu \Delta T(\hat{\mathbf n}).
\end{equation}
However, our numerical simulations show that the linear bias does not give
an accurate 
description of $\langle Q_r\rangle$ or $\langle T_r\rangle$. 
In fact, breakdown of the linear bias is precisely what is expected from
the statistics of peaks. From detailed investigations of the statistics
of peaks, \citet{desjacques:2008} found the following {\it
scale-dependent} bias: 
\begin{equation}
\delta_{\rm pk}(\hat{\mathbf n})  = \left[b_\nu -
				     b_\zeta(\partial_1^2+\partial_2^2)\right]
\Delta T(\hat{\mathbf n}). 
\label{eq:bias}
\end{equation}
While the first, constant term $b_\nu$ has been
known for a long time \citep{kaiser:1984,bardeen/etal:1986}, the second
term $b_\zeta$ has been recognized only recently. 
The presence of $b_\zeta$ is expected because, to define peaks, one
needs to use the information on the first and second derivatives of
$\Delta T$. As the first derivative must vanish at the locations of
peaks, the above equation does not contain the first derivative.

\citet{desjacques:2008} has derived the explicit forms
of $b_\nu$ and $b_\zeta$:
\begin{eqnarray}
\label{eq:b1b2}
 b_\nu = \frac1{\sigma_0}\frac{\nu-\gamma\bar{u}}{1-\gamma^2},\qquad
 b_\zeta = \frac1{\sigma_2}\frac{\bar{u}-\gamma\nu}{1-\gamma^2},
\end{eqnarray}
where $\nu\equiv \Delta T/\sigma_0$, $\gamma\equiv
\sigma_1^2/(\sigma_0\sigma_2)$, $\sigma_j$ is 
the r.m.s. of $j$th derivatives of the temperature fluctuation:
\begin{equation}
\label{eq:sigma}
 \sigma^2_j = \frac1{4\pi}\sum_l (2l+1)[l(l+1)]^jC_l^{\rm TT}(W_l^T)^2,
\end{equation}
and $W_l^T$ is the harmonic transform of a window function (which is a
combination of the experimental beam, pixel window, and any other additional
smoothing applied to the temperature data). The quantity $\bar{u}$ is called the
``mean curvature,'' and is given by $\bar{u}\equiv
G_1(\gamma,\gamma\nu)/G_0(\gamma,\gamma\nu)$, where 
\begin{equation}
\label{eq:Gint}
 G_n(\gamma,x_*) \equiv \int_0^\infty dx~x^nf(x)
\frac{\exp\left[-\frac{(x-x_*)^2}{2(1-\gamma^2)}\right]}{\sqrt{2\pi(1-\gamma^2)}}.   
\end{equation}
While \citet{desjacques:2008} applied this formalism 
to a 3-dimensional
Gaussian random field, it is straightforward to generalize his
results to a 2-dimensional case, for which $f(x)$ is given by
\citep{bond/efstathiou:1987},
\begin{equation}
 f(x) = x^2-1+\exp(-x^2).
\end{equation}

With the bias given by equation~(\ref{eq:bias}), we find
\begin{eqnarray}
\nonumber
& &\langle \delta_{\rm
pk}(\hat{\mathbf{n}})Q_r(\hat{\mathbf{n}}+\hat{\bm{\theta}})\rangle
=
\int\frac{d^2{\mathbf l}}{(2\pi)^2}
W_l^TW_l^P(b_\nu+b_\zeta l^2)\\
& &\times 
\left\{C_l^{\rm TE}\cos[2(\phi-\varphi)]
+C_l^{\rm TB}\sin[2(\phi-\varphi)]\right\}e^{i{\mathbf
 l}\cdot{\bm{\theta}}},\\ 
\nonumber
& &\langle \delta_{\rm
pk}(\hat{\mathbf{n}})U_r(\hat{\mathbf{n}}+\hat{\bm{\theta}})\rangle
=
-\int\frac{d^2{\mathbf l}}{(2\pi)^2}
W_l^TW_l^P(b_\nu+b_\zeta l^2)\\
& &\times \left\{C_l^{\rm TE}\sin[2(\phi-\varphi)]
-C_l^{\rm TB}
\cos[2(\phi-\varphi)]\right\}e^{i{\mathbf l}\cdot{\bm{\theta}}},
\end{eqnarray}
where $W_l^T$ and $W_l^P$ are spherical harmonic transforms of the smoothing
functions 
applied to the temperature and polarization data, respectively.
Recalling ${\mathbf l}\cdot{\bm{\theta}}=l\theta\cos(\phi-\varphi)$,
$\int_0^{2\pi}d\varphi\sin[2(\phi-\varphi)]e^{ix\cos(\phi-\varphi)}=0$, and
\begin{eqnarray}
 J_m(x)&=&\int_\alpha^{2\pi+\alpha}\frac{d\psi}{2\pi}e^{i(m\psi-x\sin\psi)},
\end{eqnarray}
with $m=2$, $\psi=\varphi-\phi-\pi/2$ and $\alpha=-\phi-\pi/2$, we find
\begin{eqnarray}
\nonumber
& &\langle\delta_{\rm
pk}(\hat{\mathbf{n}})Q_r(\hat{\mathbf{n}}+\hat{\bm{\theta}})\rangle\\
&=&
-\int\frac{ldl}{2\pi}
W_l^TW_l^P(b_\nu+b_\zeta l^2)C_l^{\rm TE}J_2(l\theta),\\
\nonumber
& &\langle\delta_{\rm
pk}(\hat{\mathbf{n}})U_r(\hat{\mathbf{n}}+\hat{\bm{\theta}})\rangle\\
&=&
-\int\frac{ldl}{2\pi}
W_l^TW_l^P(b_\nu+b_\zeta l^2)C_l^{\rm TB}J_2(l\theta).
\end{eqnarray}
Using these results in equations~(\ref{eq:Qr1}) and (\ref{eq:Ur1}), we
finally obtain the desired formulae for the stacked polarization profiles:
\begin{eqnarray}
\label{eq:Qrapp}
 \langle Q_r\rangle(\theta)
&=& 
-\int\frac{ldl}{2\pi}
W_l^TW_l^P(b_\nu+b_\zeta l^2)C_l^{\rm TE}J_2(l\theta),\\
\label{eq:Urapp}
 \langle U_r\rangle(\theta)
&=& 
-\int\frac{ldl}{2\pi}
W_l^TW_l^P(b_\nu+b_\zeta l^2)C_l^{\rm TB}J_2(l\theta).
\end{eqnarray}
Incidentally, the stacked profile of the temperature fluctuation can
also be calculated in a similar way:
\begin{equation}
\label{eq:Tapp}
 \langle T\rangle(\theta)
= 
\int\frac{ldl}{2\pi}
(W_l^T)^2(b_\nu+b_\zeta l^2)C_l^{\rm TT}J_0(l\theta).
\end{equation}

\subsection{A Cookbook for Computing $\langle Q_r\rangle(\theta)$ 
and $\langle U_r\rangle(\theta)$}

How can we evaluate equations~(\ref{eq:Qrapp})--(\ref{eq:Tapp})? One may
follow the following steps:
\begin{itemize}
 \item[1.] Compute $\sigma_0$, $\sigma_1$, and $\sigma_2$ from
	    equation~(\ref{eq:sigma}). For example, the \map\
	    5-year best-fitting temperature power spectrum for a
	    power-law $\Lambda$CDM model
	    \citep{dunkley/etal:2009}\footnote{We used the 5-year
	    best-fitting power spectrum to calculate the predicted
	    polarization pattern (before the
	    7-year parameter were obtained) and compare
	    it to the 7-year polarization data.}, multiplied by
	    a Gaussian smoothing of 
	    $0.5^\circ$ full-width-at-half-maximum (FWHM) and the pixel
	    window function for the HEALPix resolution of $N_{\rm
	    side}=512$, gives
	    $\sigma_0=87.9~\mu{\rm K}$, $\sigma_1=1.16\times
	    10^4~\mu{\rm K}$, and $\sigma_2=2.89\times 10^6~\mu{\rm K}$. 
 \item[2.] Compute $\gamma=\sigma_1^2/(\sigma_0\sigma_2)$. For the
	    above example, we find $\gamma=0.5306$.
 \item[3.] As we need to integrate over peak heights $\nu$, we need to
	    compute the functions $G_0(\gamma,\gamma\nu)$ and
	    $G_1(\gamma,\gamma\nu)$ for various values of $\nu$. The former
	    function, $G_0(\gamma,\gamma\nu)$, can be found analytically
	    \citep[see equation~(A1.9) of][]{bond/efstathiou:1987}. For
	    $G_1$, we need to integrate equation~(\ref{eq:Gint})
	    numerically. 
\item[4.]  Compute $\bar{u}=G_1/G_0$. For the above
	    example, we find $\bar{u}=1.596$, 1.831, 3.206, and 5.579
	    for $\nu=0$, 1, 5, and 10, respectively. 
\item[5.] Choose a threshold peak height $\nu_t$, and compute the mean
	   surface number density of peaks, 
	   $\bar{n}_{\rm pk}$, from equation~(A1.9) of
	   \citet{bond/efstathiou:1987}:
\begin{equation}
\label{eq:npk}
 \bar{n}_{\rm pk}(\nu_t)
=\frac{\sigma_2^2}{(2\pi)^{3/2}(2\sigma_1^2)}\int d\nu~
e^{-\nu^2/2}G_0(\gamma,\gamma\nu).
\end{equation} 
	   The integration boundary is taken from $\nu_{\rm
	   t}$ to $+\infty$ 
	   for temperature hot spots, and from $-\infty$ to $-|\nu_{\rm
	   t}|$ for temperature cold spots. For the above example,
	   we find $4\pi\bar{n}_{\rm pk}=15354.5$, $8741.5$, $2348.9$,
	   and $247.5$ for $\nu_t=0$, 1, 2, and 3, respectively.
\item[6.] Compute $b_\nu$ and $b_\zeta$ from equation~(\ref{eq:b1b2})
	   for various values of $\nu$.
 \item[7.] Average $b_\nu$ and $b_\zeta$ over $\nu$. We calculate the
	   averaged bias parameters, $\bar{b}_\nu$ and $\bar{b}_\zeta$,
	   by integrating $b_\nu$ and $b_\zeta$ multiplied by the number
	   density of peaks for a given $\nu$. We then divide the integral
	   by the mean number density of peaks, 
	   $\bar{n}_{\rm pk}$, to find
\begin{eqnarray}
 \bar{b}_\nu &=& \frac1{\bar{n}_{\rm pk}(\nu_t)}
\frac{\sigma_2^2}{(2\pi)^{3/2}(2\sigma_1^2)}\int d\nu~
e^{-\nu^2/2}G_0(\gamma,\gamma\nu)b_\nu,\\
\bar{b}_\zeta &=& \frac1{\bar{n}_{\rm pk}(\nu_t)}
\frac{\sigma_2^2}{(2\pi)^{3/2}(2\sigma_1^2)}\int d\nu~
e^{-\nu^2/2}G_0(\gamma,\gamma\nu)b_\zeta.
\end{eqnarray}
The integration boundary is taken from $\nu_{\rm
	   t}$ to $+\infty$ 
	   for temperature hot spots, and from $-\infty$ to $-|\nu_{\rm
	   t}|$ for temperature cold spots.
For the above example, we find $(\bar{b}_\nu,\bar{b}_\zeta)=
	   (3.289\times 10^{-3}$, $6.039\times 10^{-7})$, ($1.018\times
	   10^{-2}$, $5.393\times 10^{-7}$), ($2.006\times 10^{-2}$,
	   $4.569\times 10^{-7}$), and ($3.128\times 10^{-2}$,
	   $3.772\times 10^{-7}$)
	   for $\nu_{\rm
	   t}=0$, 1, 2, and 3, respectively (all in units of $\mu{\rm
	   K}^{-1}$). The larger the peak height
	   is, the larger the linear bias and the smaller the
	   scale-dependent bias becomes.
\item[8.] Use $\bar{b}_\nu$ and $\bar{b}_\zeta$ in
	   equations~(\ref{eq:Qrapp}) and (\ref{eq:Urapp}) to compute $\langle
	   Q_r\rangle(\theta)$ and $\langle U_r\rangle(\theta)$ for a
	   given set of $C_l^{\rm TE}$ and $C_l^{\rm TB}$, respectively.
\end{itemize}
Very roughly speaking, the bias takes on the following values:
\begin{eqnarray}
\nonumber
 \bar{b}(l)&\equiv& \bar{b}_\nu + \bar{b}_\zeta l^2\\
&\sim& 
\begin{cases}
\frac{0.3}{100~\mu{\rm K}}
\left[1+\left(\frac{l}{75}\right)^2\right] & (\mbox{for}~\nu_t=0)\\
\frac{3}{100~\mu{\rm K}}
\left[1+\left(\frac{l}{290}\right)^2\right] & (\mbox{for}~\nu_t=3)
\end{cases}.
\end{eqnarray}
The scale dependence of bias
becomes important at $l\sim 75$ for $\nu_t=0$, but the higher peaks are
closer to having a linear bias on large scales. One may also rewrite this 
using the stacked temperature values at the center, $\langle
T\rangle(0)=(107.0$, $151.4$, $216.4$, $292.1)~\mu{\rm K}$ for
$\nu_t=0$, 1, 2, and 3:
\begin{equation}
 \bar{b}(l)\simeq \frac{(0.35, 1.5, 4.3, 9.1)}{(107, 151, 216,
  292)\mu{\rm K}}
\left[1+\left(\frac{l}{(74,137,219,288)}\right)^2\right].
\end{equation}

\section{Optimal Estimator for SZ Stacking}
\label{app:sz}
\subsection{Optimal estimator}
In this Appendix, we describe an optimal likelihood-based method for estimating the stacked SZ profile
around clusters whose locations are taken from external catalogs.

Formally, we can set up the problem as follows.  We represent the \map\ data as a vector of length $(\Nchan\times \Npix)$ and denote it by $d_{\nu p}$,
where the index $\nu=1,\cdots,\Nchan$ ranges over channels, and the index $p=1,\cdots,\Npix$ ranges over sky pixels.  (We typically
take $\Nchan=6$ corresponding to V1, V2, W1, W2, W3, and W4; and $\Npix
= 12(2^{10})^2$ corresponding to a HEALPix resolution of $N_{\rm side}=1024$.)
We model the \map\ data as a sum of CMB, noise, and SZ contributions as follows:
\begin{equation}
d_{\nu p} = \sum_{\ell m} A_{\nu p, \ell m} a_{\ell m} + n_{\nu p} + \sum_{\alpha=1}^{\Nt} p_\alpha (t_{\alpha})_{\nu p}.  \label{eq:sz_dsplit}
\end{equation}
In this equation, we have written the SZ contribution as a sum of $\Nt$ template maps, $t_1,\cdots,t_{\Nt}$, whose coefficients $p_\alpha$
are free parameters to be determined.
The operator, $A_{\nu p, \ell m}$, in Eq.~(\ref{eq:sz_dsplit}) converts a harmonic-space CMB realization, $a_{\ell m}$, into a set of maps with 
black-body frequency dependence and channel-dependent beam convolution.
More formally, the matrix element $A_{\nu p, \ell m}$ is defined by
\begin{equation}
A_{\nu p, \ell m} = b_{\nu\ell} Y_{\ell m}(p),
\end{equation}
where $b_{\nu\ell}$ is the beam transfer function (including HEALPix window function) for channel $\nu$.

The specific form of the template maps, $t_\alpha$, will depend on the type of profile reconstruction which is desired.
For example, if we want to estimate a stacked amplitude for the SZ signal in $N$ angular bins, we define one template for each bin.
If the bin corresponds to angular range $\theta_{\rm min} \le \theta \le \theta_{\rm max}$, we define a map $m_p$ which is =1 if the
angular distance $\theta$ between pixel $p$ and {\em some} galaxy cluster in the catalog is in the range 
$\theta_{\rm min} \le \theta \le \theta_{\rm max}$, and zero otherwise.
We convolve this map with the beam in each channel $\nu$ and multiply by
the SZ frequency dependence to obtain the template $t_{\nu p}$.
As another example, if we want to fit for an overall multiple of a fiducial model $m_p$ for the total SZ signal (summed over all clusters) then
we define a single (i.e. $\Nt=1$) template $t_{\nu p}$ by applying beam
convolution and the SZ frequency dependence for each channel $\nu$.

Given this setup, we would like to compute the likelihood function $\L[p_\alpha | d_{\nu p}]$ for the profile $p_\alpha$, given the
noisy data $d_{\nu p}$, marginalizing over the CMB realization.
We assume a fixed fiducial model $C_\ell$ and represent the CMB signal covariance by a (diagonal) matrix $S_{\rm harm}$ in harmonic space:
\begin{equation}
(S_{\rm harm})_{\ell m, \ell' m'} = C_\ell \delta_{\ell\ell'} \delta_{mm'}.
\end{equation}
We represent the noise covariance by an (also diagonal) pixel-matrix $N_{\nu p, \nu' p'}$:
\begin{equation}
(N_{\rm pix})_{\nu p, \nu' p'} = \sigma^2_{\nu p} \delta_{\nu\nu'} \delta_{pp'}.
\end{equation}
The joint (CMB,SZ) likelihood function can now be written (up to an overall normalizing constant):
\begin{equation}
\L[a,p|d] \propto \exp\left[ -\frac{1}{2} a^T S_{\rm harm}^{-1} a -
		       \frac{1}{2} (d - Aa - p_\alpha t_\alpha)^T N^{-1}
		       (d-Aa - p_\alpha t_\alpha) \right].
\label{eq:likelihood_sz}
\end{equation}
(In this equation we have omitted some indexes for notational
compactness, e.g. $d_{\nu p} \rightarrow d$ and $a_{\ell m} \rightarrow
a$. The summation over $\alpha$ is assumed.)
We can now integrate out the CMB realization $a$ to obtain the marginalized likelihood for the profile:
\begin{eqnarray}
\L[p|d] &=& \int Da\, \L[a,p|d] \\
        &\propto& \exp\left[ -\frac{1}{2} (p_\alpha - \hp_\alpha)^T F_{\alpha\beta} (p_\beta - \hp_\beta) \right],  \label{eq:sz_p_d}
\end{eqnarray}
where we have defined the $(\Nt)$-by-$(\Nt)$ matrix
\begin{equation}
F_{\alpha\beta} = (t_\alpha)_{\nu' p'} [ N_{\rm pix} + A S_{\rm harm} A^T ]^{-1}_{\nu'p',\nu p} (t_\beta)_{\nu p},  \label{eq:f_def}
\end{equation}
and the length-$(\Nt)$ vector
\begin{equation}
\hp_\alpha = F_{\alpha\beta}^{-1} (t_\beta)_{\nu' p'} [ N_{\rm pix} + A S_{\rm harm} A^T ]^{-1}_{\nu'p',\nu p} d_{\nu p}.  \label{eq:hp_def}
\end{equation}
The likelihood function $\L[p|d]$ in Eq.~(\ref{eq:sz_p_d}) has a simple interpretation.  The likelihood for $p_\alpha$ is 
a Gaussian with mean $\hp_\alpha$ and covariance matrix $F_{\alpha\beta}^{-1}$.  This is the main result of this section, and
is the basis for all our SZ results in the body of the paper.  For example, when we reconstruct the stacked SZ profile
in angular bins, the estimated profile in each bin $\alpha$ is given by $\hp_\alpha$ and the $1\sigma$ error is given by
$\sqrt{(F^{-1})_{\alpha\alpha}}$.

It is worth mentioning that the statistic $\hp_\alpha$ also appears naturally if we use an estimator framework rather
than a likelihood formalism.
If we think of $\hp_\alpha$, defined by
Eq.~(\ref{eq:hp_def}), as an estimator for the profile given the data $d$, then one can verify that the estimator is unbiased
(i.e. $\langle \hp_\alpha \rangle = p_\alpha$, where the expectation value is taken over random CMB + noise realizations
with a fixed SZ contribution) and its covariance is $F_{\alpha\beta}^{-1}$.  
Conversely, it is not hard to show that $\hp_\alpha$ is the unbiased estimator with minimum variance, thus obtaining
$\hp_\alpha$ in a different way.
This alternate derivation also shows that the error bars on the profile obtained in our likelihood formalism are the
same as would be obtained in a direct Monte Carlo treatment.

Either from the likelihood or estimator formalism, one sees that the statistic $\hp_\alpha$ is optimal.
In the limit where all frequency channels are in the Rayleigh-Jeans regime,
the statistic $\hp_\alpha$ corresponds to $C^{-1}$-filtering the data and multiplying by each template map.
In this case, the $C^{-1}$ filter acts as a high-pass filter which optimally suppresses CMB power on scales larger than
the clusters, and also optimally weights the channels (in a way which is $\ell$-dependent if the beams differ).
When channels with higher frequency are included, the statistic $\hp_\alpha$ would get most of its information from linear
combinations of channels which contain zero CMB signal, but nonzero
response to an SZ signal.
(Such a combination of channels does not need to be high-pass filtered,
increasing its statistical weight.)

For the V+W combination in \map, {\it the $N^{-1}$-filtered (V$-$W) null map
is used to separate the SZ effect and CMB}, as CMB is canceled in this
map while the SZ is effect not. 

We conclude with a few comments on implementation.
Inspection of Eqs.~(\ref{eq:f_def}) and~(\ref{eq:hp_def}) shows that 
it would be straightforward to compute $F_{\alpha\beta}$ and $\hp_\alpha$, given an algorithm for multiplying a
set of $\Nchan$ pixel space maps $d_{\nu p}$ by the operator $[ N_{\rm pix} + A S_{\rm harm} A^T ]^{-1}$.
A fast multigrid-based algorithm for this inverse problem was found in
\citet{smith/zahn/dore:2007} but there is one small wrinkle in the
implementation: in \citet{smith/zahn/dore:2007} the problem
was formulated in harmonic space and an algorithm was given for
multiplying by the operator $[S_{\rm harm}^{-1} + A^T N_{\rm pix}^{-1} A]^{-1}$.
However, the matrix identity
\begin{equation}
[ N_{\rm pix} + A S_{\rm harm} A^T ]^{-1} = N_{\rm pix}^{-1} - A [ S_{\rm harm}^{-1} + A^T N_{\rm pix}^{-1} A ]^{-1} A^T,  \label{eq:sz_resum}
\end{equation}
allows us to relate the two inverse problems.
In fact there is another advantage to using the expression on the RHS of Eq.~(\ref{eq:sz_resum}): because the
inverse noise, $N_{\rm pix}^{-1}$, appears instead of the noise covariance $N_{\rm pix}$, a galactic mask 
can be straightforwardly included in the analysis by zeroing the matrix entries of $N_{\rm pix}^{-1}$
which correspond to masked pixels, so that the pixels are treated as
infinitely noisy.

\section{Pressure Profiles}
\label{sec:pressure}
\subsection{Pressure profile from X-ray observations}
\label{sec:arnaud}

Recently, \citet{arnaud/etal:2010} found that the following parametrized
phenomenological electron pressure profile, which is based on a ``generalized
Navarro-Frenk-White profile'' proposed by
\citet{nagai/kravtsov/vikhlinin:2007}, fits the electron pressure profiles
directly derived from X-ray data of clusters well
\citep[see equation~(13) of][]{arnaud/etal:2010}:
\begin{eqnarray}
\nonumber
 P_e(r) &=& 1.65~(h/0.7)^2~{\rm eV~cm^{-3}}\\
& \times& E^{8/3}(z)\left[\frac{M_{500}}{3\times
				    10^{14}(0.7/h)M_\sun}\right]^{2/3+\alpha_p}
 p(r/r_{500}),
\end{eqnarray}
where $\alpha_p=0.12$, $E(z)\equiv H(z)/H_0 =
\left[\Omega_m(1+z)^3+\Omega_\Lambda\right]^{1/2}$ for a $\Lambda$CDM
model, 
$r_{500}$ is the radius within which the mean overdensity is 500 times
the critical density of the universe, $\rho_c(z)=2.775\times
10^{11}E^2(z)~h^{2}~M_\sun~{\rm Mpc}^{-3}$, and $M_{500}$ is the mass
enclosed within $r_{500}$:
\begin{equation}
 M_{500}\equiv \frac{4\pi}3[500\rho_c(z)]r_{500}^3.
\label{eq:m500}
\end{equation}
The function $p(x)$ is defined by
\begin{equation}
p(x) \equiv \frac{8.403(0.7/h)^{3/2}}{(c_{500}
 x)^\gamma[1+(c_{500}x)^\alpha]^{(\beta-\gamma)/\alpha}},
\label{eq:gnfw}
\end{equation}
where $c_{500}=1.177$, $\alpha=1.051$, $\beta=5.4905$, and $\gamma=0.3081$. 

The SPT collaboration stacked the SZ maps of 11 known clusters and
fitted the stacked SZ radial profile to the above form, finding 
$c_{500}=1.0$, $\alpha=1.0$, $\beta=5.5$, and $\gamma=0.5$
\citep{plagge/etal:2010}.
While they
did not compare the overall amplitude (which is the focus of our
analysis), the {\it shape} of the pressure profile 
found by the SPT collaboration (using the SZ data) is in an excellent
agreement with that found by \citet{arnaud/etal:2010} (using the X-ray data).

\subsection{Pressure profile from hydrostatic equilibrium}
\label{sec:ks}

The KS profile builds on and extends the idea originally put forward by 
\citet{makino/sasaki/suto:1998} and \citet{suto/sasaki/makino:1998}:
(i) gas is in hydrostatic equilibrium with gravitational potential
given by a Navarro-Frenk-White (NFW) dark matter density profile
\citep{navarro/frenk/white:1997} and (ii) the equation of state of gas
is given by a polytropic form, $P\propto \rho^\gamma$. However, this
model still contains two free parameters: a polytropic index $\gamma$
and the normalization 
of $P$. \citet{komatsu/seljak:2001} found that an additional, physically
reasonable assumption that (iii) the slope of the gas density profile
and that of the dark matter density profile agree at around the virial
radius, uniquely fixes $\gamma$, leaving only one free parameter: the
normalization of $P$.
These assumptions are supported by hydrodynamical simulations of
clusters of galaxies, and the resulting shape of the KS profile indeed
agrees  with simulations reasonably well
(see, however, Section~\ref{sec:sz_discussion} for a discussion on the
shape of the profile in the outer parts of clusters).

Determining the normalization of the KS profile requires an additional
assumption, 
described below. Also note that this model does not take into account
any non-thermal pressure (such as $\rho v^2$ where $v$ is the bulk or
turbulent velocity), gas cooling, or star formation
\citep[see, e.g.,][and references therein for various attempts to
incorporate more gas
physics]{bode/ostriker/vikhlinin:2009,frederiksen/etal:2009}.

The KS {\it gas} pressure profile is given by \citep[see Section~3.3
of][for more detailed descriptions]{komatsu/seljak:2002} 
\begin{equation}
 P_{\rm gas}(r) = P_{\rm gas}(0)[y_{\rm gas}(r/r_{\rm s})]^\gamma.
\end{equation}
The {\it electron} pressure profile, $P_e$, is then given by
$P_e=[(2+2X)/(3+5X)]P_{\rm gas}=0.518P_{\rm gas}$ for $X=0.76$.
Here, $r_{\rm s}$ is the so-called ``scale radius'' of the NFW profile, and
a function $y_{\rm gas}(x)$ is defined by
\begin{equation}
 y_{\rm gas}(x)\equiv \left\{
1-B\left[1-\frac{\ln(1+x)}{x}\right]
\right\}^{1/(\gamma-1)},
\end{equation}
with 
\begin{equation}
 B\equiv 3\eta^{-1}(0)\frac{\gamma-1}{\gamma}
\left[\frac{\ln(1+c)}{c}-\frac1{1+c}\right]^{-1},
\end{equation}
\begin{equation}
 \gamma = 1.137 + 8.94\times 10^{-2}\ln(c/5) - 3.68\times 10^{-3}(c-5),
\end{equation}
and
\begin{equation}
 \eta(0)=2.235+0.202(c-5)-1.16\times 10^{-3}(c-5)^2.
\end{equation}
Here, $c$ is the so-called ``concentration parameter'' of the NFW
profile, which is related to the scale radius, $r_{\rm s}$, via $c=r_{\rm
vir}/r_{\rm s}$, and $r_{\rm vir}$ is the virial radius. 
The virial radius gives the virial mass, $M_{\rm vir}$, as
\begin{equation}
 M_{\rm vir} = \frac{4\pi}3[\Delta_c(z)\rho_c(z)]r_{\rm vir}^3.
\label{eq:mvir}
\end{equation}
Here, $\Delta_c(z)$ depends on $\Omega_m$ and $\Omega_\Lambda$ as
\citep{bryan/norman:1998}
\begin{equation}
 \Delta_c(z) = 18\pi^2+82[\Omega(z)-1]-39[\Omega(z)-1]^2,
\end{equation}
where $\Omega(z)=\Omega_m(1+z)^3/E^2(z)$
\citep[also see][for other fitting
formulae]{lacey/cole:1993,nakamura/suto:1997}. For $\Omega_m=0.277$,
one finds $\Delta_c(0)\simeq 98$.

The central gas pressure, $P_{\rm gas}(0)$, is given by 
\begin{equation}
 P_{\rm gas}(0) = 55.0~h^2~{\rm eV~cm^{-3}}
\left[\frac{\rho_{\rm gas}(0)}{10^{14}~h^2~M_\sun~{\rm Mpc}^{-3}}\right]
\left[\frac{k_BT_{\rm gas}(0)}{8~{\rm keV}}\right],
\end{equation}
where $k_B$ is the the Boltzmann constant. The 
central gas temperature, $T_{\rm gas}(0)$, is given by
\begin{equation}
 k_BT_{\rm gas}(0)=8.80~{\rm keV}~\eta(0)
\left[\frac{M_{\rm vir}/(10^{15}~h^{-1}~M_\sun)}{r_{\rm
 vir}/(1~h^{-1}~{\rm Mpc})}\right].
\end{equation}

The central gas density, $\rho_{\rm gas}(0)$, will be determined such
that the gas density at the virial radius is the cosmic mean baryon
fraction, $\Omega_b/\Omega_m$, times
the dark matter density at the same radius. {\it This is an assumption.}
In fact, the cosmic mean merely provides an upper limit on the baryon
fraction of clusters, and thus we expect the gas pressure to be less
than what is given here. How much less needs to be determined from
observations (or possibly from more detailed modeling of the intracluster
medium). In any case, with this assumption, we find  
\begin{eqnarray}
\nonumber
 \rho_{\rm gas}(0) &=& 7.96\times 10^{13}~h^2~M_\sun~{\rm Mpc}^{-3}\\
\nonumber
& &\times \left(\frac{\Omega_b}{\Omega_m}\right)
\frac{M_{\rm vir}/(10^{15}~h^{-1}~M_\sun)}{[r_{\rm
 vir}/(1~h^{-1}~{\rm Mpc})]^3}\\
& &\times \frac{c^2}{(1+c)^2}\frac1{y_{\rm gas}(c)}
\left[\ln(1+c)-\frac1{1+c}\right]^{-1}.
\end{eqnarray}
This equation fixes a typo in equation~(21) of \citet{komatsu/seljak:2002}.

The virial radius, $r_{\rm vir}$, is approximately given by $2r_{500}$;
thus, $M_{\rm vir}$ is approximately given by $8\Delta_c/500\simeq
1.6$. However, the exact relation depends on the mass \citep[see, e.g.,
Figure~1 of][]{komatsu/seljak:2001}. We calculate the mass within a
given radius, $r$, by integrating the NFW density profile
\citep{navarro/frenk/white:1997}:
\begin{equation}
 \rho_{\rm NFW}(r) = \frac{\rho_{\rm s}}{(r/r_{\rm s})(1+r/r_{\rm s})^2}.
\end{equation}
Specifically, for a given $M_{500}$ and $r_{500}$, we solve the
following non-linear equation for $M_{\rm vir}$:
\begin{equation}
 M_{\rm vir}\frac{m(cr_{500}/r_{\rm vir})}{m(c)} = M_{500},
\end{equation}
where $m(x)\equiv \ln(1+x)-x/(1+x)$.
Here, $r_{\rm vir}$ is related to $M_{\rm vir}$ via
equation~(\ref{eq:mvir}). We also need a relation between the
concentration parameter, $c$, and $M_{\rm
vir}$. \citet{komatsu/seljak:2002} used 
\begin{eqnarray}
\nonumber
 c_{\rm seljak} &=& \frac{10}{1+z}\left(\frac{M_{\rm vir}}{3.42\times
		      10^{12}~h^{-1}~M_\sun}\right)^{-0.2}\\
&=& \frac{5.09}{1+z}\left(\frac{M_{\rm
		    vir}}{10^{14}~h^{-1}~M_\sun}\right)^{-0.2},
\label{eq:c_seljak}
\end{eqnarray}
which was adopted from \citet{seljak:2000}.

Recently, \citet{duffy/etal:2008} ran larg
e $N$-body simulations with
the \map\ 5-year cosmological parameters to find a more accurate fitting
formula for the concentration parameter: 
\begin{eqnarray}
\nonumber
 c_{\rm duffy} &=& \frac{7.85}{(1+z)^{0.71}}\left(\frac{M_{\rm vir}}{2\times
		      10^{12}~h^{-1}~M_\sun}\right)^{-0.081}\\
&=& \frac{5.72}{(1+z)^{0.71}}\left(\frac{M_{\rm
		    vir}}{10^{14}~h^{-1}~M_\sun}\right)^{-0.081}.
\label{eq:c_duffy}
\end{eqnarray}
This formula makes clusters of galaxies ($M_\sun\gtrsim 10^{14}~M_\sun$)
more concentrated than $c_{\rm seljak}$ would predict. 


\end{document}